\documentclass[a4paper, 12pt]{article}
\usepackage[top=2cm, bottom=2cm, left=2.5cm, right=2.5cm]{geometry}
\usepackage{amssymb}
\usepackage{amsmath}
\usepackage{amsfonts}
\usepackage{bm}
\usepackage{mathtools}
\usepackage{bbm}
\usepackage{hyperref}
\usepackage{apacite}
\usepackage{color}
\usepackage{hyperref}
\usepackage{apacite}
\usepackage{setspace}
\usepackage{changes}
\usepackage{multirow}
\usepackage{longtable}
\usepackage{supertabular} 
\usepackage{dcolumn}
\usepackage{pdflscape}
\usepackage[shortlabels]{enumitem}
\usepackage{tabularx,calc}
\usepackage{booktabs}
\usepackage{acro}
\usepackage{afterpage}
\usepackage{lscape}
\usepackage{mathtools}
\mathtoolsset{showonlyrefs} 
\usepackage{pst-plot, pstricks} 
\usepackage{fancybox,color}
\usepackage{subfigure}
\usepackage{framed}
\usepackage{scalerel}[2016-12-29]
\usepackage{titling}
\usepackage[bottom]{footmisc} 
\usepackage[]{authblk}
\usepackage[title]{appendix}
\usepackage{caption}
\usepackage{makecell}
\usepackage{float}
\usepackage{tikz}
\usetikzlibrary{arrows,backgrounds}
\usepgflibrary{shapes.multipart}
\usepackage{xcolor}

\usepackage{siunitx}	
\usepackage{changes}
\definechangesauthor[name={Max}, color=orange]{Max}
\definechangesauthor[name={Stephan}, color=blue]{Stephan}

\newtheorem{preremark}{Remark}
\newenvironment{remark}%
  {\begin{preremark}\upshape}{\end{preremark}}
  
\DeclareMathOperator*{\argmax}{arg\,max}

\AtBeginDocument{%

}

\definecolor{Diceblue}{RGB}{52,116,181}

\hypersetup{
	colorlinks = true,
	linkcolor=Diceblue,
	citecolor=Diceblue,
	urlcolor=Diceblue
}

\begin{document}

\begin{titlingpage}

\title{Deep Learning for the Estimation of Heterogeneous Parameters in Discrete Choice Models}	
	
\author[$a$]{Stephan Hetzenecker}
\affil[$a$]{Ruhr Graduate School in Economics \& \protect\\ University of Duisburg-Essen}
\author[$b$]{Maximilian Osterhaus}
\affil[$b$]{University of Groningen}

\date{June 2022}
	

\maketitle

\begin{abstract}

This paper studies the finite sample performance of the flexible estimation approach of \citeA{farrell2020}, who propose to use deep learning for the estimation of heterogeneous parameters in economic models, in the context of discrete choice models. The approach combines the structure imposed by economic models with the flexibility of deep learning, which assures the interpretebility of results on the one hand, and allows estimating flexible functional forms of observed heterogeneity on the other hand. For inference after the estimation with deep learning, \citeA{farrell2020} derive an influence function that can be applied to many quantities of interest. We conduct a series of Monte Carlo experiments that investigate the impact of regularization on the proposed estimation and inference procedure in the context of discrete choice models.
The results show that the deep learning approach generally leads to precise estimates of the true average parameters and that regular robust standard errors lead to invalid inference results, showing the need for the influence function approach for inference.
Without regularization, the influence function approach can lead to substantial bias and large estimated standard errors caused by extreme outliers. 
Regularization reduces this property and stabilizes the estimation procedure, but at the expense of inducing an additional bias. The bias in combination with decreasing variance associated with increasing regularization leads to 
the construction of invalid inferential statements in our experiments. 
Repeated sample splitting, unlike regularization, stabilizes the estimation approach without introducing an additional bias, thereby allowing for the construction of valid inferential statements.

		

\vspace{0.4cm}
		

\end{abstract}

\vfill
\noindent\hrulefill \\
{\footnotesize{
$^a$ Ruhr Graduate School in Economics, RWI - Leibniz Institute for Economic Research, Hohenzollernstrasse 1-3, 45128 Essen, Germany and University of Duisburg-Essen, Universitätsstraße 12, 45117 Essen, Germany.   Email: \href{mailto:stephan.hetzenecker@vwl.uni-due.de}{stephan.hetzenecker@vwl.uni-due.de} \\
$^b$ University of Groningen, P.O. Box 800, 9700 AV Groningen, The Netherlands. Email: \href{mailto:m.k.osterhaus@rug.nl}{m.k.osterhaus@rug.nl}\\
Correspondence to:  \href{mailto:stephan.hetzenecker@vwl.uni-due.de}{stephan.hetzenecker@vwl.uni-due.de} (S. Hetzenecker), \href{mailto:m.k.osterhaus@rug.nl}{m.k.osterhaus@rug.nl} (M.Osterhaus).
}}

\end{titlingpage}

\onehalfspacing
\newpage

\section{Introduction}

Appropriately modeling heterogeneity across economic agents is a key challenge in many empirical economic studies. 
Often, the heterogeneity can be linked to observed characteristics of agents. This is typically achieved using parametric specifications 
in the form of linear interactions of only a few observed characteristics with the variables of interest. Even  restrictive functional forms like linear functions rapidly lead to a large number of parameters, especially if the heterogeneity is modeled as a function of multiple characteristics \cite{cranenburgh2021}. Furthermore, limiting the heterogeneity to linear functions of  only few  characteristics can  lead to  misspecification of the true shape and extent of heterogeneity, and  to potentially incorrect results for quantities of interest, such as elasticities or willingness-to-pay measures.

The increasing availability of large data sets makes it possible to reduce the reliance on parametric methods and to apply more flexible approaches to study heterogeneity. A promising tool for this task is deep learning, which is known for its ability to flexibly model functional forms  and to handle large amounts of data. While deep learning so far has been applied with great success to pure prediction tasks \cite{LeCun2015}, \citeA{farrell2020} propose to employ deep learning for the estimation of heterogeneous parameters. They incorporate the heterogeneity across economic agents into the economic model specified by the researcher through coefficients that are functions of agents' observed characteristics. The approach combines parametric approaches -- which impose structure on the model grounded in economic principles and  reasoning -- with deep learning -- which lets the data speak for itself with its flexibility.

To derive theoretically valid inferential statements after estimating the coefficient functions with deep learning, \citeA{farrell2020} extend the deep learning theory for generic regression approaches developed by \citeA{Farrell2021_econometrica} to M-estimators. Building on \citeA{cherno2018}, they derive an influence function that makes inference feasible in a wide range of settings -- the provided inferential statements cover any parameter of interest that is a function of the heterogeneous coefficient functions. \citeA{farrell2020} show  that the inference procedure allows to construct valid inferential statements under fairly weak conditions. However, they leave the role of regularization and its consequences for estimation and subsequent inference for future research.

Conducting a series of Monte Carlo experiments, we intend to fill this gap and study the finite sample properties of the proposed inference procedure in the context of discrete choice models. The results of these experiments show that deep learning generally is well suited for the estimation of heterogeneous parameters, especially if the sample size is sufficiently large, and that naive inference after estimating the parameters with deep learning leads to invalid inference. 
Further, the proposed estimation procedure is sensitive to overfitting when no regularization is used. We observe that  estimation without regularization can result in substantial bias and large estimated standard errors. The sensitivity to overfitting is more pronounced in small samples but does not completely disappear with increasing sample size. 
Regularization in the form of  $l_2$-penalties on the weights tuned in the network reduces the sensitivity to overfitting and rapidly decreases the average estimated standard errors.
However, it also appears to introduce a new source of bias, which in combination with the decreasing variance explains the poor coverage of the estimated confidence intervals observed in our experiments.
Finally, the experiments show that substantially better results are obtained when repeated sample splitting is used. Unlike regularization, repeated sample splitting substantially reduces the bias arising from overfitting  without inducing a new bias, thus leading to valid inferential results in our experiments.

Our paper contributes to a growing literature on the combination of deep learning and structural modeling in discrete choice models.\footnote{For recent surveys of the application of machine learning and deep learning for the estimation of discrete choice models, see, e.g., \citeA{karlaftis2011}, \citeA{hess2021}, and \citeA{cranenburgh2021}.} Among others, \citeA{sifringer2020} and \citeA{wong2021}  apply deep learning to estimate demand for travel modes in a logit framework. 
To avoid model misspecification in discrete choice models, \citeA{sifringer2020} propose to decompose the utility into two parts: a knowledge-driven part which includes the variables of interest and  is specified by the researcher, and a data-driven part, which is estimated with deep learning using the remaining explanatory variables that are not of primary interest. Separating those two parts of the utility assures that the parameters of interest can be interpreted. However, as the knowledge-driven part needs to be fully specified, its coefficients are constant across  agents. Therefore,  this approach seems more restrictive than the approach of   \citeA{farrell2020} which allows for heterogeneous coefficients. 
In contrast,  \citeA{wong2021} allow for a knowledge-driven part of the utility and an additional random component of the utility which can depend on the characteristics of all alternatives. That is,  their approach captures unobserved heterogeneity and  cross-effects of non-linear utilities across all alternatives. Thus, their model relaxes the IIA property. 
Both have in common that they do not provide a theoretically valid inference procedure for parameters of  interest but rely on  approximations of the confidence intervals based on the Hessian of the estimated model, which are not guaranteed to be of the correct size.
\citeA{wang2020} focus on estimating economic quantities of interest, e.g., market shares, elasticities, and changes in social welfare, with deep learning using a completely unstructured utility. Similar to \citeA{sifringer2020} and \citeA{wong2021}, they do not present a valid approach for inference on the quantities of interest.\footnote{For example, they calculate the standard deviation of the average elasticity as the standard deviation of the elasticity of each individual.} They rely on the predicted choice probabilities and the gradient of the estimated model and do not take into account that the considered quantities are accompanied with additional uncertainty when no structure is imposed on the utility.

The remainder of this paper is organized as follows. Section \ref{secDL:deep_learning_logit} illustrates how deep learning can be employed to estimate heterogeneous parameters in economic models and outlines the inference and estimation procedure. Section \ref{secDL:Monte_Carlo} presents  Monte Carlo experiments that study the inference procedure and Section \ref{secDL:Application} applies the influence function approach to real data. Section \ref{secDL:conclusion} concludes.

\section{Deep Learning for Heterogeneity}\label{secDL:deep_learning_logit}

This section introduces the methodical framework of \citeA{farrell2020} who propose to estimate heterogeneous parameters in econometric models using deep learning in the form of multi-layer feed-forward neural networks. The flexibility of deep neural networks (DNNs) makes them ideally suited for the estimation of  economic models with individual heterogeneity. Subsection \ref{secDL:network_arch} explains the design of the network which directly integrates the economic model specified by the researcher into the network architecture. Subsection \ref{secDL:inference} explains the inference approach which is based on the concept of influence functions, and Subsection \ref{secDL:inference_estimation} lays out the estimation procedure. While the estimation and inference procedure is applicable to a wide range of models, we focus on multinomial discrete choice models when introducing the estimation procedure.

\subsection{Deep Learning}\label{secDL:network_arch}

Starting point of the estimation approach is the economic model specified by the researcher. The model relates the outcome $\bm{Y}$ to the variables of interest $\bm{X}$, and to socio-demographic characteristics $\bm{W}$ that are included to capture the heterogeneity across individuals.\footnote{\textit{Notation:} The variables written in capital letters denote random variables and small letters observational units. All vectors and matrices are written in bold.}
We are interested in analyzing consumers' preferences. For that purpose, we consider a conditional logit  model to model individuals' choices over a set of $J$ mutually exclusive alternatives.
In this context, let $\bm{x}_{i,j}$ denote a $K$-dimensional real-valued vector of observed product characteristics for consumer $i=1,\ldots,N$ and alternative $j=1,\ldots,J$, $\bm{w}_i$ a  $D$-dimensional vector of observed socio-demographics of consumer $i$, and $\bm{y}_i$ a $J$-dimensional vector with entry $1$ if alternative $j$ is chosen by consumer $i$ and zero otherwise. 
Consumers choose the alternative that maximizes their utility. Given 
the unobserved individual parameters $\alpha_j(\bm{w}_i)$, $j=1,\ldots,J$, and $\bm{\beta}(\bm{w}_i) = \left(\beta_1(\bm{w}_i),\ldots,\beta_K(\bm{w}_i) \right)^{\prime}$ consumer $i$ realizes utility $u_{i,j}=\alpha_j(\bm{w}_i)+\bm{x}_{i,j}^{\prime}\bm{\beta}(\bm{w}_i)+\omega_{i,j}$ from alternative $j$, where $\omega_{i,j}$ denotes an idiosyncratic, consumer- and choice-specific error term. Thus, consumer $i$ chooses alternative $j$ if $u_{i,j}>u_{i,l}$ for all $j\neq l$. Under the assumption that $\omega_{i,j}$ is independently and identically distributed type I extreme value, the probability that consumer $i$ chooses alternative $j$ conditional on the observed product characteristics and socio-demographics is
\begin{equation}\label{eqDL:logit_model}
\mathbb{P}\left(y_{i,j}=1\vert \bm{x}_i, \bm{w}_i\right)=\frac{\exp\left(\alpha_j\left(\bm{w}_i\right) + \bm{x}^{\prime}_{i,j}\bm{\beta}\left(\bm{w}_i\right)\right)}{\sum_{m=1}^J\exp\left(\alpha_m\left(\bm{w}_i\right) + \bm{x}^{\prime}_{i,m}\bm{\beta}\left(\bm{w}_i\right)\right)}.
\end{equation}
The goal of the researcher is to estimate the unknown heterogeneous coefficient functions $\bm{\alpha}(\bm{w}_i)= \left(\alpha_1(\bm{w}_i),\ldots,\alpha_J(\bm{w}_i) \right)^{\prime}$ and $\bm{\beta}(\bm{w}_i)$, which are functions of consumers' socio-demographic characteristics that capture the observed heterogeneity across consumers. Thus, the functions capture no unobserved heterogeneity, i.e., there are no random coefficients.\footnote{The parameters $\bm{\beta}(\bm{w}_i)$ and $\bm{\alpha}(\bm{w}_i)$ can be considered as the best approximations to some unobserved individual parameters $\bm{\alpha}_i$ and $\bm{\beta}_i$ that lie in an assumed function class.} \\

\begin{figure}[h]
\captionsetup{justification=centering} 
\caption{Feedforward Neural Network for the Estimation of the Heterogeneous Parameters $\bm{\alpha}(\bm{w}_i)$ and $\bm{\beta}(\bm{w}_i)$}
\begin{center}
    \begin{tikzpicture}
    	\tikzstyle{place}=[circle, draw=black, minimum size = 10mm]
    	
    	\foreach \x in {1,...,3}
    		\draw node[fill=yellow!50] at (0, -\x*1.5-1.5/2) [place] (first_\x) {$w_{i,\x}$};
    	\foreach \x in {1,...,3}
    		\fill (0, -3.75*1.5 -\x*0.5) circle (1.3pt);
    	\draw node[fill=yellow!50] at (0, -5*1.5-1.5/2) [place] (first_n) {$w_{i, D}$};
    	
    	\foreach \x in {1,...,6}
    		\node[fill=blue!50] at (3, -\x*1.5) [place] (second_\x){};
    	
    	\foreach \x in {1,...,6}
    		\node[fill=blue!50] at (6, -\x*1.5) [place] (third_\x){};

		\node[fill=red!50] at (9, -2*1.5-1.5/2) [place] (fourth_1){$\bm{\alpha}(\bm{w}_i)$};
		\node[fill=red!50] at (9, -4*1.5-1.5/2) [place] (fourth_2){$\bm{\beta}(\bm{w}_i)$};  		
		\node[fill=yellow!50] at (12, -6*1.5) [place] (fourth_3){$\bm{x}_i$}; 
		
		\node[fill=yellow!50] at (12, -1.5) [place] (fourth_4){$\bm{y}_i$}; 
		
		\node[fill=green!50] at (12, -3*1.5-1.5/2) [place] (fifth_1){$\ell(\cdot)$}; 
  	
    		
    	\foreach \i in {1,...,3}
    		\foreach \j in {1,...,6}
    			\draw[draw=gray!50] [->] (first_\i) to (second_\j);
    	\foreach \i in {1,...,6}
    		\draw[draw=gray!50] [->] (first_n) to (second_\i);
	
    	\foreach \i in {1,...,6}
    		\foreach \j in {1,...,6}
    			\draw[draw=gray!50] [->] (second_\i) to (third_\j);
    	
    	\foreach \i in {1,...,6}
		\foreach \j in {1,...,2}
    			\draw[draw=gray!50] [->] (third_\i) to (fourth_\j);
    			
    	\foreach \i in {1,...,4}
			\draw[draw=gray!50] [->] (fourth_\i) to (fifth_1);
 	

    	\node[text width=2cm]  at (0.4, 0) [black, ] {Input\\Layer};
    	\node[text width=2cm] at (3.4, 0) [black, ] {Hidden\\Layer 1};
    	\node[text width=2cm] at (6.4, 0) [black, ] {Hidden\\Layer 2};
    	\node[text width=2cm] at (9.3, 0) [black, ] {Parameter\\Layer};
    	\node[text width=2cm] at (12.4, 0) [black, ] {Model\\Layer};
    	    	
    \end{tikzpicture}
    \label{figDL:network_architecture}
\end{center}
\end{figure}
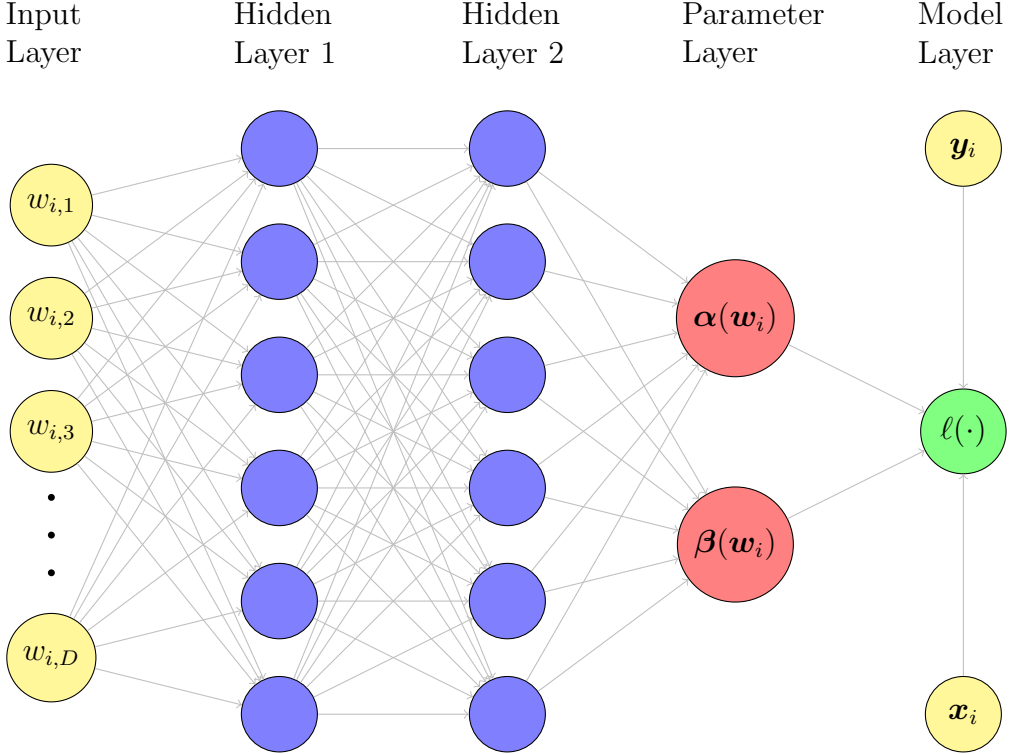

For the estimation of $\bm{\alpha}(\cdot)$ and $\bm{\beta}(\cdot)$, \citeA{farrell2020} advocate deep neural networks. The proposed network architecture allows combining a standard fully-connected feedforward neural network -- which is used to estimate the coefficient functions $\bm{\alpha}(\cdot)$ and $\bm{\beta}(\cdot)$ -- with the economic structure imposed by the conditional logit model. The key idea of the network architecture is to be fully flexible in modeling the individual heterogeneity while retaining the structure which ensures the interpretability of the results.
Figure \ref{figDL:network_architecture} illustrates such an architecture.
Given consumers' observed socio-demographics, $\bm{w}_i$, $i=1,\ldots,N$, in the input layer, the feedforward network learns the coefficient functions $\bm{\alpha}(\cdot)$ and $\bm{\beta}(\cdot)$ using two hidden layers, a parameter layer, and a model layer. The first part of the network, consisting of the input layer and the hidden layers, corresponds to the structure of a standard feedforward neural network. The number of hidden layers and the number of units per hidden layer determine the flexibility of the approach regarding the shape of the estimated coefficient functions. The coefficient functions $\bm{\alpha}(\cdot)$ and $\bm{\beta}(\cdot)$, which are returned in the parameter layer,  are then forwarded to the model layer, where they are combined with the variables of interest, $\bm{x}_i$, and the observed choices, $\bm{y}_i$, to minimize the individual loss function, $\ell\left(\bm{y}_i, \bm{x}_i, \bm{\alpha}\left(\bm{w}_i\right), \bm{\beta}\left(\bm{w}_i\right)\right)$. To be clear, the variables of interest, $\bm{x}_i$,  are additional inputs provided only to the model layer and are not used as inputs to the coefficient functions $\bm{\alpha}(\cdot)$ and $\bm{\beta}(\cdot)$. The novelty of this network architecture is the model layer, which ensures that the coefficient functions $\bm{\alpha}(\cdot)$ and $\bm{\beta}(\cdot)$ are learned within the structure imposed by the specified model. This way, the estimated results have an economically meaningful interpretation, which is typically not the case for regular machine learning applications in economics \cite{farrell2020}.

The number of hidden layers (the depth of the network) and the number of units per layer (the width of each layer) are specified by the researcher. 
According to the universal approximation theorem (\citeNP{hornik1989}, \citeNP{cybenko1989}), a feedforward network with only one hidden layer might be already sufficient to represent any function if the number of hidden units is sufficiently large. Networks with multiple hidden layers typically require fewer units per hidden layer -- and hence total parameters -- to represent the desired function, and in many circumstances generalize well in terms of out-of-sample performance. However, such networks tend to be harder to optimize \cite{goodfellow2016}. 
In Theorem 1, \citeA{farrell2020} derive error bounds for the estimated coefficient functions $\hat{\bm{\alpha}}(\cdot)$ and $\hat{\bm{\beta}}(\cdot)$, where they allow the depth of the network to increase with the sample size, and the width of the network with the sample size and the number of continuous input variables, respectively. 
Beyond the number of hidden layers and units, the researcher needs to specify the activation function at every layer. The design of hidden layers is an active area of research which does not provide definite guidelines for the choice of activation functions yet. According to \citeA{goodfellow2016}, rectified linear units are an excellent default choice and are also recommended by \citeA{farrell2020}.
Overall, specifying the network architecture is a trial-and-error process where the final architecture can be selected based on the best out-of-sample fit \cite{goodfellow2016}.

When estimating the model, the coefficient functions $\bm{\alpha}(\bm{w}_i)$ and $\bm{\beta}(\bm{w}_i)$ are learned jointly. To simplify the notation, we write $\bm{\delta}(\bm{w}_i):=(\bm{\alpha}(\bm{w}_i)^{\prime}, \bm{\beta}(\bm{w}_i)^{\prime})^{\prime}$ and $L:= J+K$  in the following. In our case, the individual loss function,  $\ell\left(\bm{y}_i, \bm{x}_i, \bm{\delta}(\bm{w}_i)\right)$, following from the economic model of interest, is the empirical log-likelihood for individual $i$, 
\begin{equation}\label{eqDL:loglik}
\ell\left(\bm{y}_i, \bm{x}_i, \bm{\delta}(\bm{w}_i)\right)=\sum\limits_{j=1}^J y_{i,j}\log\left(\mathbb{P}\left(y_{i,j}=1\vert \bm{x}_i, \bm{w}_i\right)\right),
\end{equation}
where $\mathbb{P}\left(y_{i,j}=1\vert \bm{x}_i, \bm{w}_i\right)$ is the conditional logit choice probability given in Equation \eqref{eqDL:logit_model}.
Then, $\hat{\bm{\delta}}(\bm{w}_i):= (\hat{\bm{\alpha}}(\bm{w}_i)^{\prime}, \hat{\bm{\beta}}(\bm{w}_i)^{\prime})^{\prime}$ are determined such that they simultaneously maximize the log-likelihood 
\begin{equation}
\hat{\bm{\delta}}(\bm{w}_i)=\argmax\limits_{\bm{{\delta}}} \sum\limits_{i=1}^N \ell\left(\bm{y}_i, \bm{x}_i, \bm{{\delta}}(\bm{w}_i)\right),
\end{equation} 
where we optimize over the class of DNNs which use the type of architecture described in Figure \ref{figDL:network_architecture}.
The log-likelihood loss function forces the DNN to learn the coefficient functions within the structure imposed by the conditional logit model. This has two advantages in comparison to naively applied prediction-focused machine learning methods, which predict the choice probabilities $\hat{\mathbb{P}}\left(y_{i,j}=1\vert \bm{x}_i, \bm{w}_i\right)$ using a completely unstructured nonparametric utility $\hat{u}(\bm{y}_i, \bm{w}_i,\bm{x}_i)$: 
First, it assures that the network provides economically meaningful results. For the unstructured approach, in contrast, it is not clear how estimates of $\bm{\alpha}(\bm{w}_i)$ and $\bm{\beta}(\bm{w}_i)$ can be separately recovered from  $\hat{u}(\bm{y}_i, \bm{w}_i,\bm{x}_i)$, which, however, is often necessary for interpretation. And second, even if $\bm{\alpha}(\bm{w}_i)$ and $\bm{\beta}(\bm{w}_i)$ could be separately recovered in the unstructured approach, \citeA{farrell2020} show that the additional structure of the model enables a faster rate of convergence for the estimated coefficient functions (given the model is correctly specified). For the structured approach, the rate of convergence only depends on the dimension of the socio-demographic characteristics, $\text{dim}(\bm{w}_i)$, whereas for the naive prediction focused machine learning with  unstructured  $\hat{u}(\bm{y}_i, \bm{w}_i,\bm{x}_i)$, it depends on both the dimension of the socio-demographic characteristics and the dimension of the variables of interest, i.e., $\text{dim}(\bm{w}_i)+\text{dim}(\bm{x}_i)$. While the convergence rate in the structured network is fast enough for inference, the convergence rate of the unstructured model would often be too slow for inference \cite{farrell2020}.

\subsection{Inference}\label{secDL:inference}

Inference for machine learning methods for the estimation of economic models is challenging. 
For that reason, \citeA{farrell2020} adopt the semiparametric inference procedure suggested by \citeA{cherno2018} which allows to perform inference on expected values of heterogeneous quantities using an influence function approach. Due to the structure imposed by the economic model, the proposed procedure can be applied to any  quantity of interest (e.g., expected value of coefficients, elasticities, or measures for the willingness-to-pay) which are functions of the heterogeneous coefficient functions $\bm{\delta}(\cdot)$ (and a fixed vector $\bm{x}^{*}$ containing arbitrary values of the variables of interest).

Let the real-valued function $H(\cdot)$ specified by the researcher denote the function of interest. Then, the inference procedure described in the following allows to conduct inference on the expected value of $H(\cdot)$ given some $\bm{x}^{*}$, 
\begin{equation}
\theta_0=\mathbb{E}\left[H\left(\bm{W}, \bm{\delta}\left(\bm{W}\right);\bm{x}^{*}\right)\right].
\end{equation}
Note that $H(\cdot)$ directly depends on the coefficient functions $\bm{\delta}(\cdot)$, making inference on $\theta_0$ depend on how well $\hat{\bm{\delta}}(\cdot)$ approximates its true counterpart $\bm{\delta}(\cdot)$.
Because the empirical plug-in estimator of $\theta_0$,
\begin{equation}
\hat{\theta}_{PI} = \frac{1}{N}\sum\limits_{i=1}^N H\left(\bm{w}_i,\hat{\bm{\delta}}\left(\bm{w}_i\right);\bm{x}^*\right),\notag
\end{equation}
is only valid under strong conditions on  $\hat{\bm{\delta}}(\cdot)$, which are unlikely to be satisfied if the functions are estimated with deep-neural networks, \citeA{farrell2020} propose to use the concept of influence functions for inference. The approach builds on the seminal work of \citeA{newey1994} and has the advantage that it provides results for valid inference under less restrictive conditions on the distributional approximations of $\bm{\delta}(\cdot)$. These assumptions are known to hold for many machine learning methods \cite{farrell2020}. 

The influence function for $\theta_0$ involves the gradient and Hessian corresponding to the loss function $\ell\left(\bm{y}_i, \bm{x}_i, \bm{\delta}(\bm{w}_i)\right)$ with respect to $\bm{\delta}(\bm{w}_i)$. Let $\bm{\ell}_{\bm{\delta}}\left(\bm{y}_i, \bm{x}_i, \bm{\delta}(\bm{w}_i)\right)$ denote the $L$-dimensional vector of first derivatives of $\bm{\ell}\left(\bm{y}_i, \bm{x}_i, \bm{\delta}(\bm{w}_i)\right)$ w.r.t. $\bm{\delta}(\bm{w}_i)$,
\begin{equation}
\bm{\ell}_{\bm{\delta}}\left(\bm{y}_i, \bm{x}_i, \bm{\delta}(\bm{w}_i)\right)=\frac{\partial \ell\left(\bm{y}_i, \bm{x}_i, \bm{b}\right)}{\partial\bm{b}}\bigg\vert_{\bm{b}=\bm{\delta}\left(\bm{w}_i\right)}, \notag
\end{equation}
and $\bm{\ell}_{\bm{\delta},\bm{\delta}}\left(\bm{y}_i, \bm{x}_i, \bm{\delta}(\bm{w}_i)\right)$ the $L \times L$-matrix of second order derivatives with entries $\{k_1, k_2\}$ defined as
\begin{equation}
\left[\bm{\ell}_{\bm{\delta},\bm{\delta}}\left(\bm{y}_i, \bm{x}_i, \bm{\delta}(\bm{w}_i)\right)\right]_{k_1,k_2} = \frac{\partial^2 \ell\left(\bm{y}_i, \bm{x}_i, \bm{b}\right)}{\partial b_{k_1}\partial b_{k_2}}\bigg\vert_{\bm{b}=\bm{\delta}\left(\bm{w}_i\right)}.\notag
\end{equation}
Define   $H_{\bm{\delta}}(\bm{w}_i,\bm{\delta}(\bm{w}_i);\bm{x}^*)$ as the $L$-dimensional vector of first derivatives of $H(\bm{w}_i,\bm{\delta}(\bm{w}_i);\bm{x}^*)$ w.r.t. $\bm{\delta}(\bm{w}_i)$. Further, define 
\begin{equation}
\bm{\Lambda}(\bm{w}_i):=\mathbb{E}[\ell_{\bm{\delta},\bm{\delta}}(\bm{Y},\bm{X},\bm{\delta}(\bm{W}))\vert \bm{W}=\bm{w}_i],
\end{equation}
corresponding to the expected individual Hessian for individual $i$ conditional on her socio-demographic characteristics $\bm{w}_i$. 
Then, a valid and Neyman orthogonal score for the parameter of inferential interest, $\theta_0$, is $\psi(\bm{w}_i,\bm{\delta}(\bm{w}_i),\bm{\Lambda}(\bm{w}_i))-\theta_0$, where 
\begin{equation}\label{eqDL:InfluenceFunctionCondLogit}
\psi\left(\bm{w}_i,\bm{\delta}(\bm{w}_i),\bm{\Lambda}(\bm{w}_i)\right)=H\left(\bm{w}_i,\bm{\delta}\left(\bm{w}_i\right);\bm{x}^*\right) - H_{\bm{\delta}}\left(\bm{w}_i,\bm{\delta}\left(\bm{w}_i\right);\bm{x}^*\right)^{\prime}\bm{\Lambda}\left(\bm{w}_i\right)^{-1}\bm{\ell}_{\bm{\delta}}\left(\bm{y}_i,\bm{x}_i,\bm{\delta}(\bm{w}_i)\right)
\end{equation} 
is the influence function when centered at $\theta_0$. Hence, $\theta_0$ can be identified from the condition $\mathbb{E}\left[\psi\left(\bm{W},\bm{\delta}\left(\bm{W}\right),\bm{\Lambda}\left(\bm{W}\right)\right)-\theta_0\right]=0$. 
In case of the conditional logit model stated in Equation \eqref{eqDL:logit_model}, the gradient vector $\bm{\ell}_{\bm{\delta}}(\bm{y}_i, \bm{x}_i, \bm{\delta}(\bm{w}_i))$ for individual $i$ is 
\begin{equation}
\bm{\ell}_{\bm{\delta}}(\bm{y}_i, \bm{x}_i, \bm{\delta}(\bm{w}_i))=\left(c_{i,1},\ldots,c_{i,J},\tilde{c}_{i,1},\ldots,\tilde{c}_{i,K}\right)^\prime
\end{equation}
with $j$th element $c_{i,j}=y_j-\mathbb{P}(y_{i,j}=1\vert \bm{x}_i, \bm{w}_i)$ and $(J+k)$th element $\tilde{c}_{i,k}=\sum_{j=1}^J(y_{i,j}-\mathbb{P}(y_{i,j}=1\vert \bm{x}_i, \bm{w}_i))x_{i,j,k}$. The matrix $\bm{\ell}_{\bm{\delta},\bm{\delta}}\left(\bm{y}_i, \bm{x}_i, \bm{\delta}(\bm{w}_i)\right)$ can be written as
\begin{equation*} 
\bm{\ell}_{\bm{\delta},\bm{\delta}}\left(\bm{y}_i, \bm{x}_i, \bm{\delta}(\bm{w}_i)\right)=\dot{\bm{G}}_i\tilde{\bm{x}}_{i}\tilde{\bm{x}}_i^{\prime}
\end{equation*} 
 with $\dot{\bm{G}}_i$ being the derivative of the conditional logit choice probabilities with respect to the linear index $\tilde{\bm{x}}_i^{\prime} \bm{\delta}(\bm{w}_i)$, and $\tilde{\bm{x}}_i=[\bm{e}_1,\ldots,\bm{e}_J,\bm{x}_i]$ where $\bm{e}_j$ is a unit vector with $L$ elements where the $j$th element is equal to one and zero otherwise. Thus, the $L \times L$ matrix $\dot{\bm{G}}_i$ for individual $i$ has entries $\dot{g}_{kk}=\mathbb{P}\left(y_{i,j}=1\vert \bm{x}_i, \bm{w}_i\right)(1-\mathbb{P}\left(y_{i,j}=1\vert \bm{x}_i, \bm{w}_i\right))$ on the main diagonal and $\dot{g}_{k,l}=-\mathbb{P}\left(y_{i,j}=1\vert \bm{x}_i, \bm{w}_i\right)\mathbb{P}\left(y_{i,m}=1\vert \bm{x}_i, \bm{w}_i\right)$ for all $k\neq l$ on the off-diagonal. A detailed derivation of the influence function for the conditional logit model presented in Equation \eqref{eqDL:logit_model} is given in \citeA[v1 on arXiv.org]{farrell2020}. 

The plug-in estimator $\hat{\theta}_{PI}$ takes only one source of uncertainty in  $H(\bm{w}_i,\hat{\bm{\delta}}(\bm{w}_i);\bm{x}^*)$ into account: the direct effect of perturbations in the data on $H(\bm{w}_i,\hat{\bm{\delta}}(\bm{w}_i);\bm{x}^*)$ for a given $\hat{\bm{\delta}}(\bm{w}_i)$ estimated with the sample.
In contrast, the influence function approach additionally accounts for the uncertainty in the estimated coefficient functions due to perturbations in the data when estimating $\theta_0$ with machine learning.
For illustrative purposes, assume there are estimates $\hat{\bm{\delta}}(\bm{w}_i)$ and $\hat{\bm{\Lambda}}(\bm{w}_i)$ for a given sample. Using $\hat{\bm{\delta}}(\bm{w}_i)$ and $\hat{\bm{\Lambda}}(\bm{w}_i)$ to calculate the influence function, $\psi(\bm{w}_i,\hat{\bm{\delta}}(\bm{w}_i),\hat{\bm{\Lambda}}(\bm{w}_i))$, presented in Equation \eqref{eqDL:InfluenceFunctionCondLogit}, the sample analogue of  $\mathbb{E}\big[\psi\big(\bm{W},\hat{\bm{\delta}}\left(\bm{W}\right),\hat{\bm{\Lambda}}\left(\bm{W}\right) \big)\big]$  is
 \begin{subequations} \label{eqDL:estInfluenceFunctionCondLogit}
\begin{align}
\hat{\theta}_{IF} =& \frac{1}{N}\sum\limits_{i=1}^N \psi\left(\bm{w}_i,\hat{\bm{\delta}}(\bm{w}_i),\hat{\bm{\Lambda}}(\bm{w}_i)\right)\notag\\ 
 =& \frac{1}{N}\sum\limits_{i=1}^N  H\left(\bm{w}_i,\hat{\bm{\delta}}\left(\bm{w}_i\right);\bm{x}^*\right) \label{eqDL:estInfluenceFunctionCondLogit1} \\
 & - \frac{1}{N}\sum\limits_{i=1}^N  H_{\hat{\bm{\delta}}}\left(\bm{w}_i,\hat{\bm{\delta}}\left(\bm{w}_i\right);\bm{x}^*\right)^{\prime}\hat{\bm{\Lambda}}\left(\bm{w}_i\right)^{-1}\bm{\ell}_{\hat{\bm{\delta}}}\left(\bm{y}_i,\bm{x}_i,\hat{\bm{\delta}}(\bm{w}_i)\right).  \label{eqDL:estInfluenceFunctionCondLogit2}
\end{align}
\end{subequations}
Similarly to $\hat{\theta}_{PI}$, the term in Equation \eqref{eqDL:estInfluenceFunctionCondLogit1} captures the changes in the function $H(\bm{w}_i, \hat{\bm{\delta}}(\bm{w}_i);\bm{x}^*)$ in response to perturbations in the data, treating the coefficient functions $\hat{\bm{\delta}}(\bm{w}_i)$ as if they were known. This way, the term accounts for the uncertainty in the parameter of inferential interest due to changes in $H(\bm{w}_i, \hat{\bm{\delta}}(\bm{w}_i);\bm{x}^*)$. The term in Equation \eqref{eqDL:estInfluenceFunctionCondLogit2} is an additional correction term that includes an estimate of the nuisance function $\bm{\Lambda}(\bm{w}_i)$ and, thereby, accounts for the uncertainty in  the functional forms of the coefficient functions $\bm{\delta}(\bm{w}_i)$ arising from perturbations in the data. The correction term isolates the impact of the nonparametric estimation on the estimated parameters of inferential interest, which is enabled through the imposed structure of the economic model relating the outcome $\bm{Y}$ to the covariates $\bm{X}$ in a known way.

The correction terms $H_{\bm{\delta}}(\bm{w}_i)$, $\bm{\ell}_{\bm{\delta}}\left(\bm{y}_i, \bm{x}_i, \bm{\delta}(\bm{w}_i)\right)$ and $\bm{\ell}_{\bm{\delta},\bm{\delta}}\left(\bm{y}_i, \bm{x}_i, \bm{\delta}(\bm{w}_i)\right)$ can be calculated  analytically and do not need to be estimated.
In contrast, the matrix $\bm{\Lambda}(\bm{w}_i)$ consists of regression-type objects which must be estimated, i.e., the individual Hessian $\ell_{\bm{\delta},\bm{\delta}}(\bm{Y},\bm{X},\bm{\delta}(\bm{W}))$ is projected on $\bm{W}$. For this projection, DNNs can be used as well. 
Further, note that the product $\Lambda(\bm{w}_i)^{-1}\bm{\ell}_{\bm{\delta}}(\bm{w}_i,\bm{\delta}(\bm{w}_i))$ does not depend on the function $H(\cdot)$, which simplifies calculations if multiple parameters are of inferential interest.

An important assumption of the inference procedure is that the matrix $\bm{\Lambda}(\bm{w}_i)$ is invertible with bounded inverse. With respect to the conditional logit model in Equation \eqref{eqDL:logit_model}, the assumption implies that the choice probabilities are bounded away from zero and one.\footnote{In order to assure the numerical stability of the approach, \citeA{farrell2020} propose trimming or regularization of $\Lambda(\bm{w}_i)$ by adding a positive constant to the main diagonal, e.g., $\Lambda(\bm{w}_i)+I$.}


\subsection{Estimation}\label{secDL:inference_estimation}

With the influence function in Equation \eqref{eqDL:InfluenceFunctionCondLogit}, the estimator $\hat{\theta}$ of $\theta_0$ and a corresponding estimator $\hat{\Psi}$ of its asymptotic variance can be formed using the semiparametric inference procedure of \citeA{cherno2018}. For the estimation, the influence function $\psi(\bm{w}_i,\bm{\delta}(\bm{w}_i),\bm{\Lambda}(\bm{w}_i))$ needs to be evaluated at every data point in the sample. In order to obtain a properly centered limiting distribution under weaker conditions on the first-stage estimates $\hat{\bm{\delta}}(\bm{w}_i)$, the estimation procedure for $\theta_0$ is based on sample splitting \cite{farrell2020}. 

For the conditional expected individual Hessian  of the conditional logit model, $\bm{\Lambda}(\bm{w}_i)$,  the dependent  variable $\bm{Z}:=\dot{\bm{G}}\bm{XX}'$ is regressed on the socio-demographic characteristics $\bm{W}$. Because $\dot{\bm{G}}$, and hence $\bm{Z}$, depend on the coefficient functions $\bm{\delta}(\bm{W})$, the estimation of the influence function requires three-way splitting of the sample. The first sub-sample is used to estimate the heterogeneous parameter functions $\hat{\bm{\delta}}(\bm{w}_i)$. These are subsequently treated as the inputs to calculate the ``observed'' matrix $\bm{z}_i$ of $\bm{Z}$, using $\bm{w}_i$ and $\bm{x}_i$ of the second sub-sample. Using $\bm{z}_i$ as the dependent variable and $\bm{w}_i$ as the independent variable, $\hat{\bm{\Lambda}}(\bm{w}_i)$ is estimated with the second sub-sample. The influence function is then calculated with the third sub-sample \cite{farrell2020}. The procedure thus consists of the following steps:
\begin{itemize}
\item[1.] Split the observation units $\{1, \ldots, n\}$ into $S$ subsets, denoted by $\mathcal{S}_s\subset\{1, \ldots, n\}$, $s=1, \ldots, S$.
\item[2.] For each $s=1, \ldots, S$, let $\mathcal{S}^c_s$ denote the complement  of $\mathcal{S}_s$.  For nonlinear models like the conditional logit model, the functions $\bm{\delta}_s(\bm{w}_i)$ and $\bm{\Lambda}_s(\bm{w}_i)$, corresponding to split $s$, cannot be estimated simultaneously.   Instead, the complement $\mathcal{S}_s^c$ is split into two pieces to first estimate $\hat{\bm{\delta}}_s(\bm{w}_i)$ using the first piece, and then $\hat{\bm{\Lambda}}_s(\bm{w}_i)$ using the second piece together with the fixed functions $\hat{\bm{\delta}}_s(\bm{w}_i)$.
\item[3.] The final estimator of $\theta_0$ is then
\begin{equation}\label{eq:DLestTheta}
\hat{\theta}=\frac{1}{S}\sum\limits\hat{\theta}_s,\quad\quad\hat{\theta}_s=\frac{1}{\vert\mathcal{S}_s\vert}\sum\limits_{i\in\mathcal{S}_s}\psi\left(\bm{w}_i, \hat{\bm{\delta}}_s(\bm{w}_i), \hat{\bm{\Lambda}}_s(\bm{w}_i)\right),
\end{equation}
where $\vert\mathcal{S}_s\vert$ is the cardinality of $\mathcal{S}_s$ and is assumed to be proportional to the sample size.

Furthermore, an estimator $\hat{\Psi}$ of the asymptotic variance of $\hat{\theta}$  is given by the variance-analogue of Equation \eqref{eq:DLestTheta}
\begin{equation}  \label{eq:DLestPsi}
\hat{\Psi}=\frac{1}{S}\sum\limits_{s=1}^S\hat{\Psi}_s,\quad\quad\hat{\Psi}_s=\frac{1}{\vert\mathcal{S}_s\vert}\sum\limits_{i\in\mathcal{S}_s}\left(\psi\left(\bm{w}_i, \hat{\bm{\delta}}_s(\bm{w}_i), \hat{\bm{\Lambda}}_s(\bm{w}_i)\right) - \hat{\theta}\right)^2.
\end{equation}
\end{itemize}
For $\hat{\theta}$ and $\hat{\Psi}$, \citeA{farrell2020} provide inference results that establish asymptotic normality and validity of standard errors, 
\begin{equation} \label{eq:psiNormal}
\sqrt{n}\hat{\Psi}^{-1/2}\left(\hat{\theta}-\theta\right)\rightarrow_d\mathcal{N}\left(0, 1\right).
\end{equation} 
This allows a simple construction of confidence intervals for $\theta$. \citeA{cherno2018} prove that these are uniformly valid but not necessarily semi-parametrically efficient.\footnote{However, the constructed standard confidence intervals for $\theta$ can be semi-parametrically efficient, and \citeA{farrell2020} also conjecture that they are semi-parametrically  efficient but do not prove it.}

A central input to the influence function, and hence to the estimated inference results, is the  conditional expected individual Hessian $\bm{\Lambda}(\bm{w}_i)$ which is a nuisance function as it is required only for the calculation of the influence functions but not of interest per se. Estimating $\bm{\Lambda}(\bm{w}_i)$ is a prediction problem for which different machine learning methods can be used.
In the Monte Carlo experiments and application presented below, we estimate $\bm{\Lambda}(\bm{w}_i)$ by another neural network  using the mean squared error (MSE) as loss function. Because the matrix $\bm{\Lambda}(\bm{w}_i)$ is symmetric, we only need to estimate $L(L+1)/2$ entries. To keep the estimation procedure as simple as possible, we estimate the entries of $\bm{\Lambda}(\bm{w}_i)$ using a single network with $L(L+1)/2$ output units. Alternatively, one could estimate each entry with a separate network, which is more flexible but has the disadvantage that it is computationally more expensive.

The estimation procedure described above has some potential weaknesses that can lead to misleading results. The first one is potential overfitting when predicting the choice probability for each alternative, which can lead to estimated probabilities close to zero and one, respectively. 
As a consequence, the matrix $\hat{\bm{\Lambda}}(\bm{w}_i)$ might not be invertible (or close to not being invertible, leading to extremely large entries of the inverse) if the entries are estimated precisely.
Related to the overfitting problem, a practical disadvantage of the sample splitting -- beyond the computational cost -- is that small sub-samples potentially provide imprecise estimates, which is particularly relevant for applications with small sample sizes \cite{farrell2020}.\footnote{For the asymptotic results of the sample splitting procedure,  \citeA{farrell2020}  treat  $S$  as fixed and therefore,  the sample splitting  is asymptotically negligible.}
\begin{remark} \label{rem:repeatedCV}
To increase finite sample precision,  \citeA{cherno2018} suggest to repeat the sample splitting procedure outlined above $R$ times. To this end, let $\hat{\theta}_r$ and  $\hat{\Psi}_r$  denote the estimators shown in Equation \eqref{eq:DLestTheta} and \eqref{eq:DLestPsi} for repetition $r= 1, \ldots, R$. Then, the final estimator is the median over the repetitions,\footnote{\citeA{cherno2018} also consider taking the average across repetitions instead of the median. However, they recommend to use the median since it is less dependent on the outcome of a single repetition.} i.e,
\begin{align*}
\hat{\theta}^{med} = median\left\{\hat{\theta}_r\right\}_{r=1}^{R}, \quad\text{and} \quad
\hat{\Psi}^{med} = median\left\{\hat{\Psi}_r + \left(\hat{\theta}_r - \hat{\theta}^{med}  \right)^2\right\}_{r=1}^{R}.
\end{align*}
\citeA{cherno2018} note that the choice of $R\geq1$ does not affect the asymptotic distribution of $\hat{\theta}^{med}$.  By Equation \eqref{eq:psiNormal}, each $\hat{\theta}_k$ is asymptotically normal and therefore,  $\hat{\theta}^{med}$ is asymptotically normal, too. In our simulations, we set $R=5$ and find that repeated sample splitting substantially improves the precision of the estimates.
\end{remark}

\section{Monte Carlo Experiments}\label{secDL:Monte_Carlo}

This section presents different Monte Carlo experiments that study the performance of the deep learning estimation procedure and, in particular, the inference procedure presented in Section \ref{secDL:deep_learning_logit}. 
To study the performance in a realistic setup, we use semi-synthetic data for the experiments. The data is taken from the Swissmetro dataset \cite{bierlaire2001}, which is an openly available dataset collected in Switzerland during March 1998.\footnote{We downloaded the test and training data from the github repository \hyperlink{https://github.com/BSifringer/EnhancedDCM/tree/master/ready_example/swissmetro_paper}{github.com/BSifringer/EnhancedDCM}.} The data consists of survey data from $1,191$ car and train travelers. It was collected to analyze the impact of a new innovative transportation mode, represented by the Swissmetro, against usual transportation modes, namely car and regular train connections.\footnote{The Swissmetro is a revolutionary mag-lev underground system operating at speeds up to 500 km/h in partial vacuum.} For every respondent, nine stated choice situations were generated in which the respondents could choose between three travel mode alternatives: Swissmetro (abbreviated as sm), train, and car (only for car owners). In total, the data consists of $10,719$ choice situations \cite{antonini2007swissmetro}. When preparing the data, we follow the instructions of \citeA{sifringer2020} and remove all observations for which not all three alternatives -- Swissmetro, train, car -- are available. This reduces the number of travelers to $1,683$ and thus, the final data set to $9,036$ observations.\footnote{For the estimation, we follow \citeA{sifringer2020} and ignore the panel structure of the data.} 

For the data generation, we consider an individual-level discrete choice demand model of the form presented in Equation \ref{eqDL:logit_model}. 
The variables of interest in our Monte Carlo experiments are the travel cost (\textit{cost}), the travel time (\textit{time}), and the frequency (\textit{freq}) of the train and Swissmetro connections (frequency is zero for car).\footnote{The travel cost, travel time, and frequency variables are scaled downwards by factor one hundred \cite{sifringer2020}. For those travelers that have an annual season pass, we set the travel cost of the train and Swissmetro to zero.}
Each traveler chooses the travel mode among the three alternatives car, Swissmetro, and train that provides her with the highest utility,
\begin{equation}
u_{i,j}=\alpha_j\left(\bm{w}_i\right) + cost_{i,j}\beta^{\text{cost}}\left(\bm{w}_i\right) + time_{i,j}\beta^{\text{time}}\left(\bm{w}_i\right) + freq_{i,j}\beta^{\text{freq}}\left(\bm{w}_i\right) + \omega_{i,j},\notag
\end{equation}
for $j=\{\text{car}, \text{train}, \text{sm}\}$.  
We specify the true coefficients  as functions of travelers' yearly income (\textit{income}), age (\textit{age}), gender (\textit{male}), and a variable indicating who paid for the ticket (\textit{who}). Income and age are categorical variables that assign travelers' income and age into four and six groups, respectively. The gender variable is equal to one if the traveler is male and zero otherwise.
The variable \textit{who} is a categorical variable that takes four values ($0$ if it is unknown who pays, $1$ if the traveler paid herself, $2$ if the employer pays, and $3$ if the traveler and employer split half-half). In order to make the information represented by the categorical variable more easily accessible for the network, we transform  \textit{who}  into three dummy variables denoted by $who^1$, $who^2$, and $who^3$, leaving out the category $0$ as reference category.\footnote{A detailed description of the data and summary statistics can be found \href{https://transp-or.epfl.ch/documents/technicalReports/CS_SwissmetroDescription.pdf}{here}.}
We specify the observed consumer socio-demographics as $\bm{w}_i := (age_i,\, income_i,\,male_i,\, who^1_i,\, who^2_i,\, who^3_i)^{\prime}$.
The intercept functions for each alternative are 
\begin{align*}
\alpha_{\text{train}}\left(\bm{w}_i\right)&=-1 + 1\cdot\textit{income}_i,\\
 \alpha_{\text{sm}}\left(\bm{w}_i\right)&=-3 + 1\cdot\textit{age}_i,
\end{align*}
and $\alpha_{\text{car}}\left(\bm{w}_i\right) = 0$, i.e., the alternative car serves as reference. The coefficient functions for the covariates of interest are specified as
\begin{align}
\beta^{\text{cost}}\left(\bm{w}_i\right)&=-6 + \textit{income}_i - 0.8 \cdot who^1_i- 1 \cdot who^2_i  - 1.2 \cdot who^3_i \\
\beta^\text{freq}\left(\bm{w}_i\right)&=-5  + \textit{income}_i + 0.9\cdot\textit{male}_i \notag\\ 
 \beta^{\text{time}}\left(\bm{w}_i\right)&=-6 + 1\cdot \textit{age}_i.\notag
\end{align}

To study the finite sample performance of the proposed inference procedure, we consider  the expected value of the heterogeneous coefficients     $\beta^{\text{cost}}\left(\bm{w}_i\right)$,  $\beta^\text{freq}\left(\bm{w}_i\right)$, and $\beta^{\text{time}}\left(\bm{w}_i\right)$ as the parameters of inferential interest, i.e., $\theta_0^k = E[\beta^{k}(\bm{w}_i)]$, $k \in \{\text{cost}, \text{freq}, \text{time}  \}$. Accordingly,  the function $H(\cdot)$ corresponds to
\begin{equation}
{H}(\bm{w}_i, \bm{\delta}(\bm{w}_i);\bm{x}^*)=\beta^{k}(\bm{w}_i),\notag
\end{equation}
where $\bm{\delta}(\bm{w}_i)= \left(\alpha_{\text{train}}\left(\bm{w}_i\right), \alpha_{\text{sm}}\left(\bm{w}_i\right), \beta^{\text{cost}}\left(\bm{w}_i\right),\beta^\text{freq}\left(\bm{w}_i\right),  \beta^{\text{time}}\left(\bm{w}_i\right) \right)^{\prime}$.
Thus, the gradient vector $H_{\bm{\delta}}\left(\bm{w}_i, \bm{\delta}(\bm{w}_i);\bm{x}^*\right)$  is equal  to one for the element corresponding to the derivative with respect to  $\beta^{k}$, and zero for all other entries.

\subsection{Small Data Set}

We conduct $1000$ Monte Carlo repetitions. In every repetition, we use the individual coefficients, the covariates, and an idiosyncratic error term $\omega_{i,j}$ to calculate the utility for each alternative and each individual. For that purpose, we draw $\omega_{i,j}$ from a Type I extreme value distribution for every traveler and alternative in every replicate and select the alternative that provides the largest utility. 

To simulate deviations between the sample and the population values of the covariates, we split the data into two sets. We use all observations to calculate the true values, $\theta_0^k$, $k \in \{\text{cost}, \text{freq}, \text{time}  \}$, but use only three quarters of the data for the estimation. This way, we can test whether the proposed inference procedure adequately accounts for  the uncertainty related to $H(\cdot)$, and for the uncertainty related to the functional form of the heterogeneous coefficient functions $\bm{\delta}(\bm{w}_i)$ which arises due to deviations between observations in the sample and the population.

We use the same network architecture  to estimate the heterogeneous coefficient functions and to estimate the conditional expected individual Hessian $\bm{\Lambda}(\bm{w}_i)$ -- except for the number of output units in both networks. More precisely, we choose one hidden layer with $100$ units and rectified linear activation functions. For the units in the output layer, we use linear activation functions. The number of output units  is  five in the network for  the heterogeneous coefficient functions, and $15$ in the network for $\bm{\Lambda}(\bm{w}_i)$. Both networks use travelers' income, age, gender, and the dummy variables indicating who is paying for the ticket as inputs. When estimating the coefficient functions, we set the dropout rate to $0.2$. 
For the network used to estimate $\bm{\Lambda}(\bm{w}_i)$, we  test different regularizers to account for the difficulty of  projecting $\bm{\Lambda}(\bm{w}_i)$. We consider the $l_2$-regularizers $\lambda= 0, 10^{-5}, 10^{-4},2 \cdot 10^{-3}$ which we use to avoid overfitting $\bm{z}_i$ and, thereby, to ensure that the predicted individual Hessian  $\hat{\bm{\Lambda}}(\bm{w}_i)$ does not become collinear for any individual $i$.  While using a $l_2$-regularizer $\lambda > 0 $ ensures that we can invert $\hat{\bm{\Lambda}}(\bm{w}_i)$, we note that $\lambda > 0$ potentially introduces a bias in the estimation and is not covered by the inference results of \citeA{farrell2020}.
When training the networks, we set the maximum number of epochs to $20,000$, and  the batch size to $50$. During the training, we track the in-sample log-likelihood and the in-sample mean squared error, respectively, and stop the training if the change in the loss function does not exceed $10^{-8}$ across epochs (with a patience of $100$ epochs). We select the network with the best in-sample fits. For the estimation with the influence function approach, we split the training data into $S=5$ folds. Furthermore, we split $\mathcal{S}^c_s$ into two equally sized pieces, using the first one to estimate $\bm{\delta}_s(\bm{w}_i)$, and the second one to estimate $\bm{\Lambda}_s(\bm{w}_i)$.

As a benchmark, we estimate the model with maximum likelihood using the true specification. We refer to this estimator as oracle logit estimator. 
In addition, we also estimate a conditional logit model where we do not account for any type of heterogeneity but instead include only two alternative-specific intercepts and the slope coefficients for $\textit{cost}, \textit{freq}$, and $\textit{time}$. This allows us to study the potential consequences when one does not account for heterogeneity across travelers even though it is present in the data. Finally, we also use a neural network to estimate the heterogeneous coefficient functions without the outlined inference procedure of \citeA{farrell2020}. Instead, we conduct naive inference using the average heterogeneous coefficient functions and the corresponding estimated Fisher information matrix to calculate robust standard errors. This allows us to assess the importance of an appropriate inference procedure after the estimation of the model parameters with machine learning.

Table \ref{tab:MC_swissmetroMean} reports the coverage of the estimated $95\%$ confidence intervals, the average estimated standard errors, and estimated bias across  Monte Carlo replicates for all three covariates of interest. 
Furthermore, we present the share of Monte Carlo replicates in which the false null hypotheses that the coefficients are zero are correctly rejected at a significance level of $0.05$. This is supposed to serve as an indicator for the power of the hypothesis tests when calculated with the different inference procedures. 
For the influence function approach, we additionally calculate the in-sample and out-of-sample MSE of the neural network for $\bm{\Lambda}_s(\bm{w}_i)$, and  track the share of outliers across Monte Carlo replicates. We calculate the in-sample MSE with the part of $\mathcal{S}^c_s$ used for the estimation of $\bm{\Lambda}_s(\bm{w}_i)$, and the out-of-sample MSE with the left out fold. We treat a Monte Carlo replicate as outlier if the estimated standard error is larger than $5$ for at least one of the three estimated parameters.

\begin{table}[h!]
\centering	
\captionsetup{justification=centering} 
	\caption{Average Summary Statistics of 1000 Monte Carlo Replicates for Small Data and without Repeated Sample Splitting}
	\label{tab:MC_swissmetroMean}
	\small
\begin{tabular}{  l@{\hspace{-0.5em}}  *{7}S[table-format=2.4] } 
\toprule 
 & \multicolumn{2}{c}{Conditional} &  \multicolumn{4}{c}{Influence Function Approach} & \\ 
& \multicolumn{2}{c}{Logit} &  \multicolumn{4}{c}{with $\lambda$ equal to} & \\  
\cmidrule(l{5pt}r{5pt}){2-3}   \cmidrule(l{5pt}r{5pt}){4-7} 
& \multicolumn{1}{c}{Oracle} & \multicolumn{1}{c}{Basic} & \multicolumn{1}{c}{$0$} & \multicolumn{1}{c}{$10^{-5}$} & 
\multicolumn{1}{c}{$10^{-4}$}  & \multicolumn{1}{c}{$2 \cdot 10^{-3}$} & \multicolumn{1}{c}{NN} \\ 
\cmidrule(l{5pt}r{5pt}){2-2}  \cmidrule(l{5pt}r{5pt}){3-3}  \cmidrule(l{5pt}r{5pt}){4-4}  \cmidrule(l{5pt}r{5pt}){5-5}  \cmidrule(l{5pt}r{5pt}){6-6} 
 \cmidrule(l{5pt}r{5pt}){7-7}  \cmidrule(l{5pt}r{5pt}){8-8} 
$ \theta_{cost} \in \widehat{CI}_{cost} $ & 0.95 & 0.00 & 0.93 & 0.92 & 0.83 & 0.40 & 0.99 \\ 
$ \theta_{freq} \in \widehat{CI}_{freq} $ & 0.95 & 0.00 & 0.93 & 0.92 & 0.89 & 0.68 & 1.00 \\ 
$ \theta_{time} \in \widehat{CI}_{time} $ & 0.94 & 0.00 & 0.93 & 0.94 & 0.88 & 0.54 & 1.00 \\ [0.5em] 
$\widehat{se}_{cost} $ & 0.07 & 0.05 & 6.85 & 3.67 & 1.70 & 0.75 & 0.61 \\ 
$\widehat{se}_{freq} $ & 0.10 & 0.07 & 5.25 & 8.19 & 2.26 & 1.24 & 3.56 \\ 
$\widehat{se}_{time} $ & 0.07 & 0.06 & 5.52 & 4.91 & 1.53 & 1.05 & 3.08 \\ [0.5em] 
Bias$_{cost} $ & -0.01 & 0.65 & -4.95 & 0.07 & -0.09 & -0.51 & -0.17 \\ 
Bias$_{freq} $ & -0.00 & 0.59 & 0.61 & -4.45 & -0.77 & -0.61 & -0.18 \\ 
Bias$_{time} $ & -0.01 & 0.80 & -2.58 & -1.49 & -0.27 & -0.57 & -0.17 \\ [0.5em] 
Rej. $\theta_{cost}=0$ & 1.00 & 1.00 & 0.47 & 0.56 & 0.78 & 0.93 & 1.00 \\ 
Rej. $\theta_{freq}=0$ & 1.00 & 1.00 & 0.27 & 0.35 & 0.50 & 0.79 & 0.00 \\ 
Rej. $\theta_{time}=0$ & 1.00 & 1.00 & 0.60 & 0.69 & 0.84 & 0.93 & 0.03 \\ [0.5em] 
$MSE(\Lambda)^{Train}$ & {.}  & {.}  & 5.04 & 5.18 & 5.41 & 5.99 & {.}  \\ 
$MSE(\Lambda)^{Test}$ & {.}  & {.}  & 5.30 & 5.38 & 5.51 & 6.04 & {.}  \\ 
Share Outlier & 0.00 & 0.00 & 0.26 & 0.18 & 0.11 & 0.04 & 0.12 \\ 
  \bottomrule  
 \end{tabular}

 \begin{minipage}{0.827\textwidth} %
		{   \footnotesize \begin{singlespace}
\textit{Note:}		The table reports the average summary statistics over all Monte Carlo replicates for the conditional logit using the true specification (Oracle), the conditional logit using the three variables of interest for the estimation (Basic),  the influence function approach using five different values for $\lambda$ for the estimation of  $\Lambda_s(\bm{w}_i)$, and the neural network (NN), which uses robust standard erros and does not rely on the influence function approach. 
			\end{singlespace}
			\par}
	\end{minipage}
\end{table}
The reported average results for the oracle logit estimator across Monte Carlo replicates reveal that accounting for the correct (functional) form of heterogeneity provides precise estimates of the true average coefficients, and correct coverage of the true average coefficients through the estimated $95\%$ confidence intervals. In addition, the hypotheses tests with the nulls that the average coefficients are zero have high power when calculated with the oracle logit estimator, as the null hypotheses are correctly rejected in every Monte Carlo replicate. 
In contrast, the basic logit estimator, which does not account for any heterogeneity across consumers, performs poorly both in terms of the estimated coefficients and in terms of the coverage of the  confidence intervals. The estimated standard errors of the  oracle logit and the basic logit seem similar but the confidence intervals do not cover the true values of interest in any of the Monte Carlo replicates when estimated with the basic logit. The poor coverage can be explained by the bias of the estimated coefficients, which implies confidence intervals centered around biased estimates. 

The results for the influence function approach depend on the regularization parameter $\lambda$ used for the estimation of $\bm{\Lambda}(\bm{w}_i)$. For $\lambda=0$, the confidence intervals for all three parameters have a coverage of $93\%$, giving the impression that the influence function approach is a valid inference procedure when the heterogeneous coefficient functions are estimated with deep learning and without regularization in the network used to estimate $\bm{\Lambda}(\bm{w}_i)$.  
However, the estimated average coefficients deviate quite substantially from the true values -- especially for the travel cost and travel time coefficients --, and the estimated standard errors are substantially larger than in the oracle logit estimator. The large estimated standard errors explain the correct coverage of the confidence intervals despite of the biased average coefficient estimates. Even though the confidence intervals are centered around biased estimates, they are so large that they cover the true parameters in about $93\%$ of the replicates for all three variables of interest. Moreover, the large estimated standard errors lead to  low power of the hypotheses tests with the nulls that the true coefficients are zero as shown by the small share of rejections of the null hypotheses -- at most in only about $60\%$ of the Monte Carlo replicates.

\begin{figure}[H]%
	\centering
	\captionsetup{justification=centering} 
	\caption{Boxplots of $\hat{\theta}_{freq}$ across Monte Carlo Replicates for Small Data and Different $\lambda$-values}
	\subfigure[No repeated sample splitting ($R=1$)]{\includegraphics[width=1\textwidth]{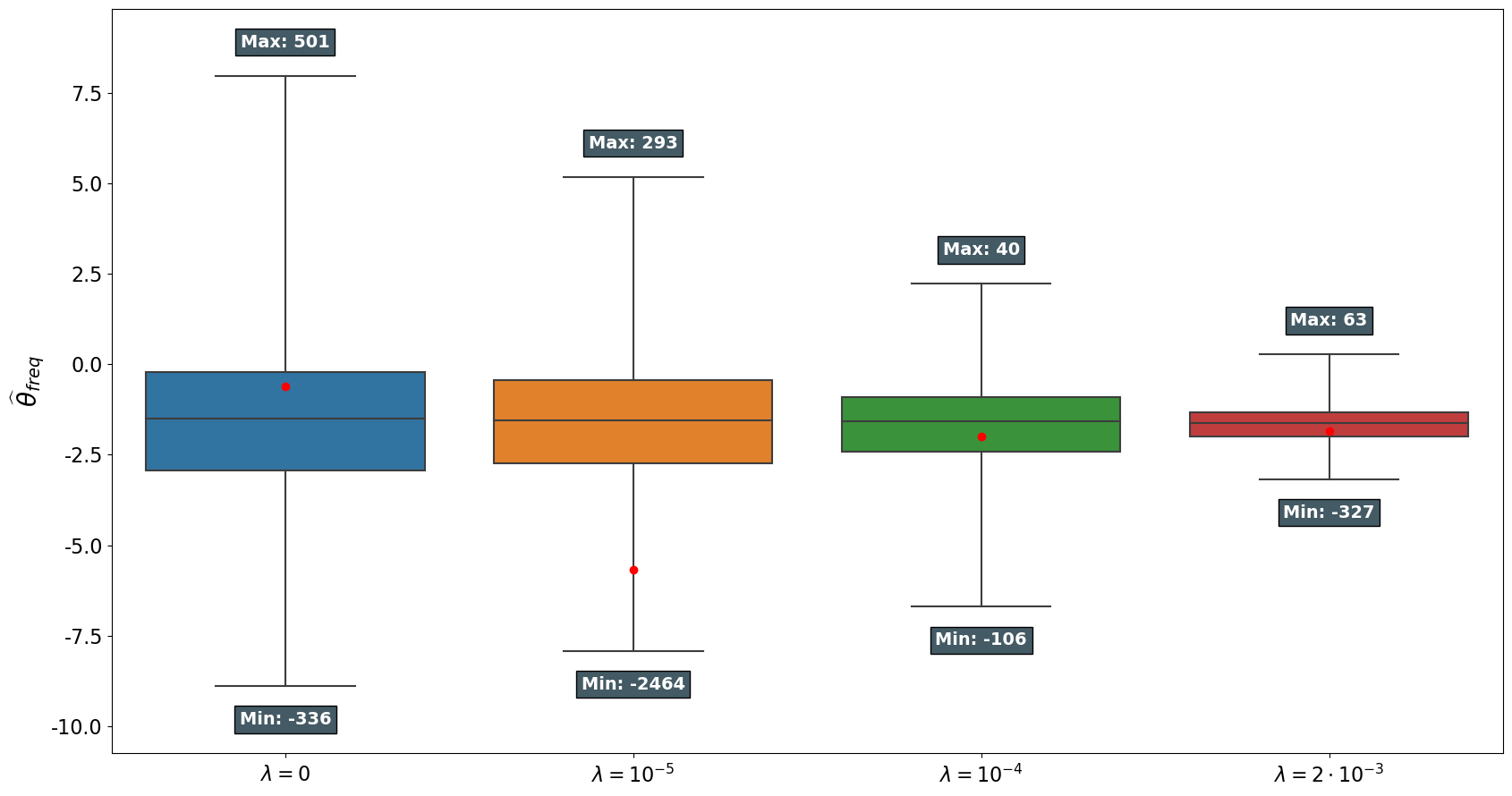}} 
	\subfigure[Repeated sample splitting ($R=5$)]{\includegraphics[width=1\textwidth]{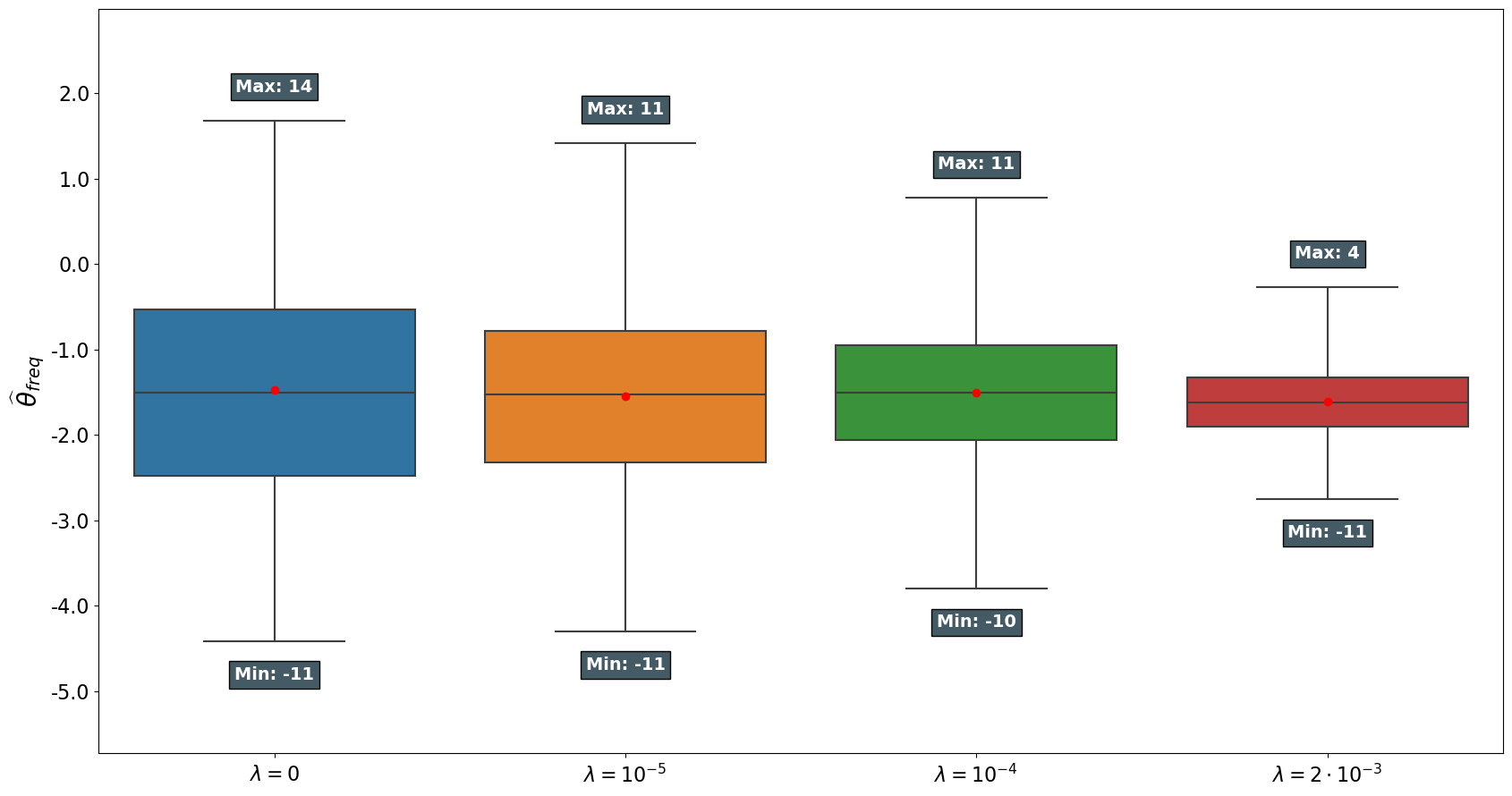}} 
	\label{fig:BoxplotsSmallMC}
	 \begin{minipage}{1\textwidth} %
		{   \footnotesize \begin{singlespace}
\textit{Note:} Panel (a) and (b) show boxplots for the influence function approach, using four different values of $\lambda$.	The colored region within each boxplots highlights the interquartile range (IQR), the horizontal line within the IQR corresponds to the median, and the whiskers indicate the $0.05$ and $0.95$ quantile, respectively. The red dot is the mean across Monte Carlo replicates.
			\end{singlespace}
			\par}
				\end{minipage}
\end{figure}

Overall, choosing $\lambda>0$ leads to more precise estimates of the true average coefficients (considering all three coefficients together, the estimates are most precise for $\lambda=10^{-4}$), and to smaller estimated standard errors. However, the bias of the estimated average coefficients remains relatively large, so that the coverage of the confidence intervals gradually declines with increasing $\lambda$ due to the smaller estimated standard errors with increasing $\lambda$. For instance for $\lambda=2\cdot 10^{-3}$, the confidence intervals have a coverage of only about $68\%$ or less. 
The fact that the estimated coefficients tend to become more precise and the share of outliers decreases with increasing $\lambda$ indicates that the large deviation of the estimated coefficients from the true values for $\lambda=0$ are driven by outliers. This is illustrated by the boxplot of $\hat{\theta}_{freq}$  in Panel (a) of Figure \ref{fig:BoxplotsSmallMC}. The mean (red point) and median (horizontal line inside the colored boxes) values deviate quite substantially, which is due to the high minimum and maximum values of $\hat{\theta}_{freq}$ across Monte Carlo replicates.
The median biases and estimated standard errors across Monte Carlo replicates reported in Table \ref{tab:MC_swissmetroMedian} in Appendix \ref{app:addTables} confirm this impression. The results show that the median of the estimated coefficients across Monte Carlo replicates are closer to the true values for $\lambda=0$ and become less precise with increasing $\lambda$. More importantly, the median of the estimated standard errors are substantially smaller than the mean values across Monte Carlo replicates for each $\lambda$ value. Overall, the median results for different values of $\lambda$ are in line with the expected effect of regularization: The bias increases and the estimated standard errors decrease with increasing $\lambda$. The average of the $MSE$s of the neural network for $\bm{\Lambda}_s(\bm{w}_i)$ in the training and test sample are lowest for $\lambda=0$ and therefore, the $MSE$ may be used to choose an appropriate $\lambda$ value.\footnote{Note that  the $MSE$ in the test sample is also available to the researcher since it is calculated with the left out fold.}

Estimating the average heterogeneous coefficients with a neural network without the influence function approach provides more accurate estimates than the influence function approach. However, the confidence intervals are too wide (the coverage is at least $99\%$ for all three variables), implying that the naive inference procedure with the regular robust standard errors is not valid. This is also indicated by the poor power of the hypotheses tests with the nulls that the average travel time and frequency coefficients are zero, which are rejected in only $3\%$ and $0\%$ of the Monte Carlo replicates, respectively.
The results on the share of outliers reveal that the issue is not unique to the influence function approach but also appears when the parameters are estimated with a neural network and without sample splitting. However, the share is substantially smaller in comparison to the influence function approach with $\lambda=0$, indicating that the smaller samples used for the estimation of the networks due to sample splitting might be one of the reasons causing the issue. The Monte Carlo experiment in Subsection \ref{subsecDL:largeMC} studies the performance of the influence function approach for a larger sample size.\\

To resolve the sensitivity of the estimated results to potential outliers, we apply the repeated sample splitting procedure outlined in Remark \ref{rem:repeatedCV}. Table \ref{tab:MC_swissmetroMean_cross} reports the results for the sample splitting procedure with $R=5$ repetitions.\footnote{To reduce computation time, we only employ repeated sample splitting if we observe an outlier in the first repetition of each Monte Carlo run.}
\begin{table}[h!]
\centering	
\captionsetup{justification=centering}
	\caption{Average Summary Statistics of 1000 Monte Carlo Replicates for Small Data and Repeated Sample Splitting with $R=5$}
	\label{tab:MC_swissmetroMean_cross}
	\small
\begin{tabular}{  l@{\hspace{-0.5em}}  *{7}S[table-format=2.4] } 
\toprule 
 & \multicolumn{2}{c}{Conditional} &  \multicolumn{4}{c}{Influence Function Approach} & \\ 
& \multicolumn{2}{c}{Logit} &  \multicolumn{4}{c}{with $\lambda$ equal to} & \\  
\cmidrule(l{5pt}r{5pt}){2-3}   \cmidrule(l{5pt}r{5pt}){4-7} 
& \multicolumn{1}{c}{Oracle} & \multicolumn{1}{c}{Basic} & \multicolumn{1}{c}{$0$} & \multicolumn{1}{c}{$10^{-5}$} & 
\multicolumn{1}{c}{$10^{-4}$}  & \multicolumn{1}{c}{$2 \cdot 10^{-3}$} & \multicolumn{1}{c}{NN} \\ 
\cmidrule(l{5pt}r{5pt}){2-2}  \cmidrule(l{5pt}r{5pt}){3-3}  \cmidrule(l{5pt}r{5pt}){4-4}  \cmidrule(l{5pt}r{5pt}){5-5}  \cmidrule(l{5pt}r{5pt}){6-6} 
 \cmidrule(l{5pt}r{5pt}){7-7}  \cmidrule(l{5pt}r{5pt}){8-8} 
$ \theta_{cost} \in \widehat{CI}_{cost} $ & 0.94 & 0.00 & 0.94 & 0.93 & 0.82 & 0.42 & 0.99 \\ 
$ \theta_{freq} \in \widehat{CI}_{freq} $ & 0.96 & 0.00 & 0.94 & 0.92 & 0.90 & 0.66 & 1.00 \\ 
$ \theta_{time} \in \widehat{CI}_{time} $ & 0.96 & 0.00 & 0.95 & 0.95 & 0.87 & 0.59 & 1.00 \\ [0.5em] 
$\widehat{se}_{cost} $ & 0.07 & 0.05 & 1.61 & 1.34 & 0.72 & 0.32 & 0.61 \\ 
$\widehat{se}_{freq} $ & 0.10 & 0.07 & 1.91 & 1.64 & 1.10 & 0.58 & 3.65 \\ 
$\widehat{se}_{time} $ & 0.07 & 0.06 & 1.59 & 1.19 & 0.68 & 0.43 & 3.13 \\ [0.5em] 
Bias$_{cost} $ & -0.01 & 0.65 & -0.29 & -0.30 & -0.32 & -0.41 & -0.18 \\ 
Bias$_{freq} $ & -0.00 & 0.59 & -0.26 & -0.32 & -0.28 & -0.39 & -0.19 \\ 
Bias$_{time} $ & -0.00 & 0.81 & -0.02 & -0.16 & -0.23 & -0.36 & -0.17 \\ [0.5em] 
Rej. $\theta_{cost}=0$ & 1.00 & 1.00 & 0.48 & 0.61 & 0.84 & 0.96 & 0.99 \\ 
Rej. $\theta_{freq}=0$ & 1.00 & 1.00 & 0.25 & 0.32 & 0.54 & 0.81 & 0.00 \\ 
Rej. $\theta_{time}=0$ & 1.00 & 1.00 & 0.64 & 0.78 & 0.92 & 0.96 & 0.02 \\ [0.5em] 
$MSE(\Lambda)^{Train}$ & {.}  & {.}  & 5.07 & 5.23 & 5.46 & 6.04 & {.}  \\ 
$MSE(\Lambda)^{Test}$ & {.}  & {.}  & 5.30 & 5.37 & 5.51 & 6.03 & {.}  \\ 
Share Outlier & 0.00 & 0.00 & 0.05 & 0.02 & 0.00 & 0.00 & 0.12 \\ 
  \bottomrule  
 \end{tabular}

 \begin{minipage}{0.827\textwidth} %
		{   \footnotesize \begin{singlespace}
\textit{Note:}		The table reports the average summary statistics over all Monte Carlo replicates for the conditional logit using the true specification (Oracle), the conditional logit using the three variables of interest for the estimation (Basic),  the influence function approach, using five different values of $\lambda$ for the estimation of  $\Lambda_s(\bm{w}_i)$, and the neural network (NN), which uses robust standard erros and does not rely on the influence function approach. 
			\end{singlespace}
			\par}
	\end{minipage}
\end{table}
The repeated sample splitting reduces the share of outliers substantially in comparison to the approach without repeated sample splitting. In fact, for $\lambda\geq 10^{-4}$, there are no outliers anymore. Comparing Panel (a) and (b) in Figure \ref{fig:BoxplotsSmallMC} illustrates that the estimates vary less across Monte Carlo replicates when estimated with repeated sample splitting. Furthermore, the less extreme minimum and maximum values indicate that the extreme outliers are removed. Accordingly, the mean and median values are closer to each other when the coefficient functions are estimated with repeated sample splitting. The reduced share of outliers leads to more precise estimates of the average coefficients and to smaller estimated standard errors. In contrast to the influence function approach without repeated sample splitting, the overall average bias of the estimated average coefficients is smallest for $\lambda=0$ and increases with increasing $\lambda$. With respect to the confidence intervals, the coverage for $\lambda=0$ is $94\%$ for the travel cost and frequency coefficients, and $95\%$ for the travel time coefficient. The coverage of the confidence intervals gradually decreases with $\lambda$. While for $\lambda=10^{-5}$ the coverage is below but still close to $95\%$ (for the travel time it is exactly $95\%$), the coverage for $\lambda=2\cdot 10^{-3}$ is at most $66\%$ (for the travel cost coefficient, the coverage of the confidence interval is just $42\%$).
Thus, the influence function approach with repeated sample splitting and regularizer $\lambda=0$ allows to precisely estimate average effects across travelers and provides a valid inference procedure. Using a regularizer $\lambda>0$ increases the average bias and decreases the estimated variance of the coefficients.
The combination of increasing bias and decreasing magnitude of the estimated standard errors with increasing $\lambda$ leads to inappropriately small confidence intervals centered around biased estimates and, hence, to a poor coverage of the true values. Based on these results, we do not recommend using regularization in the form of a $l_2$-penalty with $\lambda > 0$ in the network used to estimate $\bm{\Lambda}(\bm{w}_i)$ to stabilize the inference procedure but to rather rely on repeated sample splitting.
However, even for the repeated sample splitting, the estimated standard errors are substantially larger than those in the oracle logit model. This leads to a poor power as indicated by the rare rejection of the false null hypotheses that the true average coefficients are zero, which are rejected in only about $48\%$, $25\%$, and $64\%$ of the Monte Carlo replicates for the travel cost parameter, the frequency parameter, and  the travel time parameter, respectively, for $\lambda=0$.

\begin{figure}[H]%
	\centering
	\captionsetup{justification=centering} 
	\caption{Density of Estimated $t$-Statistic of $\hat{\theta}_{cost}$ for Different Estimators}
	\subfigure[No repeated sample splitting ($R=1$)]{\includegraphics[width=1\textwidth]{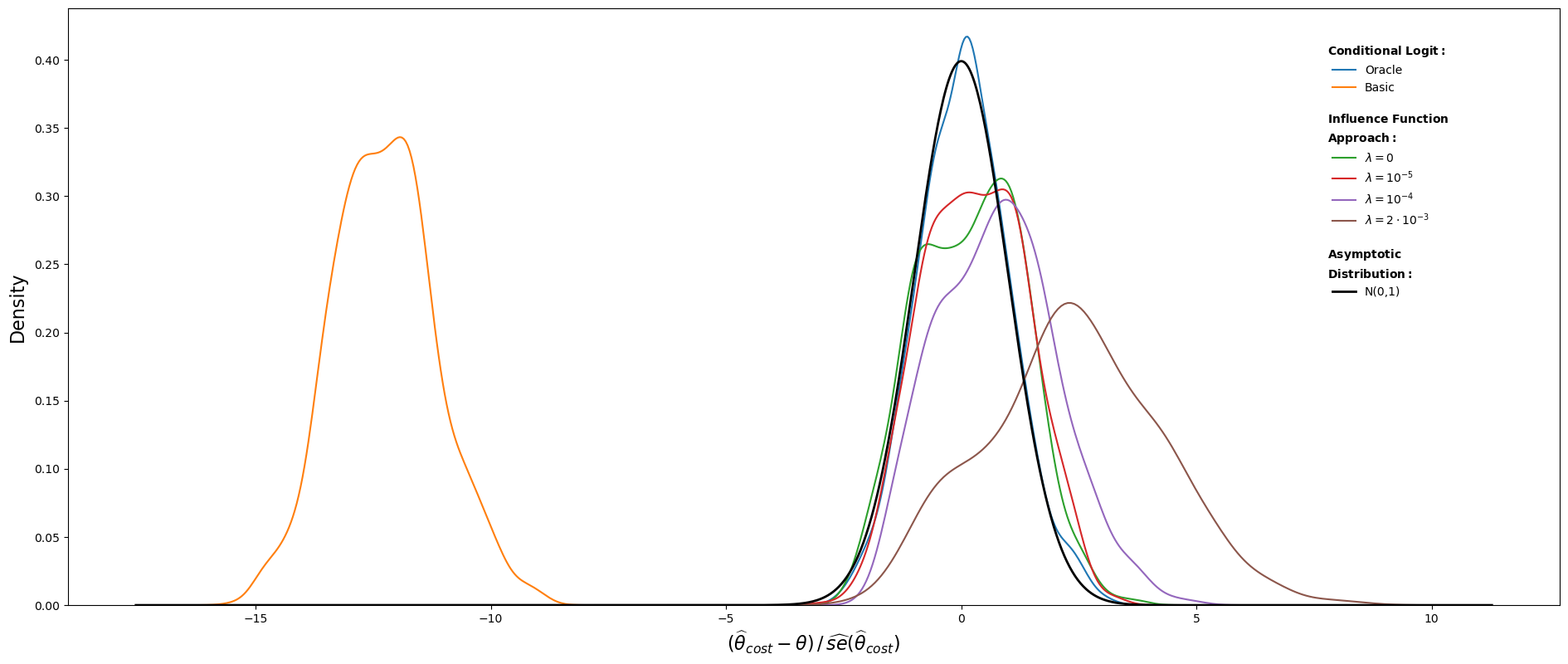}}
	\subfigure[Repeated sample splitting ($R=5$)]{\includegraphics[width=1\textwidth]{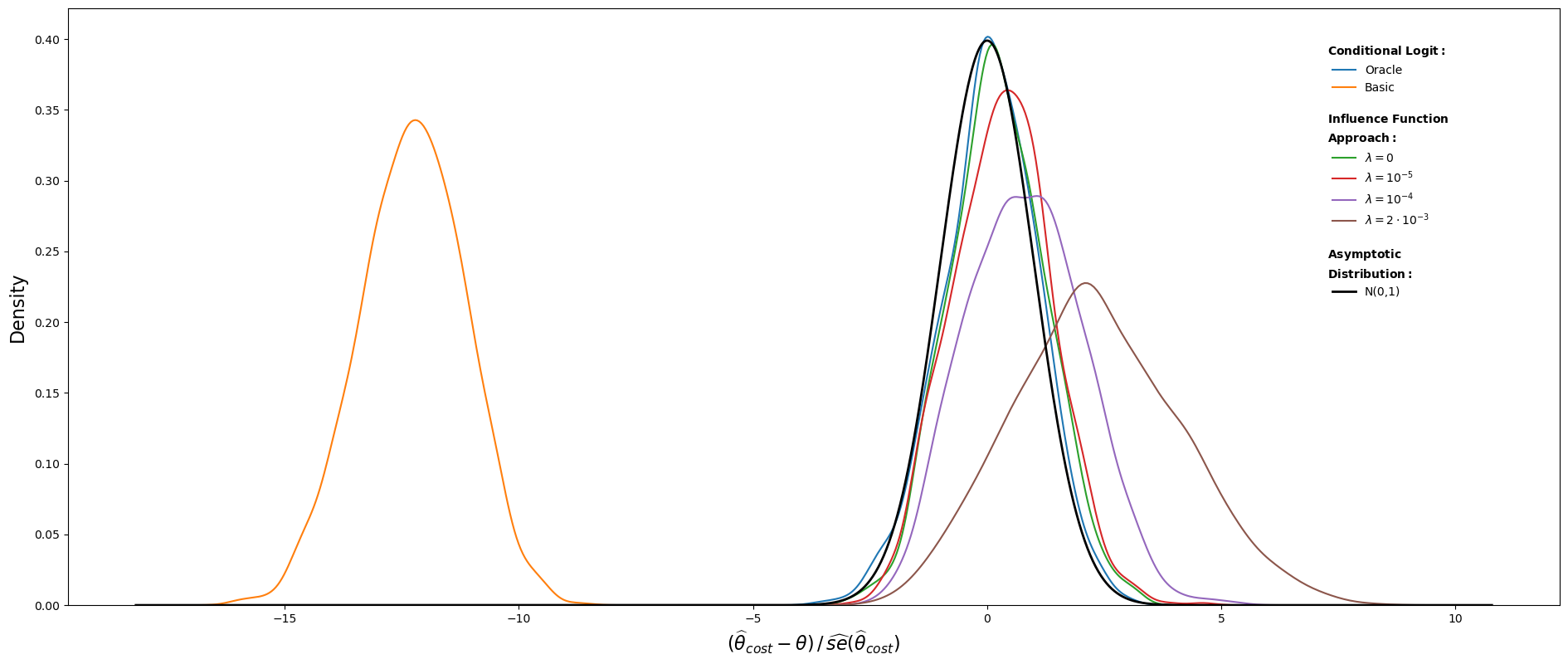}}
	\label{fig:asymptDistrSmallMC}
	 \begin{minipage}{1\textwidth} %
		{   \footnotesize \begin{singlespace}
\textit{Note:} The plot shows kernel density estimates of the estimated $t$-statistic for the conditional logit using the true specification (Oracle), the conditional logit using the three variables of interest for the estimation (Basic),  the influence function approach, using four different values for $\lambda$ for the estimation of  $\Lambda_s(\bm{w}_i)$. Additionally, the standard normal distribution is included.
			\end{singlespace}
			\par}
				\end{minipage}
\end{figure}

Figure \ref{fig:asymptDistrSmallMC} shows the estimated densities of $ (\hat{\theta}_{cost}-\theta) / \hat{se}(\hat{\theta}_{cost})$ for the oracle logit  estimator, the basic logit estimator, and the influence function approach for different values of $\lambda$. The limiting distribution of the influence function approach is the standard normal as stated in Equation \eqref{eq:psiNormal}. First, the figure illustrates the bias of the basic logit estimator and illustrates that the estimated  $t$-statistics of the oracle logit estimator are well approximated by a standard normal distribution. Second, comparing Panel (a) and Panel (b) reveals that the estimates obtained with the influence function approach only seem to be close to the standard normal distribution when repeated sample splitting is used and $\lambda= 0$ or  $\lambda=10^{-5}$.

\begin{remark} 
Beyond  repeated sample splitting, we conduct several other adjustments of the estimation procedure that are intended to reduce  outliers observed in some Monte Carlo replicates. We considered taking the median instead of the average  in Equation \eqref{eq:DLestTheta}  and \eqref{eq:DLestPsi}, i.e., replacing $\hat{\theta}=\frac{1}{S}\sum\limits\hat{\theta}_s$  by $\hat{\theta}=median\left\{\hat{\theta}_s \right\}_{s=1}^{S}$ and  $\hat{\Psi}=\frac{1}{S}\sum\limits_{s=1}^S\hat{\Psi}_s$ by $\hat{\Psi}=median\left\{\hat{\Psi}_s \right\}_{s=1}^{S}$. This leads to smaller estimated standard errors but also to a lower average coverage across Monte Carlo replicates ( $ < 0.85$), indicating that the bias remains large. Furthermore, we also apply the modification suggested by \citeA{farrell2020} and add a constant $c$ to the diagonal elements of $\hat{\bm{\Lambda}}_s(\bm{w}_i)$. For $c=1$, the coverage is quite poor, and $c=10^{-5}$ seems to have no impact on the results. That is, the choice of the constant $c$ seems to require further tuning which we did not investigate further.
\end{remark}

\subsection{Large Data Set}  \label{subsecDL:largeMC}
The following Monte Carlo experiment aims to analyze whether the results of the previous experiment persist for larger sample sizes. 
For that purpose, we revisit the Swissmetro data set and use the same specification as before. However, we now sample the socio-demographic characteristics and the covariates of interest with replacement from the original data set such that we obtain $50,000$ travelers choosing among the three alternatives. With respect to the socio-demographic characteristics, we randomly generate new travelers by drawing from the values of \textit{income}, \textit{age}, \textit{gender}, and \textit{who}.
Because we sample independently across characteristics, we create new types of travelers characterized through new combinations of socio-demographic variables.

With respect to the covariates of interest, we make sure that we randomly draw the travel time, travel cost, and frequency for a specific alternative only from the the values for the specific alternative existing in the data (e.g., the cost variable for alternative car can only take values of existing values of the cost variable for cars). However, for a given alternative, we draw the covariates independently across variables from different choice situations.
Otherwise, the Monte Carlo study is the same as the one presented above.\\ 
\begin{table}[b!]
\centering
\captionsetup{justification=centering}
	\caption{Average Summary Statistics of 1000 Monte Carlo Replicates for Large Data and Repeated Sample Splitting with $R=5$}
	\label{tab:MC_swissmetroLargeMean_cross}
	\small
\begin{tabular}{  l@{\hspace{-0.5em}}  *{7}S[table-format=2.4] } 
\toprule 
 & \multicolumn{2}{c}{Conditional} &  \multicolumn{4}{c}{Influence Function Approach} & \\ 
& \multicolumn{2}{c}{Logit} &  \multicolumn{4}{c}{with $\lambda$ equal to} & \\  
\cmidrule(l{5pt}r{5pt}){2-3}   \cmidrule(l{5pt}r{5pt}){4-7} 
& \multicolumn{1}{c}{Oracle} & \multicolumn{1}{c}{Basic} & \multicolumn{1}{c}{$0$} & \multicolumn{1}{c}{$10^{-5}$} & 
\multicolumn{1}{c}{$10^{-4}$}  & \multicolumn{1}{c}{$2 \cdot 10^{-3}$} & \multicolumn{1}{c}{NN} \\ 
\cmidrule(l{5pt}r{5pt}){2-2}  \cmidrule(l{5pt}r{5pt}){3-3}  \cmidrule(l{5pt}r{5pt}){4-4}  \cmidrule(l{5pt}r{5pt}){5-5}  \cmidrule(l{5pt}r{5pt}){6-6} 
 \cmidrule(l{5pt}r{5pt}){7-7}  \cmidrule(l{5pt}r{5pt}){8-8} 
$ \theta_{cost} \in \widehat{CI}_{cost} $ & 0.95 & 0.00 & 0.94 & 0.92 & 0.91 & 0.76 & 1.00 \\ 
$ \theta_{freq} \in \widehat{CI}_{freq} $ & 0.95 & 0.00 & 0.95 & 0.93 & 0.90 & 0.83 & 1.00 \\ 
$ \theta_{time} \in \widehat{CI}_{time} $ & 0.94 & 0.00 & 0.95 & 0.94 & 0.92 & 0.86 & 1.00 \\ [0.5em] 
$\widehat{se}_{cost} $ & 0.02 & 0.02 & 0.50 & 0.41 & 0.35 & 0.12 & 0.49 \\ 
$\widehat{se}_{freq} $ & 0.04 & 0.04 & 0.80 & 0.59 & 0.44 & 0.14 & 1.72 \\ 
$\widehat{se}_{time} $ & 0.03 & 0.02 & 0.45 & 0.36 & 0.29 & 0.14 & 1.27 \\ [0.5em] 
Bias$_{cost} $ & 0.00 & 0.60 & -0.05 & -0.02 & -0.05 & -0.05 & -0.03 \\ 
Bias$_{freq} $ & 0.00 & 0.53 & -0.09 & -0.07 & -0.06 & -0.05 & -0.03 \\ 
Bias$_{time} $ & 0.00 & 0.78 & 0.01 & 0.02 & -0.03 & -0.04 & -0.02 \\ [0.5em] 
Rej. $\theta_{cost}=0$ & 1.00 & 1.00 & 0.89 & 0.91 & 0.93 & 0.98 & 1.00 \\ 
Rej. $\theta_{freq}=0$ & 1.00 & 1.00 & 0.62 & 0.75 & 0.84 & 0.96 & 0.00 \\ 
Rej. $\theta_{time}=0$ & 1.00 & 1.00 & 0.95 & 0.96 & 0.97 & 0.99 & 0.94 \\ [0.5em] 
$MSE(\Lambda)^{Train}$ & {.}  & {.}  & 8.80 & 8.84 & 8.90 & 9.23 & {.}  \\ 
$MSE(\Lambda)^{Test}$ & {.}  & {.}  & 8.88 & 8.87 & 8.91 & 9.23 & {.}  \\ 
Share Outlier & 0.00 & 0.00 & 0.00 & 0.00 & 0.00 & 0.00 & 0.00 \\ 
  \bottomrule  
 \end{tabular}

 \begin{minipage}{0.827\textwidth} %
		{   \footnotesize \begin{singlespace}
\textit{Note:}		The table reports the average summary statistics over all Monte Carlo replicates for the conditional logit using the true specification (Oracle), the conditional logit using the three variables of interest for the estimation (Basic),  the influence function approach, using five different values for $\lambda$ for the estimation of  $\bm{\Lambda}_s(\bm{w}_i)$, and the neural network (NN), which uses robust standard erros and does not rely on the influence function approach.
			\end{singlespace}
			\par}
	\end{minipage}
\end{table}

Table \ref{tab:MC_swissmetroLargeMean_cross} reports the average Monte Carlo results for $N=50,000$ and when the influence function approach is estimated with repeated sample splitting. 
The results for the oracle logit and the basic logit are similar to those obtained for the small sample size. For the oracle logit, the average estimated bias across Monte Carlo replicates is (almost) zero and the confidence intervals cover the true frequency and travel time coefficients in $95\%$ of the Monte Carlo replicates, and the true travel cost coefficient in $94\%$. For the basic logit model, the standard errors of the estimated coefficients are  similar to those of the oracle logit. Nevertheless, the confidence intervals have zero coverage due to the substantial bias of the estimated average coefficients. 

\begin{figure}[H]%
	\centering
	\captionsetup{justification=centering}
	\caption{Boxplots of $\hat{\theta}_{freq}$ across Monte Carlo Replicates for Large Data and Different $\lambda$-values}
	\subfigure[No repeated sample splitting ($R=1$)]{\includegraphics[width=1\textwidth]{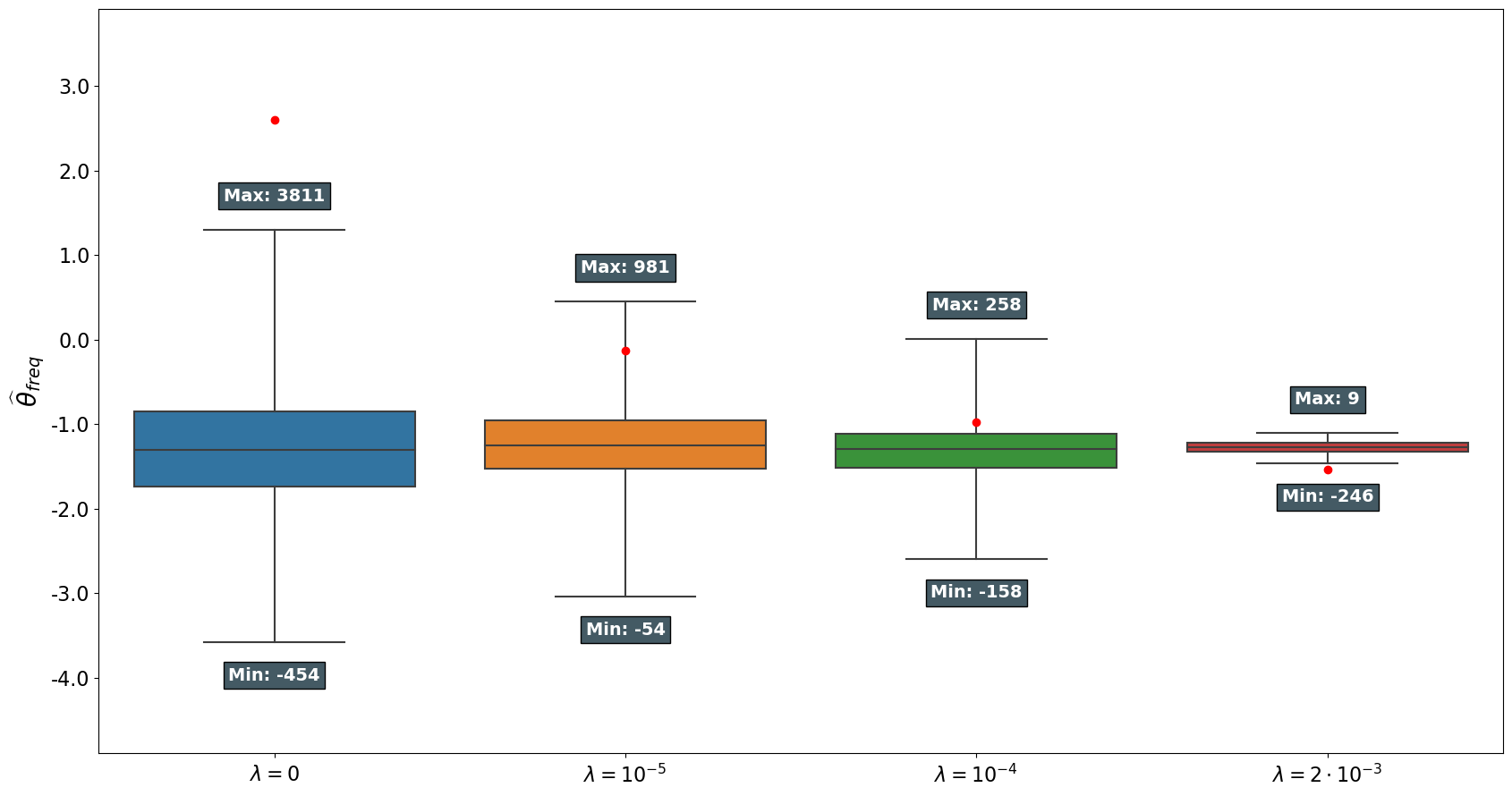}} 
	\subfigure[Repeated sample splitting ($R=5$)]{\includegraphics[width=1\textwidth]{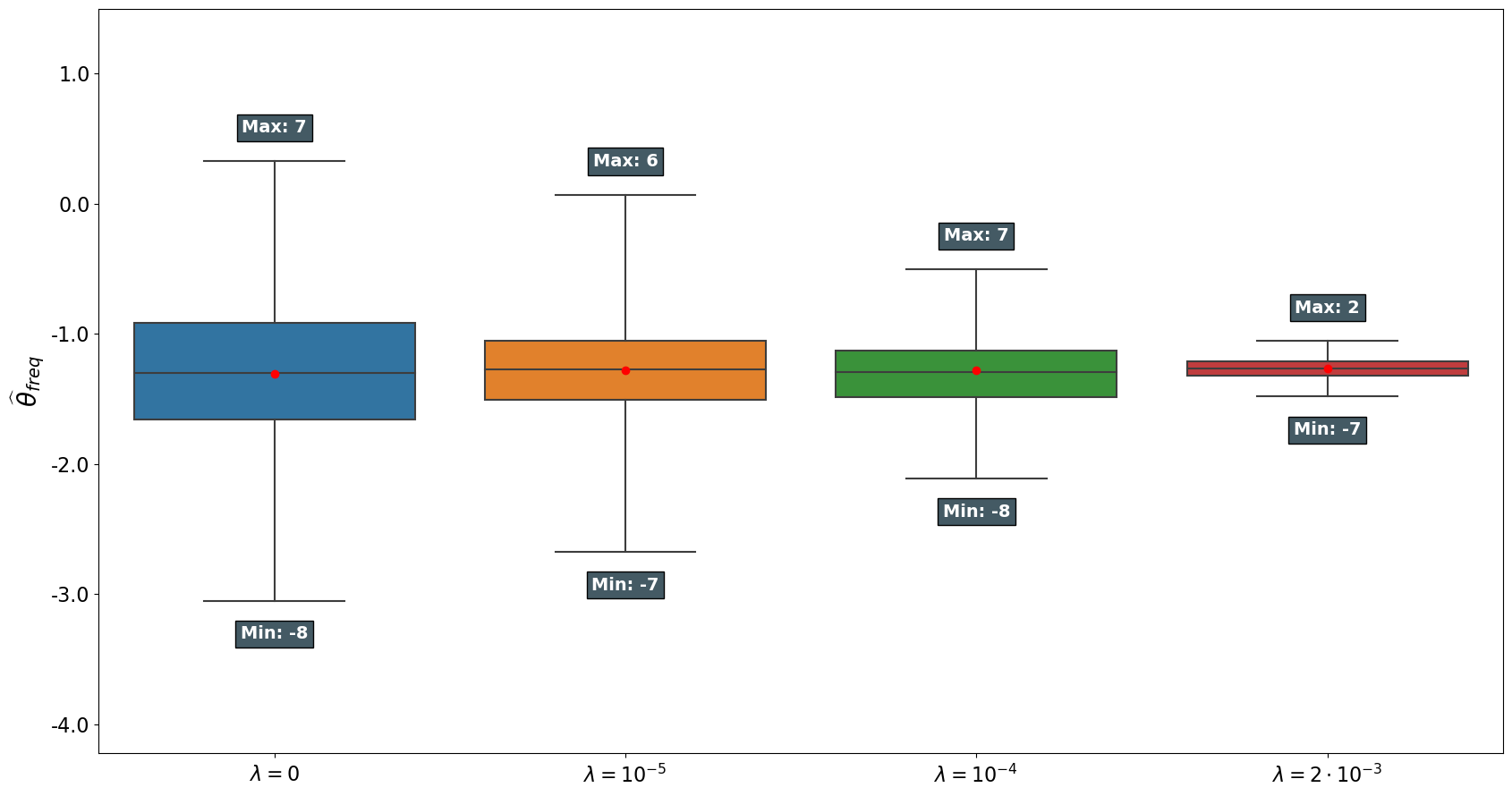}} 
	\label{fig:BoxplotsLargeMC}
	 \begin{minipage}{1\textwidth} %
		{   \footnotesize \begin{singlespace}
\textit{Note:} Panel (a) and (b) show boxplots for the influence function approach, using four different values of $\lambda$.	The colored region within each boxplots highlights the interquartile range (IQR), the horizontal line within the IQR corresponds to the median, and the whiskers indicate the $0.05$ and $0.95$ quantile, respectively. The red dot is the mean across Monte Carlo replicates.
			\end{singlespace}
			\par}
				\end{minipage}
\end{figure}

For the influence function approach with repeated sample splitting, the estimated coefficients are almost as precise as those estimated with the oracle logit model, independent of $\lambda$ (i.e., the average values vary only slightly across different values for $\lambda$), and the estimated standard errors are substantially smaller in comparison to the results for the small sample size. However, they are still larger than those estimated with the oracle logit estimator.
For $\lambda=0$, the confidence intervals have the correct coverage (they cover the true travel cost parameter in $94\%$, and the true frequency and travel time parameters in $95\%$ of the Monte Carlo replicates). For $\lambda>0$, the coverage of the confidence intervals decreases below $95\%$, which is the result of the declining estimated standard errors with increasing $\lambda$. However, the coverage declines not as rapidly with increasing $\lambda$ as observed for the small sample size. 
With respect to the power of the hypotheses tests with the nulls that the coefficients are  zero, the percentage of rejections of the incorrect null hypothesis are substantially larger for $\lambda=0$ than for the small sample size -- in $89\%$ of the Monte Carlo replicates for the travel time coefficient, $62\%$ for the frequency coefficient, and $95\%$ for the travel time coefficient.
Even though the share of outliers for the influence function approach decreases substantially compared to the Monte Carlo experiment with the small sample size, repeated sample splitting seems still necessary as the mean deviates substantially from the median when no repeated sample splitting is used (cf. Table \ref{tab:MC_swissmetroLargeMean} and Table \ref{tab:MC_swissmetroLargeMedian} and Figure  \ref{fig:BoxplotsLargeMC}).
With respect to the estimation of the coefficient functions with a deep neural network and naive inference, we observe a similar improvement when increasing the sample size as for the influence function approach. The estimated average coefficients become more precise -- they are similarly precise as those obtained with the oracle logit -- and the estimated standard errors become smaller. A potential explanation for the more precise coefficient estimates and the smaller standard errors might be the fact that the issue with the outlier disappears completely, both for the influence function approach with repeated sample splitting (even for $\lambda=0$) and when only the coefficient functions are estimated with the neural network.
However, the confidence intervals remain too wide, confirming the impression from the experiments with the small sample size that regular robust standard errors calculated with parameters estimated with deep learning are not a valid inference procedure.

\section{Application} \label{secDL:Application}

This section applies the estimation procedure presented in Section \ref{secDL:deep_learning_logit} to the Swissmetro dataset. 
We consider the same utility specification as in the Monte Carlo experiments. That is, we include alternative-specific constants (car remains the reference category) along with the travel cost, frequency, and travel time, i.e.,  
\[ \bm{\delta}(\bm{w}_i)= \left(\alpha_{\text{train}}\left(\bm{w}_i\right), \alpha_{\text{sm}}\left(\bm{w}_i\right), \beta^{\text{cost}}\left(\bm{w}_i\right),\beta^\text{freq}\left(\bm{w}_i\right),  \beta^{\text{time}}\left(\bm{w}_i\right) \right)^{\prime}. \] 
We estimate the model with the influence function approach using 
\begin{equation}
\bm{w}_i := (age_i,\, income_i,\, who^1_i,\, who^2_i,\, who^3_i,\,luggage_i)^{\prime}\notag 
\end{equation}
as the set of input variables to the network. The variable $luggage$ is an ordinal variable with information on the pieces of luggage a traveler carries on her trip. It is zero if the traveler carries no luggage, $1$ if she carries one piece, and $3$ if she carries several pieces. 

As a benchmark, we estimate a conditional logit model and a nested logit model. In comparison to the conditional logit model, the nested logit allows for more realistic substitution patterns across alternatives (it does not exhibit the IIA property with respect to alternatives across nests). For the nested logit model, we follow \citeA{bierlaire2001} and group the alternatives car and train in one nest (representing existing alternatives), and Swissmetro in another  nest (representing the newly introduced alternative).\footnote{Since the nest including the alternative Swissmetro is a degenerate nest, we estimate an unscaled version of the nested logit in order to make the identification of the dissimilarity parameter feasible  \cite<see, e.g.,>{heiss2002}.}  
For both models, we use the same utility specification as for the influence function approach, except that we model the coefficients as linear functions of the input variables $\bm{w}_i$. More precisely, in addition to alternative-specific constants and the variables travel cost, frequency, and travel time, we include interactions of the alternative-specific constants and the variables of interest with each of the variables in $\bm{w}_i$.\footnote{Interacting the alternative-specific constants with  $\bm{w}_i$ yields multinomial coefficients for each variable in $\bm{w}_i$.}
Similarly to the Monte Carlo experiments, we also include  a neural network estimated with the full training sample as a benchmark. For the neural network, we conduct naive inference using robust standard errors for the estimated coefficient  functions.
For the influence function approach and for the neural network approach with naive inference, we use the same network architectures as in the Monte Carlo experiment. In line with the results from the Monte Carlo experiments, we use repeated sample splitting with $R=5$ repetitions and set $\lambda=0$ in the network for the estimation of $\bm{\Lambda}(\bm{w}_i)$ when estimating the model with the influence function approach, as $\lambda > 0$ provides incorrect coverage of the confidence intervals in the Monte Carlo experiments.

For the estimation, we follow \citeA{sifringer2020} and split the $9,036$ observations into a training and a test set which consist of three and one quarter of the total observations, respectively. 
We use the test set to compare the out-of-sample performance of the influence function approach to the benchmark models. 
Table \ref{tabDL:MC_swissmetroMean_cross} reports the average heterogeneous coefficient functions for the travel cost, frequency, and travel time and their corresponding estimated standard errors.
\begin{table}[h!]
\centering	
\captionsetup{justification=centering}
	\caption{Estimated Average Travel Cost, Frequency and Travel Time Parameters and Corresponding Estimated Standard Errors}
	\label{tabDL:MC_swissmetroMean_cross} 
	\tabcolsep=0.5cm 
	
\begin{tabular}{  l@{\hspace{1em}}  *{4}S[table-format=2.4] } 
\toprule 
& \multicolumn{1}{c}{CL} & \multicolumn{1}{c}{NL} & \multicolumn{1}{c}{IFA} & \multicolumn{1}{c}{NN}   \\ 
\cmidrule(l{5pt}r{5pt}){2-2}  \cmidrule(l{5pt}r{5pt}){3-3}  \cmidrule(l{5pt}r{5pt}){4-4}  \cmidrule(l{5pt}r{5pt}){5-5}  
$ \hat{\theta}_{cost} $ 	& -1.144 & -1.418 & -1.849  &  -1.943 \\ 
$ \hat{\theta}_{freq}  $ 	& -0.891 & -0.966 & -1.040  &  -1.106 \\ 
$ \hat{\theta}_{time}  $ 	& -1.368 & -1.728 & -1.797  &  -2.172 \\ [0.5em] 
$\hat{se}_{cost} $ 			& 0.061  &  0.078 &  0.954  &  1.343 \\ 
$\hat{se}_{freq} $ 			& 0.129  &  0.154 &  2.440  &  2.476 \\ 
$\hat{se}_{time} $			& 0.085  &  0.099 &  2.119  &  1.375 \\ [0.5em] 
LL$^{Train}$ 				& -0.763 & -0.762 & -0.655  &  -0.638 \\ 
LL$^{Test}$  				& -0.777 & -0.772 & -0.753  &  -0.695 \\ 
  \bottomrule  
 \end{tabular}

 \begin{minipage}{0.69\textwidth} %
		{   \footnotesize \begin{singlespace}
\textit{Note:} The table reports the estimated average coefficients and the standard errors three variables of interest, and the in- and out-of-sample log-likelihood for the conditional logit (CL), the nested logit (NL), the influence function approach with $\lambda=0$ and repeated sample splitting with $R=5$ (IFA), and the neural network with naive inference using robust standard errors (NN).
			\end{singlespace}
			\par}
	\end{minipage}
\end{table}

Additionally, we calculate the in- and out-of-sample log-likelihood per observation. Both the in- and out-of-sample log-likelihood increases with increasing flexibility of the estimation approach. While there is only a slight improvement when going from the conditional logit to the nested logit model, the influence function approach has a substantially higher in-sample as well as out-of-sample log-likelihood. 
With respect to the estimated average coefficients, all four estimators estimate the same sign. Travelers find alternatives with higher travel cost, frequency, and travel time less attractive.\footnote{Frequency is calculated as average minutes of waiting time  for a given transportation mode, i.e.,  a higher frequency variable implies less frequent connections.} 
The estimated average coefficients are smallest in magnitude when the model is estimated with the conditional logit model and increase in magnitude with increasing out-of-sample log-likelihood, which is especially the case for the travel cost coefficient.
The results for the estimated standard errors are in line with the results from the Monte Carlo experiments, as the estimated standard errors of the influence function approach are substantially larger than those of the conditional and nested logit model. In fact, for the influence function approach, none of the estimated average coefficients is significantly different from zero, highlighting that larger samples might be needed for the influence function approach compared to traditional logit models.

\begin{figure}[H]%
	\centering
	\captionsetup{justification=centering}
	\caption{Histograms of  Estimated Coefficient Functions for Influence Function Approach  and Nested Logit Model}
	\subfigure[Cost (IFA)]{\includegraphics[width=0.49\textwidth, height=0.13\textheight]{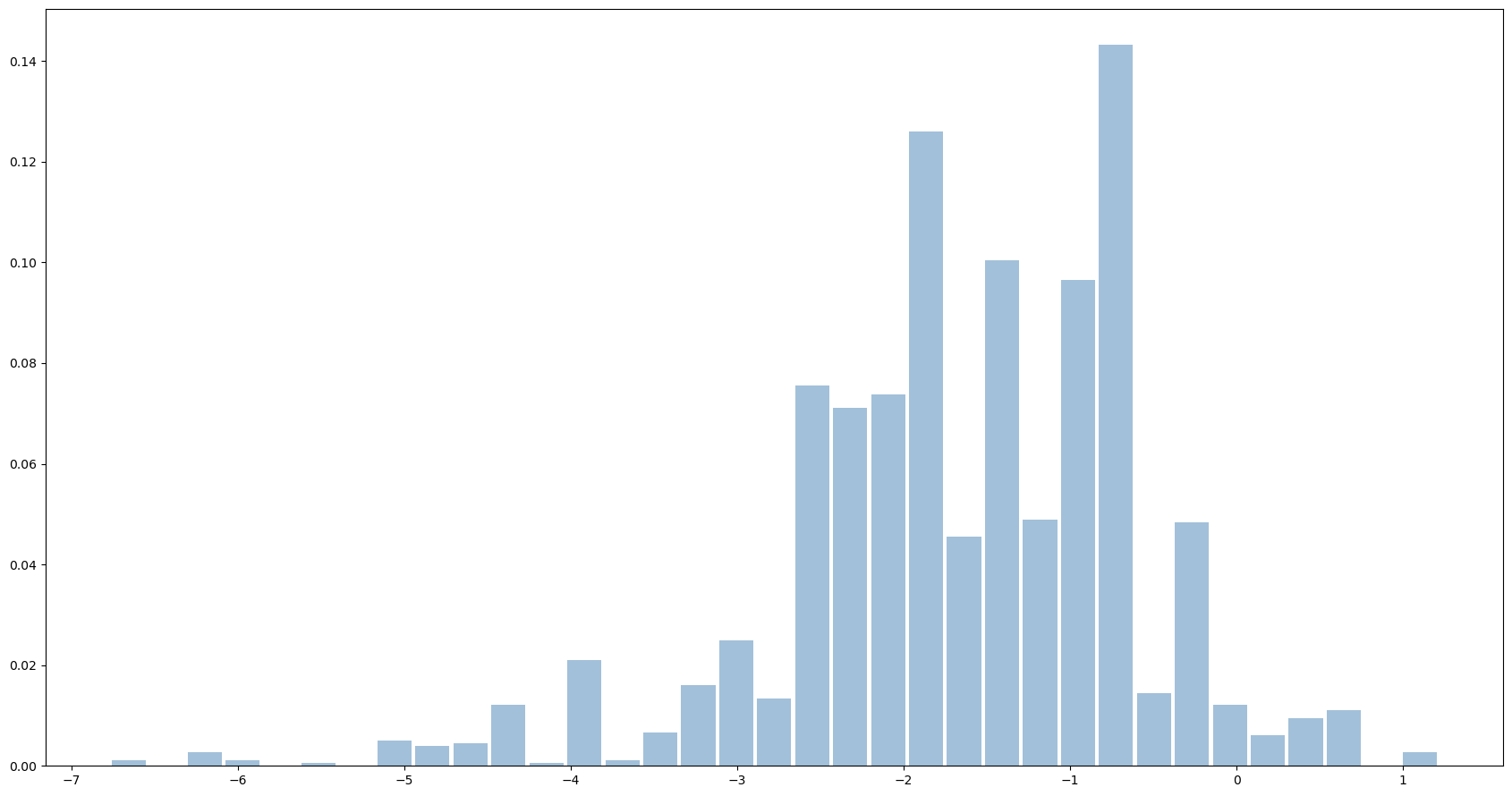}} 
		\subfigure[Cost (Nested Logit)]{\includegraphics[width=0.49\textwidth, height=0.13\textheight]{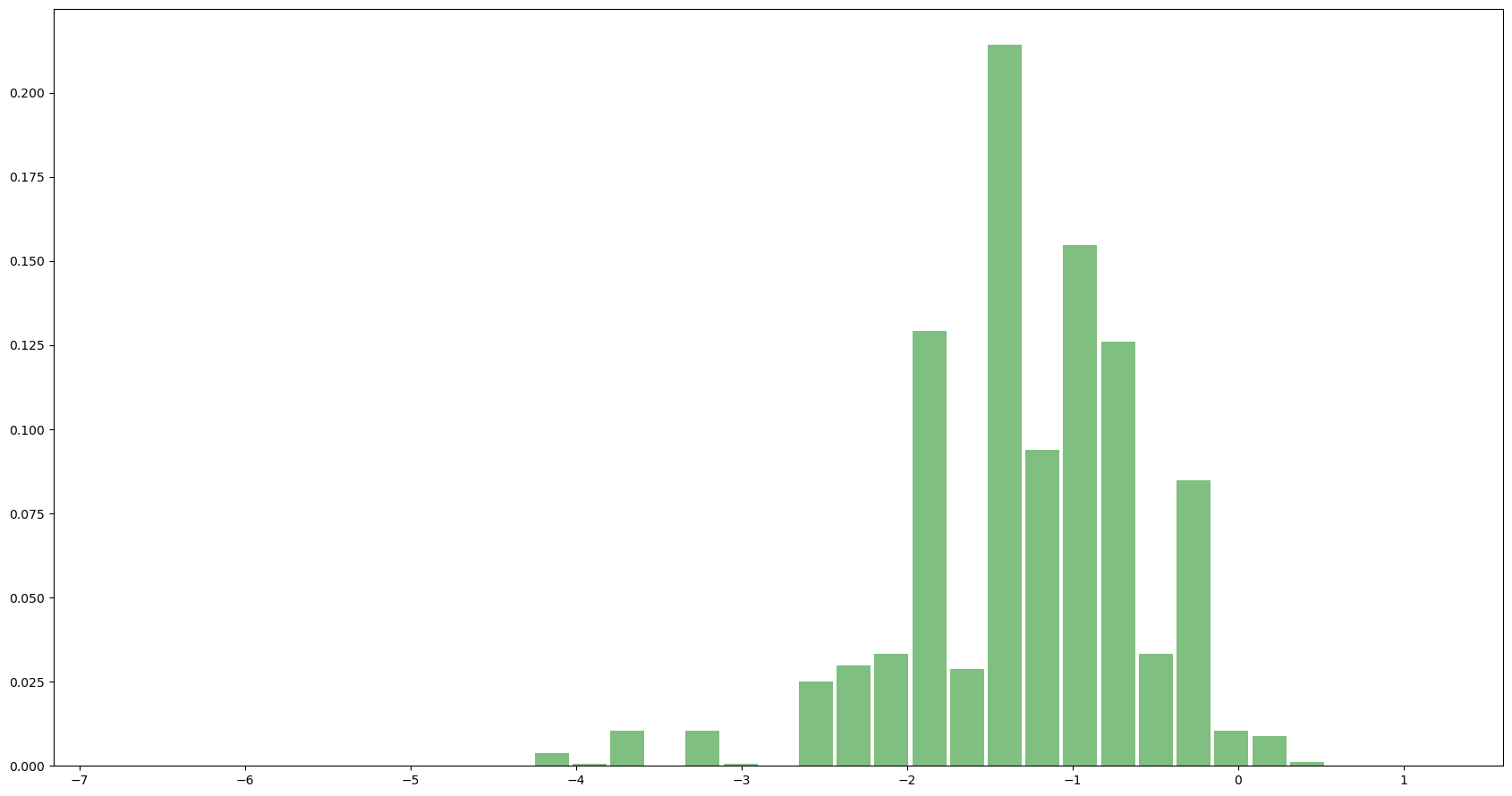}} 
	\subfigure[Frequency (IFA)]{\includegraphics[width=0.49\textwidth, height=0.13\textheight]{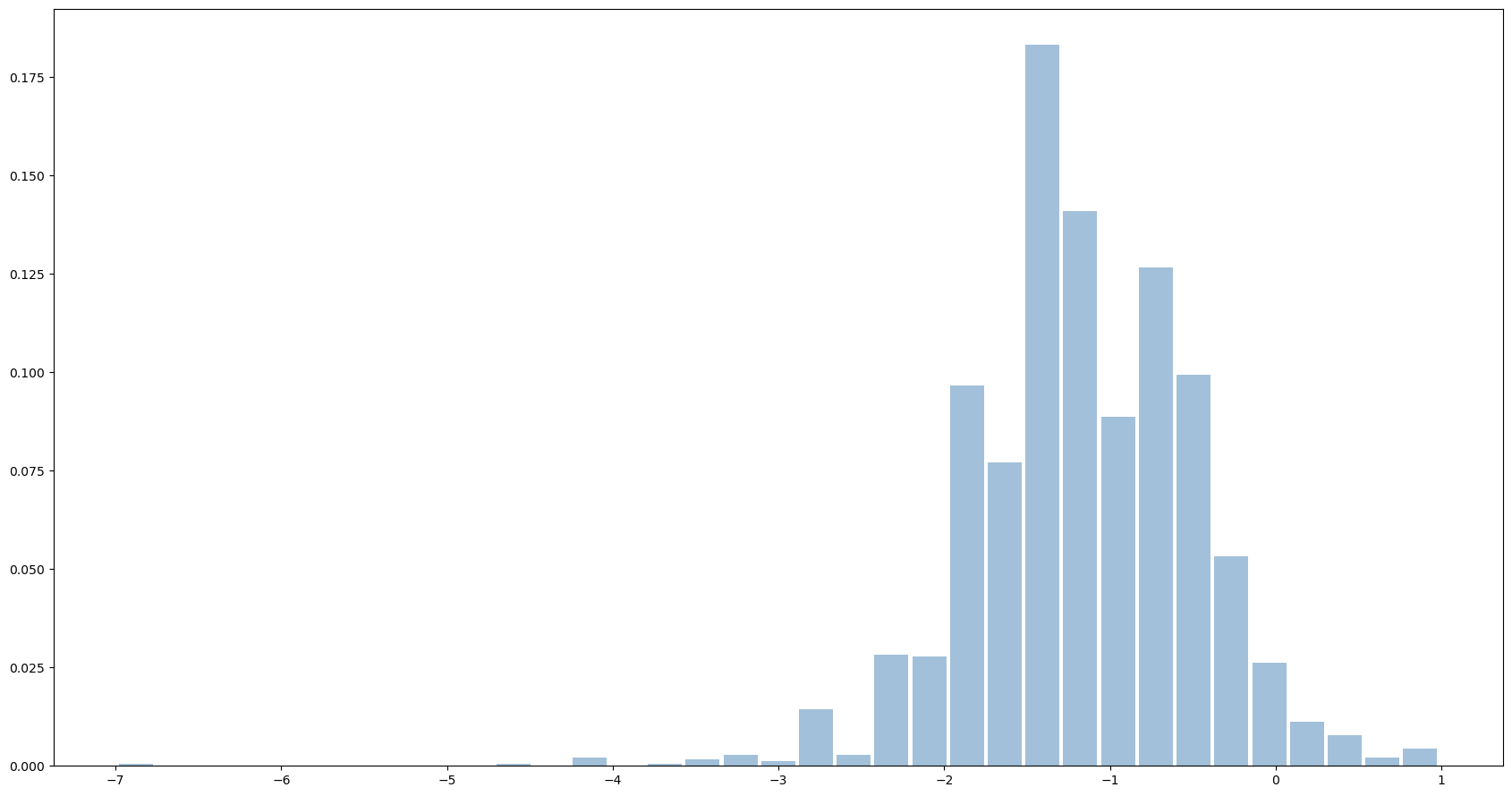}} 
		\subfigure[Frequency (Nested Logit)]{\includegraphics[width=0.49\textwidth, height=0.13\textheight]{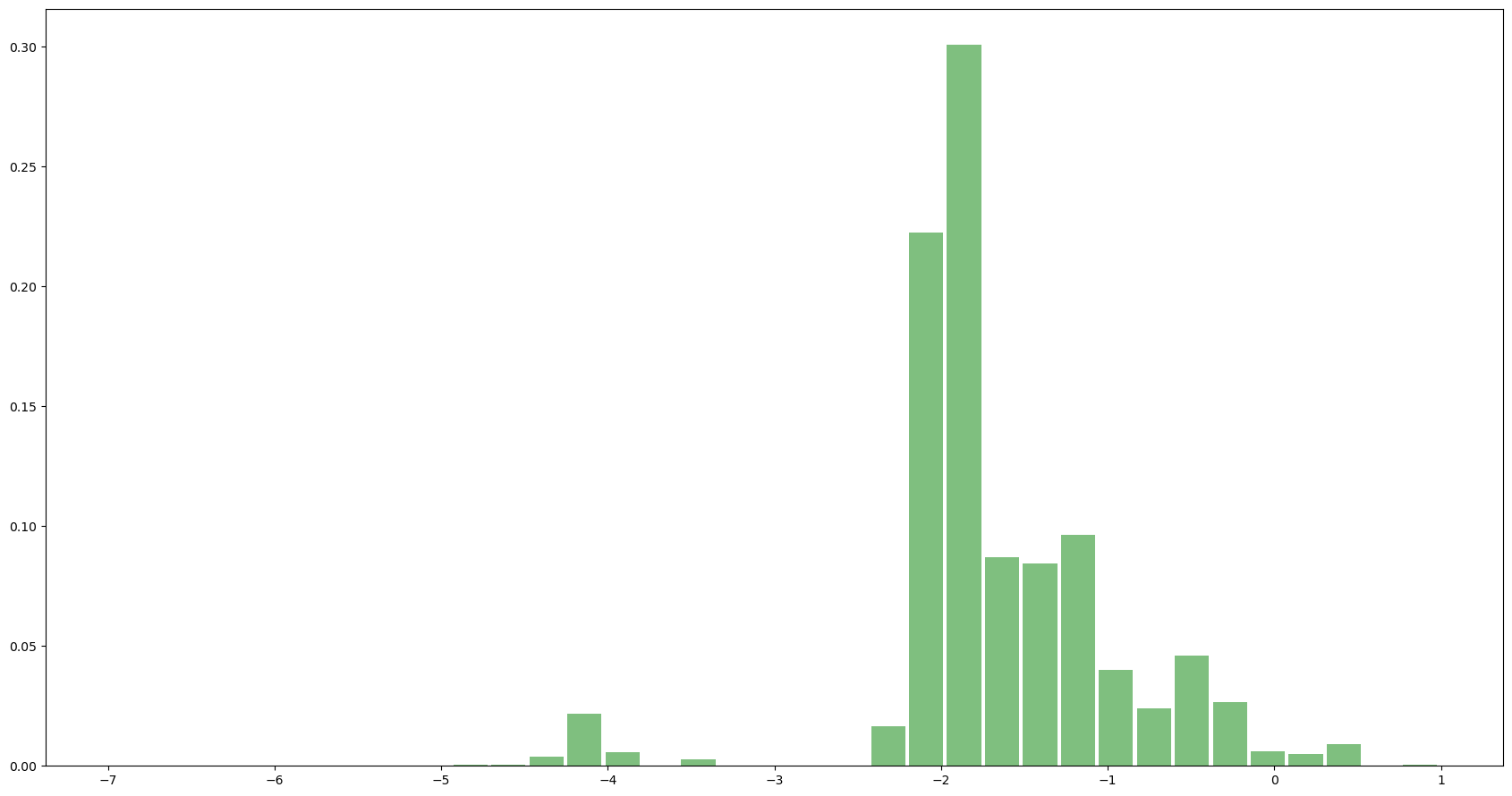}} 
	\subfigure[Travel Time (IFA)]{\includegraphics[width=0.49\textwidth, height=0.13\textheight]{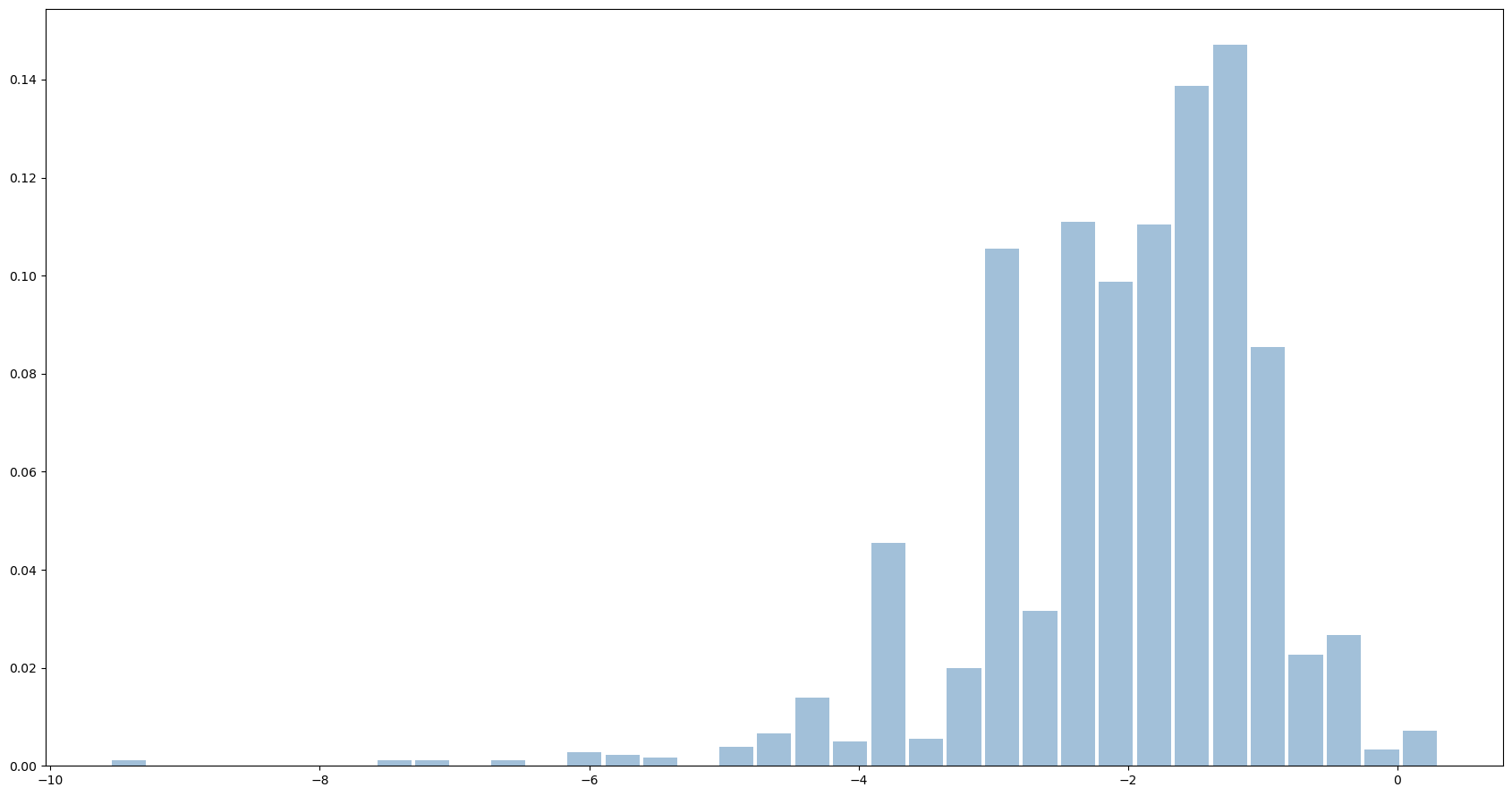}} 
		\subfigure[Travel Time (Nested Logit)]{\includegraphics[width=0.49\textwidth, height=0.13\textheight]{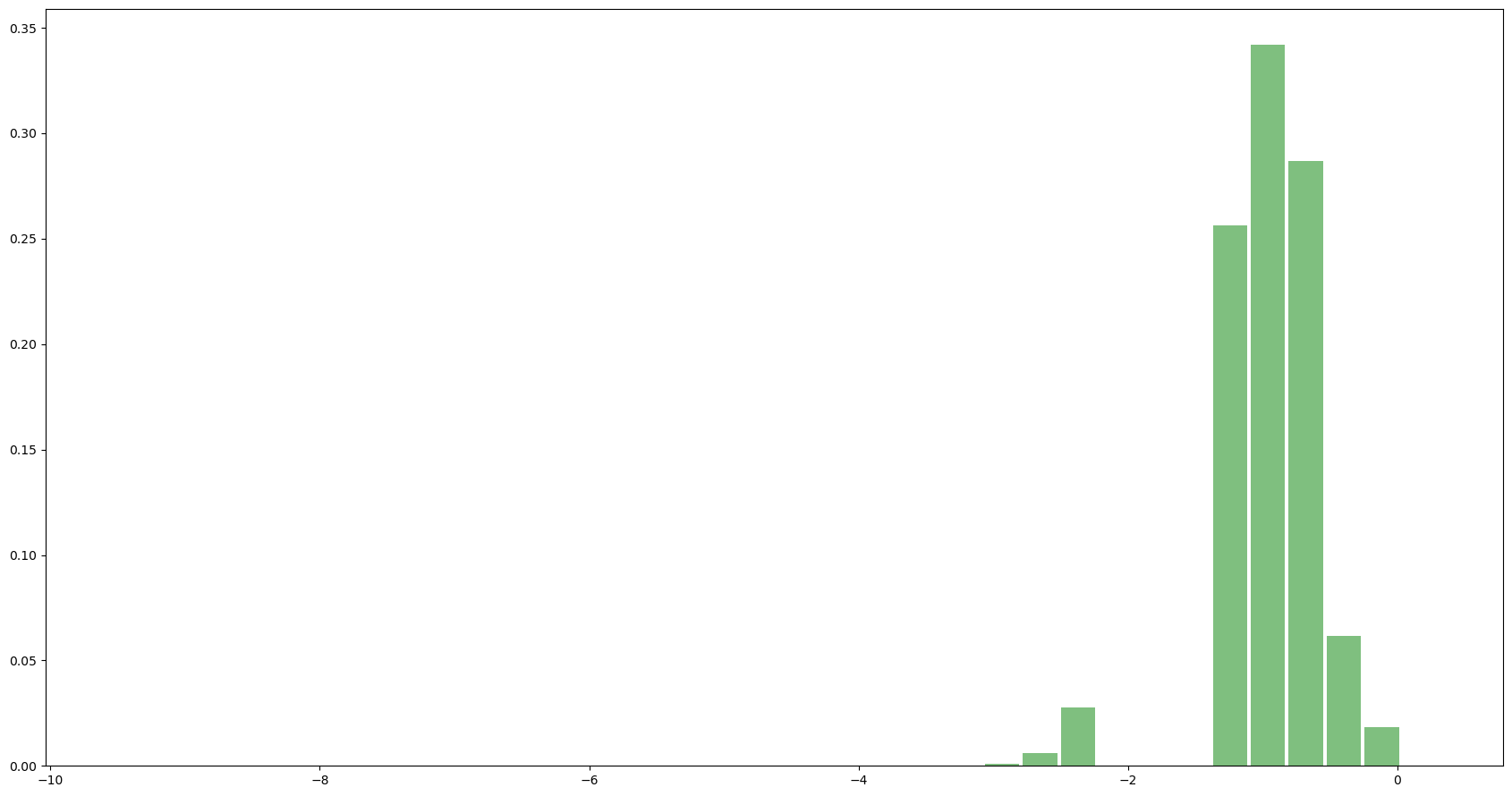}} 
			\subfigure[Intercept Train (IFA)]{\includegraphics[width=0.49\textwidth, height=0.13\textheight]{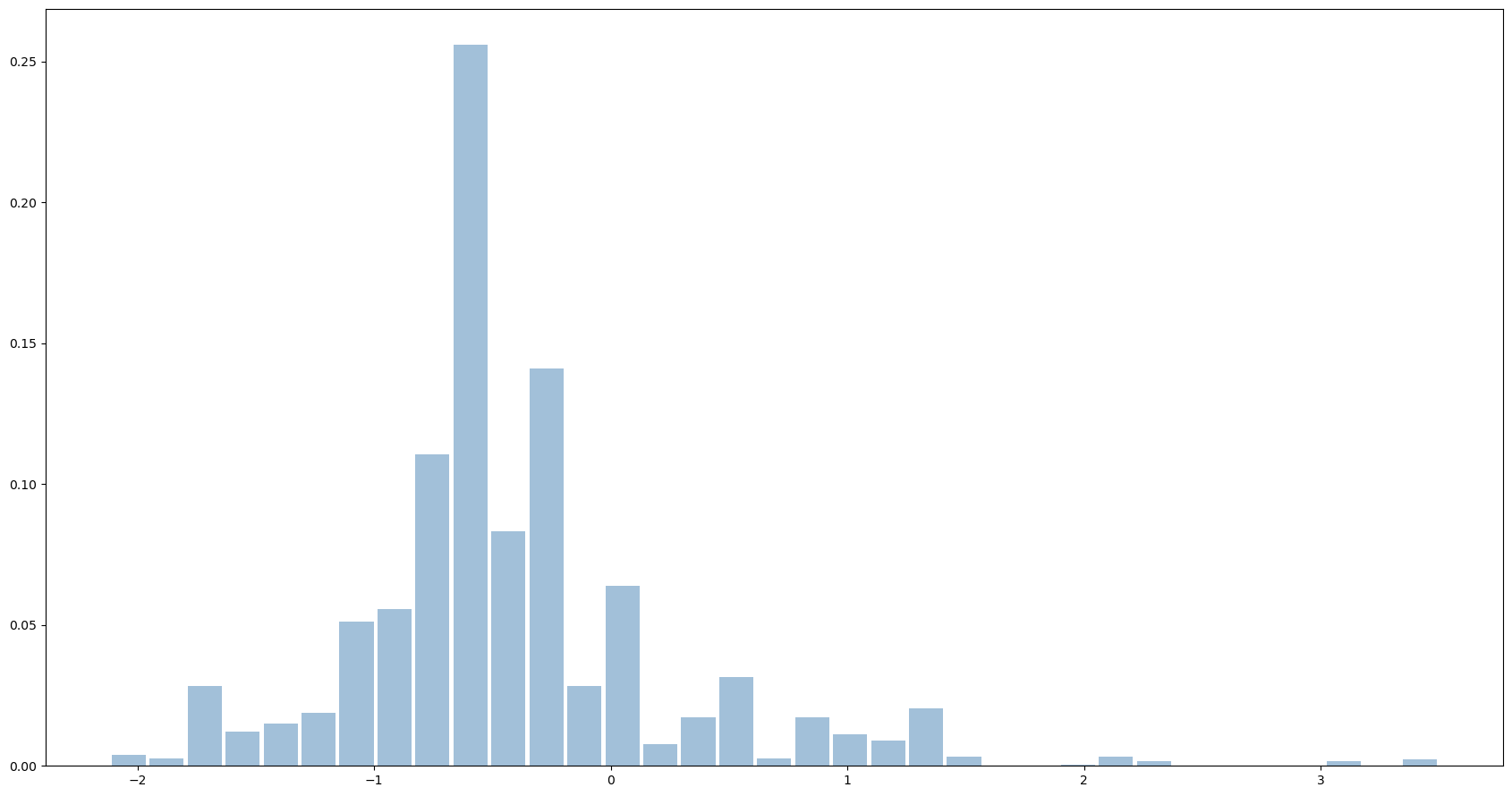}} 
		\subfigure[Intercept Train (Nested Logit)]{\includegraphics[width=0.49\textwidth, height=0.13\textheight]{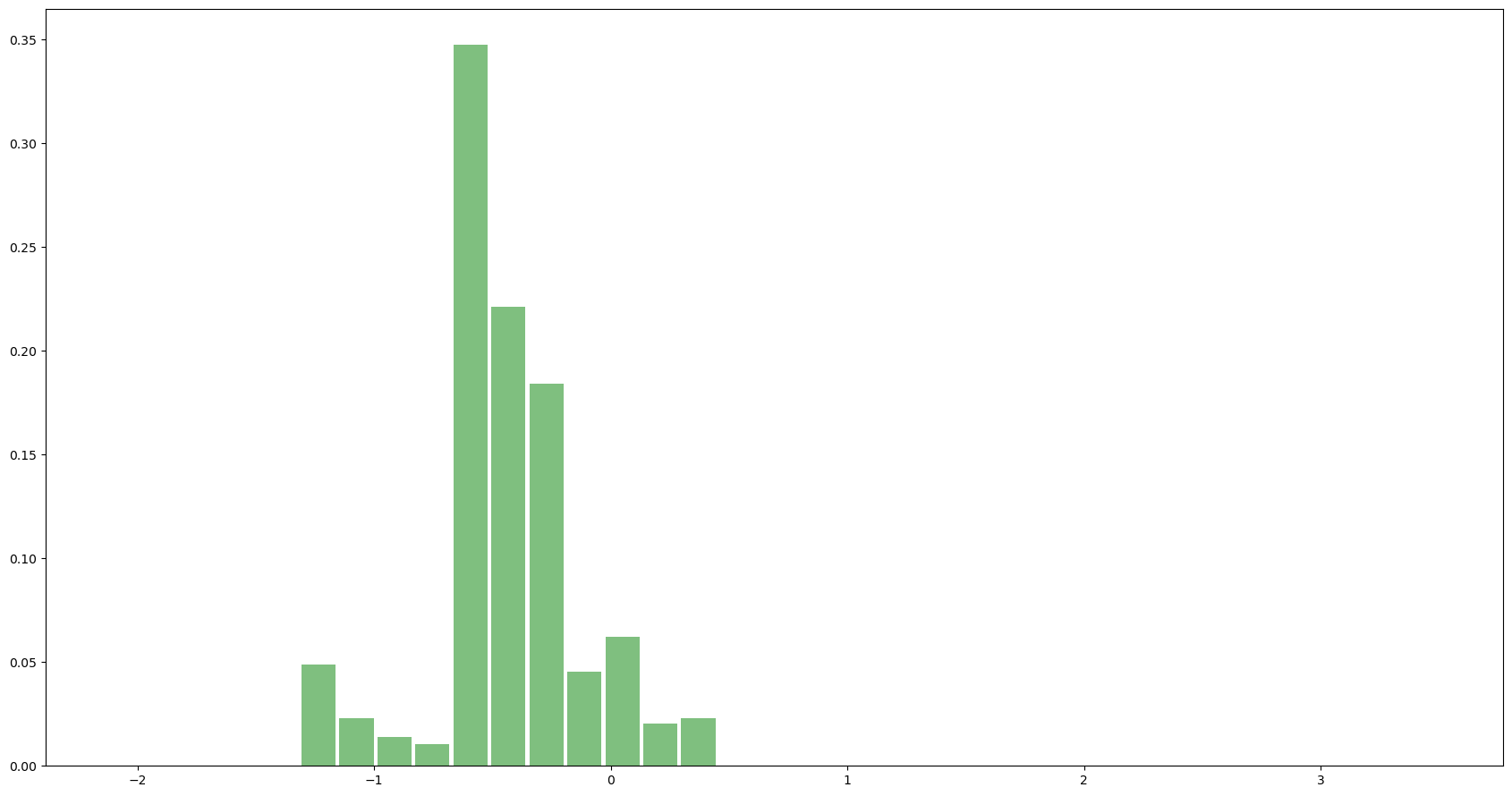}} 
					\subfigure[Intercept SM (IFA)]{\includegraphics[width=0.49\textwidth, height=0.13\textheight]{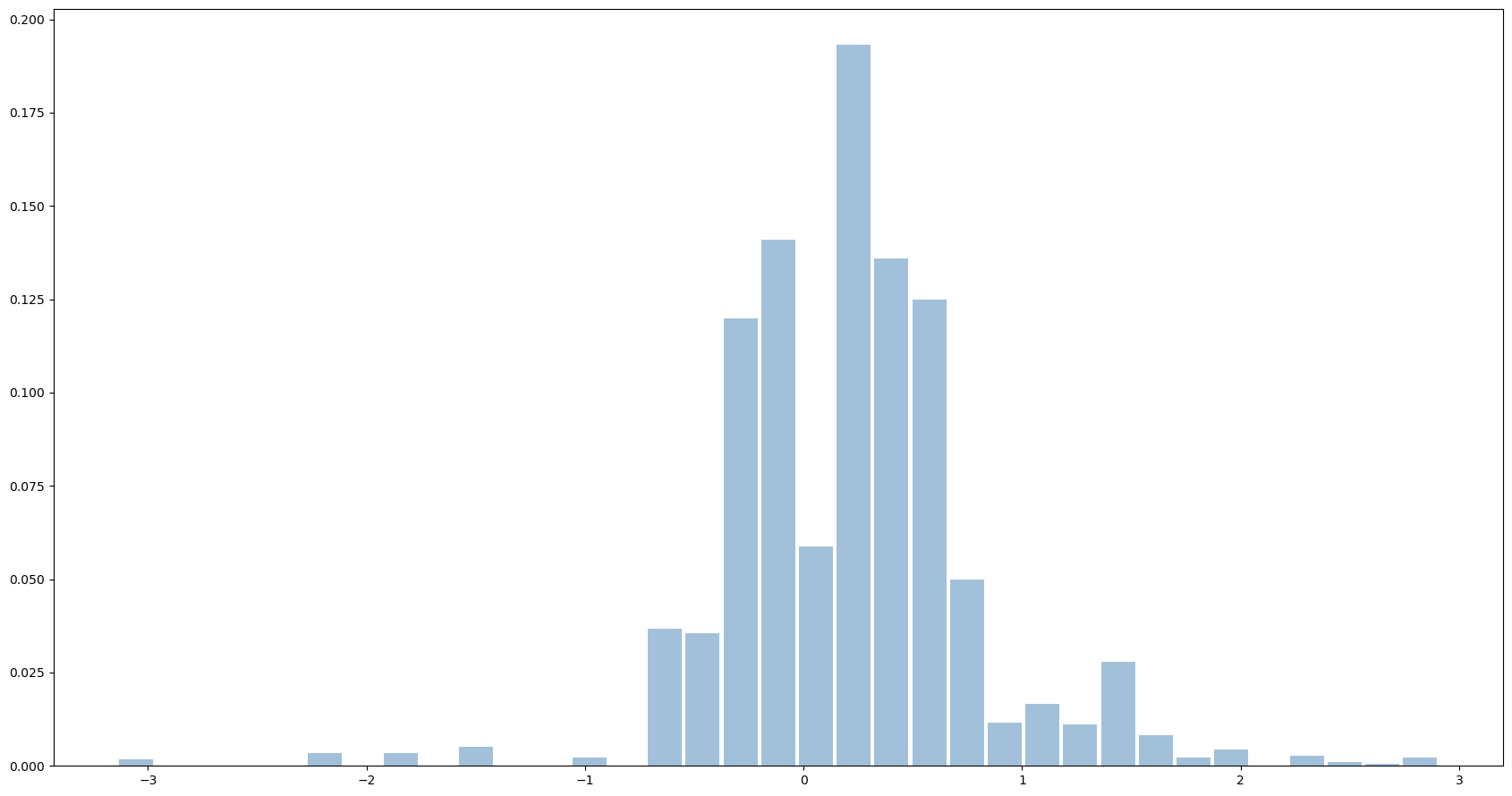}} 
		\subfigure[Intercept SM (Nested Logit)]{\includegraphics[width=0.49\textwidth, height=0.13\textheight]{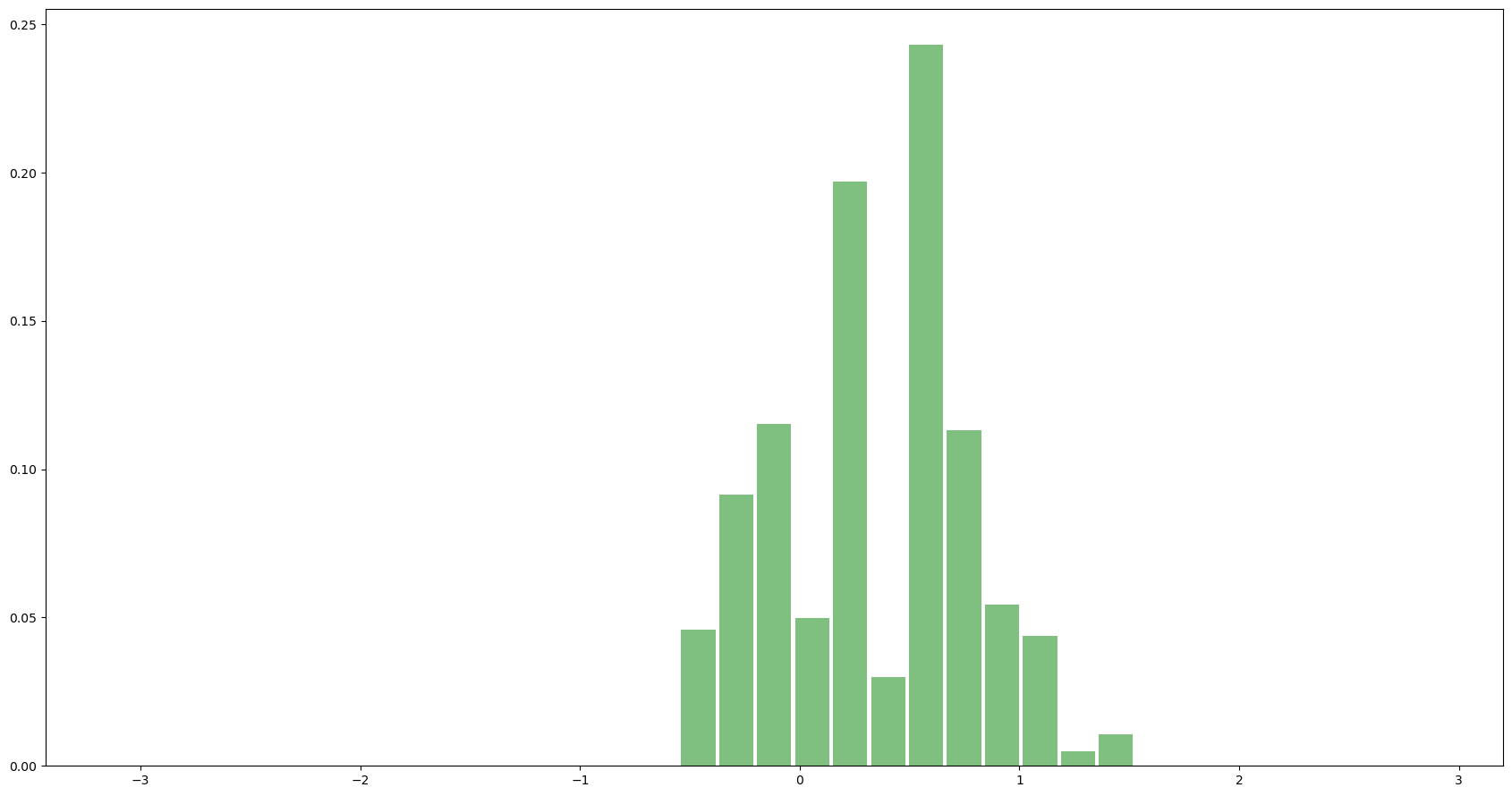}} 
	\label{figDL:HistsLargeSM}
	 \begin{minipage}{1\textwidth} %
		{   \footnotesize \begin{singlespace}
\textit{Note:} The green bars represent the heterogeneous coefficients in the test set predicted with the nested logit model, and the blue bars the heterogeneous coefficients in the test set predicted with the influence function approach (IFA) with repeated sample splitting with $R=5$.
			\end{singlespace}
			\par}
				\end{minipage}
\end{figure}

Figure \ref{figDL:HistsLargeSM} plots the histograms of the estimated coefficient functions predicted by the influence function approach (blue bars) and the nested logit model (green bars) using the test set. First, the plots reveal that there is substantial heterogeneity across travelers.  Second, the heterogeneity in the intercept for Swissmetro and in the coefficient for travel cost   across travelers appears to be similar when estimated with the influence function approach and the nested logit model, implying that the heterogeneity can be well captured by the linear approximation employed by the nested logit model. In contrast, the heterogeneity in the intercept for train and the coefficients for  frequency and travel time predicted by the more flexible  influence function approach deviates to a larger extent from the heterogeneity predicted by the nested logit model. 

One advantage of the influence function approach is that it can be easily applied to any parameter of inferential interest that is a function of the heterogeneous coefficient functions. In addition to the estimated average coefficient for the travel time, travel cost, and frequency, we are interested in estimating mean elasticities. More precisely, we focus on the expected own- and cross-travel time elasticities with respect to changes in the travel time evaluated at the mean values of travel cost, frequency, and travel time of every alternative.  Thus, the parameters of inferential interest calculated with the influence function approach are \[\theta_0^{l,m}=E\left[H^{l,m}(\bm{w_i}, \bm{\delta}(\bm{w}_i);\bm{x}^*)\right]\] where  $\bm{x}^{*}$ is a matrix with row entries $\bar{\bm{x}}^{\prime}_j$  which contain the average travel time, travel cost and frequency for alternative $j\in\{\text{car},\text{train},\text{sm}\}$, and 
\begin{equation}
H^{l,m}(\bm{w_i}, \bm{\delta}(\bm{w}_i);\bm{x}^*) =\beta^{\text{time}}\left(\bm{w}_i\right)\,\bar{x}_{m,\text{time}}\left(\mathbb{I}_{m,l}-\mathbb{P}\left(y_{i,m}=1\vert \bm{x}^*, \bm{w}_i\right)\right) \notag
\end{equation}
where $\mathbb{I}_{l,m}$ is an indicator that is equal to one when $l$ is equal to $m$ and zero otherwise for $l,m \in\{\text{car},\text{train},\text{sm}\}$.

Hence, $H^{l,m}(\bm{w_i}, \bm{\delta}(\bm{w}_i);\bm{x}^*)$ is  the individual  own- and cross-travel time elasticity calculated at the average travel cost, frequency, and travel time of every alternative, indicating the percentage change of choosing  alternative $l$ after a one percentage increase in the average travel time of alternative $m$. Consequently, $\theta_0^{l,m}$ corresponds to the expected own- and cross-travel time elasticity across individuals.

For the conditional logit, nested logit, and naive neural network approach, we use Efron's Bootstrap \cite{efron1979} with $1000$ bootstraps iterations to calculate the estimated standard errors of the own- and cross-travel time elasticities evaluated at the means.\footnote{For the nested logit model, we estimate the own- and cross-travel time elasticities at the mean using numerical derivatives of the choice probabilities with respect to the travel time. }
\begin{table}[H] 
\centering 
\small
\captionsetup{justification=centering}
\caption{Estimated Own- \& Cross-Travel Time Elasticities} 
  \label{tab:elasticitiesSmall} \small 
 \tabcolsep=0.15cm 
\begin{tabular}{lcccccc}
  \toprule \noalign{\smallskip}
&  \multicolumn{3}{c}{\textbf{Influence Function:}} & \multicolumn{3}{c}{\textbf{Neural Network:}} \\
 \cmidrule(l{0.5pt}r{9pt}){2-4}\cmidrule(l{0.5pt}r{9pt}){5-7}
  & Car & SM & Train & Car & SM & Train \\ [0.2ex] 
  \hline \\ [-2ex] 
Car & -2.385 \scriptsize{(0.274)} & 1.338 \scriptsize{(0.167)} & 0.143 \scriptsize{(0.105)} & -2.313 \scriptsize{(0.172)} & 1.315 \scriptsize{(0.098)} & 0.237 \scriptsize{(0.039)} \\ [0.5ex] 
 SM & 0.605 \scriptsize{(0.274)} & -0.463 \scriptsize{(0.167)} & 0.143 \scriptsize{(0.105)} & 0.876 \scriptsize{(0.066}) & -0.670 \scriptsize{(0.054}) & 0.237 \scriptsize{(0.039)} \\ [0.5ex] 
 Train & 0.605 \scriptsize{(0.274)} & 1.338 \scriptsize{(0.167)} & -3.466 \scriptsize{(0.107}) & 0.876 \scriptsize{(0.066)} & 1.315 \scriptsize{(0.098)} & -3.526 \scriptsize{(0.231)}  \\ [0.5ex] 
\hline \\ 

& \multicolumn{3}{c}{\textbf{Conditional Logit:}} & \multicolumn{3}{c}{\textbf{Nested Logit:}} \\
 \cmidrule(l{0.5pt}r{9pt}){2-4}\cmidrule(l{0.5pt}r{9pt}){5-7}
   & Car & SM & Train & Car & SM & Train\\ [0.2ex] \hline \\ [-2ex] 
Car & -1.71 \scriptsize{(0.388)} & 0.791 \scriptsize{(0.127)} & 0.211 \scriptsize{(0.265)} &  -0.936 \scriptsize{(0.051)} & 0.715 \scriptsize{(0.061)} & 0.145 \scriptsize{(0.02)}  \\ [0.25ex]
 SM &0.559 \scriptsize{(0.327)} & -0.46 \scriptsize{(0.126)} & 0.211 \scriptsize{(0.265)}  & 0.398 \scriptsize{(0.023)} & -0.45 \scriptsize{(0.032)} &  0.075 \scriptsize{(0.01)}  \\ [0.25ex]
 Train & 0.559 \scriptsize{(0.327)} & 0.791 \scriptsize{(0.127)} & -1.83 \scriptsize{(0.254)} &  1.097 \scriptsize{(0.288)} &  0.715 \scriptsize{(0.061)} & -1.255 \scriptsize{(0.068)}  \\ [0.25ex]

 \bottomrule
 \end{tabular} 
\begin{minipage}{0.95\textwidth} %
		{   \footnotesize \begin{singlespace}
\textit{Note:} The table reports estimated mean and the standard errors (in brackets) over  individuals' own- and cross-travel time elasticities evaluated at the mean for the influence function approach, the neural network, the conditional logit, and the nested logit model. The reported numbers  correspond to the   percentage change of the choice probability of an alternative in a row after a one percent increase in the travel time of an alternative in a column.
			\end{singlespace}
			\par}
	\end{minipage}
\end{table}
Overall, the own- and cross-travel time elasticities estimated with the influence function approach and the neural network are quite similar. With respect to the own-travel time elasticities, both the influence function approach and the neural network predict that travelers respond more sensitively to an increase in the travel time than predicted by the conditional and nested logit model.

A disadvantage  of the influence function approach, the neural network, and the conditional logit model is the restriction  of the cross-elasticities through the IIA property imposed by the conditional logit model and the model specified in Equation \eqref{eqDL:logit_model}, which restricts the cross-elasticities to be identical across alternatives. In contrast, the nested logit model, which allows for different cross-elasticities across alternatives in different nests, predicts that travelers are substantially more likely to substitute from car to train and vice versa in response to an increase in the travel time of either of the alternatives.

Moreover, the standard errors of the own- and cross-travel time elasticities estimated with the influence function approach remain larger than those of the nested logit model estimated with Efron's bootstrap -- though the difference is not as large as for the estimated average coefficients -- and are only slightly larger than in the conditional logit model and even smaller for some own- and cross-elasticities.

\section{Conclusion}\label{secDL:conclusion}

This paper investigates the finite sample performance of the estimation approach of \citeA{farrell2020} in the context of discrete choice models, who propose deep learning for the estimation of heterogeneous parameters in econometric models. For the construction of valid second-stage inferential statements after the first-stage estimation of the heterogeneous parameters with deep learning, they provide an influence function approach that builds on Neyman orthogonal scores in combination with sample splitting.

To study the proposed estimation and inference procedure, we conduct several Monte Carlo experiments. First, the experiments reveal that deep learning generally allows to recover precise estimates of the true average heterogeneous parameters -- especially if the number of observations is sufficiently large -- and that naive inference with robust standard errors leads to incorrect inferential statements. Second, we observe that the influence function approach proposed for the construction of valid inferential statements is sensitive to overfitting when no $l_2$-regularization is employed. Overfitting results in substantial average estimated bias and extremely large average estimated standard errors across Monte Carlo replicates. The sensitivity to overfitting is more pronounced for small samples but does not disappear with increasing sample size in our experiments. 
Using $l_2$-regularization appears to stabilize the estimation as it reduces the number of Monte Carlo replicates with extreme outliers, but leads to poor coverage of the confidence intervals. This is a consequence of the decreasing magnitude of the estimated standard errors and the increasing bias induced with increasing regularization, which in combination lead to tighter confidence intervals that are centered around biased estimates. 
A tool that achieves substantially better results in our Monte Carlo experiments than regularization is repeated sample splitting. Unlike $l_2$-regularization, it substantially reduces the number of outliers across Monte Carlo replicates without inducing additional bias, enabling the construction of valid inferential statements. 
However, repeated sample splitting appears to have a less drastic effect on the estimated variance than $l_2$-regularization, which causes relatively large estimated standard errors.

Due to the complexity of neural networks, we restrict our Monte Carlo experiments to the impact of $l_2$-regularization on the inference procedure. An interesting avenue for future research is to consider different forms of regularization, such as dropout rates, and varying complexities of the network architecture used to estimate the influence function approach (e.g.,  to vary the number of neurons and hidden layers). 
In addition, both the influence function approach and the neural network combined with naive inference exploit that the variables of interest enter the utility linearly. An interesting comparison, however, is the estimation with a completely unstructured network (e.g., with Efron's bootstrap for inference) which could potentially further illustrate the advantage of the influence function approach.

\newpage

\renewcommand\refname{References}
\bibliography{Literature_Mendeley}

\newpage
\begin{appendix}
	
\onehalfspacing	

\section*{Appendix}	

\setcounter{equation}{0}
\setcounter{table}{0}
\numberwithin{equation}{section}
\numberwithin{table}{section}
\numberwithin{figure}{section}
\numberwithin{subsection}{section}
\renewcommand{\thetable}{A.\arabic{table}}
\renewcommand{\theequation}{A.\arabic{equation}}
\renewcommand{\thesubsection}{A.\arabic{subsection}}
\captionsetup[figure]{list=no}
\captionsetup[table]{list=no}
\captionsetup[subsection]{list=no}

\section[Appendix A: \\ Additional Tables]{Additional Tables}\label{app:addTables}

\begin{table}[H]
\centering
\captionsetup{justification=centering}
	\caption{Median Summary Statistics of 1000 Monte Carlo Replicates for Small Data and without Repeated Sample Splitting}
	\label{tab:MC_swissmetroMedian}
	\small
\begin{tabular}{  l@{\hspace{-0.5em}}  *{7}S[table-format=2.4] } 
\toprule 
 & \multicolumn{2}{c}{Conditional} &  \multicolumn{4}{c}{Influence Function Approach} & \\ 
& \multicolumn{2}{c}{Logit} &  \multicolumn{4}{c}{with $\lambda$ equal to} & \\  
\cmidrule(l{5pt}r{5pt}){2-3}   \cmidrule(l{5pt}r{5pt}){4-7} 
& \multicolumn{1}{c}{Oracle} & \multicolumn{1}{c}{Basic} & \multicolumn{1}{c}{$0$} & \multicolumn{1}{c}{$10^{-5}$} & 
\multicolumn{1}{c}{$10^{-4}$}  & \multicolumn{1}{c}{$2 \cdot 10^{-3}$} & \multicolumn{1}{c}{NN} \\ 
\cmidrule(l{5pt}r{5pt}){2-2}  \cmidrule(l{5pt}r{5pt}){3-3}  \cmidrule(l{5pt}r{5pt}){4-4}  \cmidrule(l{5pt}r{5pt}){5-5}  \cmidrule(l{5pt}r{5pt}){6-6} 
 \cmidrule(l{5pt}r{5pt}){7-7}  \cmidrule(l{5pt}r{5pt}){8-8} 
$ \theta_{cost} \in \widehat{CI}_{cost} $ & 0.95 & 0.00 & 0.93 & 0.92 & 0.83 & 0.40 & 0.99 \\ 
$ \theta_{freq} \in \widehat{CI}_{freq} $ & 0.95 & 0.00 & 0.93 & 0.92 & 0.89 & 0.68 & 1.00 \\ 
$ \theta_{time} \in \widehat{CI}_{time} $ & 0.94 & 0.00 & 0.93 & 0.94 & 0.88 & 0.54 & 1.00 \\ [0.5em] 
$\widehat{se}_{cost} $ & 0.07 & 0.05 & 1.36 & 1.02 & 0.53 & 0.18 & 0.60 \\ 
$\widehat{se}_{freq} $ & 0.10 & 0.07 & 1.62 & 1.34 & 0.84 & 0.34 & 3.29 \\ 
$\widehat{se}_{time} $ & 0.07 & 0.06 & 1.28 & 0.96 & 0.46 & 0.24 & 2.98 \\ [0.5em] 
Bias$_{cost} $ & -0.01 & 0.65 & -0.23 & -0.21 & -0.31 & -0.44 & -0.16 \\ 
Bias$_{freq} $ & -0.00 & 0.59 & -0.28 & -0.32 & -0.35 & -0.40 & -0.19 \\ 
Bias$_{time} $ & -0.01 & 0.80 & -0.09 & -0.09 & -0.23 & -0.41 & -0.17 \\ [0.5em] 
Rej. $\theta_{cost}=0$ & 1.00 & 1.00 & 0.47 & 0.56 & 0.78 & 0.93 & 1.00 \\ 
Rej. $\theta_{freq}=0$ & 1.00 & 1.00 & 0.27 & 0.35 & 0.50 & 0.79 & 0.00 \\ 
Rej. $\theta_{time}=0$ & 1.00 & 1.00 & 0.60 & 0.69 & 0.84 & 0.93 & 0.03 \\ [0.5em] 
$MSE(\Lambda)^{Train}$ & {.}  & {.}  & 4.99 & 5.13 & 5.35 & 5.93 & {.}  \\ 
$MSE(\Lambda)^{Test}$ & {.}  & {.}  & 5.20 & 5.28 & 5.40 & 5.93 & {.}  \\ 
Share Outlier & 0.00 & 0.00 & 0.26 & 0.18 & 0.11 & 0.04 & 0.12 \\ 
  \bottomrule  
 \end{tabular}

\begin{minipage}{0.825\textwidth} %
		{   \footnotesize \begin{singlespace}
\textit{Note:} The table reports the median of the variables $\hat{se}_i$, $BIAS_i$,  $MSE(\Lambda)^{Train}$, and $MSE(\Lambda)^{Test}$  and the average of the variables $ \theta_{i} \in \hat{CI}_{i}$, Rej. $\theta_{i}=0$, and Share Outlier, $i \in \{cost,\, freq,\, time\}$,  over all Monte Carlo replicates for the conditional logit using the true specification (Oracle), the conditional logit using the three variables of interest for the estimation (Basic),  the influence function approach, using five different values of $\lambda$ for the estimation of  $\Lambda_s(\bm{w})$, and the neural network (NN), which uses robust standard erros and does not rely on the influence function approach. 
			\end{singlespace}
			\par}
	\end{minipage}
\end{table}

\begin{table}[H]
\centering
\captionsetup{justification=centering}
	\caption{Median Summary Statistics of 1000 Monte Carlo Replicates for Small Data and Repeated Sample Splitting with $R=5$}
	\label{tab:MC_swissmetroMedian}
	\small
\begin{tabular}{  l@{\hspace{-0.5em}}  *{7}S[table-format=2.4] } 
\toprule 
 & \multicolumn{2}{c}{Conditional} &  \multicolumn{4}{c}{Influence Function Approach} & \\ 
& \multicolumn{2}{c}{Logit} &  \multicolumn{4}{c}{with $\lambda$ equal to} & \\  
\cmidrule(l{5pt}r{5pt}){2-3}   \cmidrule(l{5pt}r{5pt}){4-7} 
& \multicolumn{1}{c}{Oracle} & \multicolumn{1}{c}{Basic} & \multicolumn{1}{c}{$0$} & \multicolumn{1}{c}{$10^{-5}$} & 
\multicolumn{1}{c}{$10^{-4}$}  & \multicolumn{1}{c}{$2 \cdot 10^{-3}$} & \multicolumn{1}{c}{NN} \\ 
\cmidrule(l{5pt}r{5pt}){2-2}  \cmidrule(l{5pt}r{5pt}){3-3}  \cmidrule(l{5pt}r{5pt}){4-4}  \cmidrule(l{5pt}r{5pt}){5-5}  \cmidrule(l{5pt}r{5pt}){6-6} 
 \cmidrule(l{5pt}r{5pt}){7-7}  \cmidrule(l{5pt}r{5pt}){8-8} 
$ \theta_{cost} \in \widehat{CI}_{cost} $ & 0.94 & 0.00 & 0.94 & 0.93 & 0.82 & 0.42 & 0.99 \\ 
$ \theta_{freq} \in \widehat{CI}_{freq} $ & 0.96 & 0.00 & 0.94 & 0.92 & 0.90 & 0.66 & 1.00 \\ 
$ \theta_{time} \in \widehat{CI}_{time} $ & 0.96 & 0.00 & 0.95 & 0.95 & 0.87 & 0.59 & 1.00 \\ [0.5em] 
$\widehat{se}_{cost} $ & 0.07 & 0.05 & 1.20 & 0.96 & 0.46 & 0.18 & 0.60 \\ 
$\widehat{se}_{freq} $ & 0.10 & 0.07 & 1.48 & 1.16 & 0.73 & 0.33 & 3.45 \\ 
$\widehat{se}_{time} $ & 0.07 & 0.06 & 1.21 & 0.85 & 0.42 & 0.24 & 3.04 \\ [0.5em] 
Bias$_{cost} $ & -0.01 & 0.65 & -0.26 & -0.29 & -0.33 & -0.41 & -0.18 \\ 
Bias$_{freq} $ & -0.00 & 0.59 & -0.28 & -0.31 & -0.28 & -0.40 & -0.18 \\ 
Bias$_{time} $ & -0.00 & 0.81 & -0.03 & -0.13 & -0.24 & -0.38 & -0.17 \\ [0.5em] 
Rej. $\theta_{cost}=0$ & 1.00 & 1.00 & 0.48 & 0.61 & 0.84 & 0.96 & 0.99 \\ 
Rej. $\theta_{freq}=0$ & 1.00 & 1.00 & 0.25 & 0.32 & 0.54 & 0.81 & 0.00 \\ 
Rej. $\theta_{time}=0$ & 1.00 & 1.00 & 0.64 & 0.78 & 0.92 & 0.96 & 0.02 \\ [0.5em] 
$MSE(\Lambda)^{Train}$ & {.}  & {.}  & 5.01 & 5.15 & 5.37 & 5.94 & {.}  \\ 
$MSE(\Lambda)^{Test}$ & {.}  & {.}  & 5.22 & 5.30 & 5.43 & 5.95 & {.}  \\ 
Share Outlier & 0.00 & 0.00 & 0.05 & 0.02 & 0.00 & 0.00 & 0.12 \\ 
  \bottomrule  
 \end{tabular}

\begin{minipage}{0.825\textwidth} %
		{   \footnotesize \begin{singlespace}
\textit{Note:}	The table reports the median of the variables $\hat{se}_i$, $BIAS_i$,  $MSE(\Lambda)^{Train}$, and $MSE(\Lambda)^{Test}$  and the average of the variables $ \theta_{i} \in \hat{CI}_{i}$, Rej. $\theta_{i}=0$, and Share Outlier, $i \in \{cost,\, freq,\, time\}$, over all Monte Carlo replicates for the conditional logit using the true specification (Oracle), the conditional logit using the three variables of interest for the estimation (Basic),  the influence function approach, using five different values of $\lambda$ for the estimation of  $\Lambda_s(\bm{w})$, and the neural network (NN), which uses robust standard erros and does not rely on the influence function approach.
			\end{singlespace}
			\par}
	\end{minipage}
\end{table}

\begin{table}[H]
\centering
\captionsetup{justification=centering}
	\caption{Average Summary Statistics of 1000 Monte Carlo Replicates for Large Data and without Repeated Sample Splitting}
	\label{tab:MC_swissmetroLargeMean}
	\small
\begin{tabular}{  l@{\hspace{-0.5em}}  *{7}S[table-format=2.4] } 
\toprule 
 & \multicolumn{2}{c}{Conditional} &  \multicolumn{4}{c}{Influence Function Approach} & \\ 
& \multicolumn{2}{c}{Logit} &  \multicolumn{4}{c}{with $\lambda$ equal to} & \\  
\cmidrule(l{5pt}r{5pt}){2-3}   \cmidrule(l{5pt}r{5pt}){4-7} 
& \multicolumn{1}{c}{Oracle} & \multicolumn{1}{c}{Basic} & \multicolumn{1}{c}{$0$} & \multicolumn{1}{c}{$10^{-5}$} & 
\multicolumn{1}{c}{$10^{-4}$}  & \multicolumn{1}{c}{$2 \cdot 10^{-3}$} & \multicolumn{1}{c}{NN} \\ 
\cmidrule(l{5pt}r{5pt}){2-2}  \cmidrule(l{5pt}r{5pt}){3-3}  \cmidrule(l{5pt}r{5pt}){4-4}  \cmidrule(l{5pt}r{5pt}){5-5}  \cmidrule(l{5pt}r{5pt}){6-6} 
 \cmidrule(l{5pt}r{5pt}){7-7}  \cmidrule(l{5pt}r{5pt}){8-8} 
$ \theta_{cost} \in \widehat{CI}_{cost} $ & 0.95 & 0.00 & 0.94 & 0.93 & 0.90 & 0.75 & 1.00 \\ 
$ \theta_{freq} \in \widehat{CI}_{freq} $ & 0.95 & 0.00 & 0.92 & 0.94 & 0.91 & 0.83 & 1.00 \\ 
$ \theta_{time} \in \widehat{CI}_{time} $ & 0.95 & 0.00 & 0.94 & 0.94 & 0.92 & 0.85 & 1.00 \\ [0.5em] 
$\widehat{se}_{cost} $ & 0.02 & 0.02 & 3.43 & 1.45 & 1.15 & 0.24 & 0.49 \\ 
$\widehat{se}_{freq} $ & 0.04 & 0.04 & 8.24 & 2.03 & 1.60 & 0.30 & 1.73 \\ 
$\widehat{se}_{time} $ & 0.03 & 0.02 & 3.02 & 3.24 & 0.98 & 0.26 & 1.28 \\ [0.5em] 
Bias$_{cost} $ & -0.00 & 0.60 & 1.38 & 0.87 & 0.05 & -0.24 & -0.03 \\ 
Bias$_{freq} $ & -0.00 & 0.53 & 3.82 & 1.08 & 0.24 & -0.32 & -0.03 \\ 
Bias$_{time} $ & -0.00 & 0.78 & 0.67 & 3.13 & -0.01 & -0.24 & -0.02 \\ [0.5em] 
Rej. $\theta_{cost}=0$ & 1.00 & 1.00 & 0.83 & 0.88 & 0.89 & 0.98 & 1.00 \\ 
Rej. $\theta_{freq}=0$ & 1.00 & 1.00 & 0.58 & 0.69 & 0.81 & 0.97 & 0.00 \\ 
Rej. $\theta_{time}=0$ & 1.00 & 1.00 & 0.90 & 0.93 & 0.94 & 0.98 & 0.93 \\ [0.5em] 
$MSE(\Lambda)^{Train}$ & {.}  & {.}  & 8.81 & 8.85 & 8.90 & 9.23 & {.}  \\ 
$MSE(\Lambda)^{Test}$ & {.}  & {.}  & 8.89 & 8.88 & 8.92 & 9.24 & {.}  \\ 
Share Outlier & 0.00 & 0.00 & 0.07 & 0.05 & 0.04 & 0.01 & 0.00 \\ 
  \bottomrule  
 \end{tabular}

 \begin{minipage}{0.825\textwidth} %
		{   \footnotesize \begin{singlespace}
\textit{Note:}	The table reports the average summary statistics over all Monte Carlo replicates for the conditional logit using the true specification (Oracle), the conditional logit using the three variables of interest for the estimation (Basic),  the influence function approach, using five different values of $\lambda$ for the estimation of  $\Lambda_s(\bm{w})$, and the neural network (NN), which uses robust standard erros and does not rely on the influence function approach.
			\end{singlespace}
			\par}
	\end{minipage}
\end{table}

\newpage

\begin{table}[H]
\centering
\captionsetup{justification=centering}
	\caption{Median Summary Statistics of 1000 Monte Carlo Replicates for Large Data and without Repeated Sample Splitting}
	\label{tab:MC_swissmetroLargeMedian}
	\small
\begin{tabular}{  l@{\hspace{-0.5em}}  *{7}S[table-format=2.4] } 
\toprule 
 & \multicolumn{2}{c}{Conditional} &  \multicolumn{4}{c}{Influence Function Approach} & \\ 
& \multicolumn{2}{c}{Logit} &  \multicolumn{4}{c}{with $\lambda$ equal to} & \\  
\cmidrule(l{5pt}r{5pt}){2-3}   \cmidrule(l{5pt}r{5pt}){4-7} 
& \multicolumn{1}{c}{Oracle} & \multicolumn{1}{c}{Basic} & \multicolumn{1}{c}{$0$} & \multicolumn{1}{c}{$10^{-5}$} & 
\multicolumn{1}{c}{$10^{-4}$}  & \multicolumn{1}{c}{$2 \cdot 10^{-3}$} & \multicolumn{1}{c}{NN} \\ 
\cmidrule(l{5pt}r{5pt}){2-2}  \cmidrule(l{5pt}r{5pt}){3-3}  \cmidrule(l{5pt}r{5pt}){4-4}  \cmidrule(l{5pt}r{5pt}){5-5}  \cmidrule(l{5pt}r{5pt}){6-6} 
 \cmidrule(l{5pt}r{5pt}){7-7}  \cmidrule(l{5pt}r{5pt}){8-8} 
$ \theta_{cost} \in \widehat{CI}_{cost} $ & 0.95 & 0.00 & 0.94 & 0.93 & 0.90 & 0.75 & 1.00 \\ 
$ \theta_{freq} \in \widehat{CI}_{freq} $ & 0.95 & 0.00 & 0.92 & 0.94 & 0.91 & 0.83 & 1.00 \\ 
$ \theta_{time} \in \widehat{CI}_{time} $ & 0.95 & 0.00 & 0.94 & 0.94 & 0.92 & 0.85 & 1.00 \\ [0.5em] 
$\widehat{se}_{cost} $ & 0.02 & 0.02 & 0.35 & 0.25 & 0.19 & 0.04 & 0.49 \\ 
$\widehat{se}_{freq} $ & 0.04 & 0.04 & 0.58 & 0.37 & 0.23 & 0.06 & 1.72 \\ 
$\widehat{se}_{time} $ & 0.03 & 0.02 & 0.27 & 0.19 & 0.16 & 0.05 & 1.27 \\ [0.5em] 
Bias$_{cost} $ & -0.00 & 0.60 & -0.05 & -0.01 & -0.07 & -0.05 & -0.03 \\ 
Bias$_{freq} $ & -0.00 & 0.53 & -0.09 & -0.04 & -0.08 & -0.06 & -0.03 \\ 
Bias$_{time} $ & -0.00 & 0.78 & -0.00 & 0.00 & -0.04 & -0.03 & -0.02 \\ [0.5em] 
Rej. $\theta_{cost}=0$ & 1.00 & 1.00 & 0.83 & 0.88 & 0.89 & 0.98 & 1.00 \\ 
Rej. $\theta_{freq}=0$ & 1.00 & 1.00 & 0.58 & 0.69 & 0.81 & 0.97 & 0.00 \\ 
Rej. $\theta_{time}=0$ & 1.00 & 1.00 & 0.90 & 0.93 & 0.94 & 0.98 & 0.93 \\ [0.5em] 
$MSE(\Lambda)^{Train}$ & {.}  & {.}  & 8.81 & 8.85 & 8.90 & 9.24 & {.}  \\ 
$MSE(\Lambda)^{Test}$ & {.}  & {.}  & 8.89 & 8.88 & 8.91 & 9.24 & {.}  \\ 
Share Outlier & 0.00 & 0.00 & 0.07 & 0.05 & 0.04 & 0.01 & 0.00 \\ 
  \bottomrule  
 \end{tabular}

 \begin{minipage}{0.825\textwidth} %
		{   \footnotesize \begin{singlespace}
\textit{Note:}	The table reports the median of the variables $\hat{se}_i$, $BIAS_i$,  $MSE(\Lambda)^{Train}$, and $MSE(\Lambda)^{Test}$  and the average of the variables $ \theta_{i} \in \hat{CI}_{i}$, Rej. $\theta_{i}=0$, and Share Outlier, $i \in \{cost,\, freq,\, time\}$, over all Monte Carlo replicates for the conditional logit using the true specification (Oracle), the conditional logit using the three variables of interest for the estimation (Basic),  the influence function approach, using five different values of $\lambda$ for the estimation of  $\Lambda_s(\bm{w})$, and the neural network (NN), which uses robust standard erros and does not rely on the influence function approach. 
			\end{singlespace}
			\par}
	\end{minipage}
\end{table}

\newpage

\begin{table}[H]
\centering
\captionsetup{justification=centering}
	\caption{Median Summary Statistics of 1000 Monte Carlo Replicates for Large Data and Repeated Sample Splitting with $R=5$}
	\label{tab:MC_swissmetroLargeMedian}
	\small
\begin{tabular}{  l@{\hspace{-0.5em}}  *{7}S[table-format=2.4] } 
\toprule 
 & \multicolumn{2}{c}{Conditional} &  \multicolumn{4}{c}{Influence Function Approach} & \\ 
& \multicolumn{2}{c}{Logit} &  \multicolumn{4}{c}{with $\lambda$ equal to} & \\  
\cmidrule(l{5pt}r{5pt}){2-3}   \cmidrule(l{5pt}r{5pt}){4-7} 
& \multicolumn{1}{c}{Oracle} & \multicolumn{1}{c}{Basic} & \multicolumn{1}{c}{$0$} & \multicolumn{1}{c}{$10^{-5}$} & 
\multicolumn{1}{c}{$10^{-4}$}  & \multicolumn{1}{c}{$2 \cdot 10^{-3}$} & \multicolumn{1}{c}{NN} \\ 
\cmidrule(l{5pt}r{5pt}){2-2}  \cmidrule(l{5pt}r{5pt}){3-3}  \cmidrule(l{5pt}r{5pt}){4-4}  \cmidrule(l{5pt}r{5pt}){5-5}  \cmidrule(l{5pt}r{5pt}){6-6} 
 \cmidrule(l{5pt}r{5pt}){7-7}  \cmidrule(l{5pt}r{5pt}){8-8} 
$ \theta_{cost} \in \widehat{CI}_{cost} $ & 0.95 & 0.00 & 0.94 & 0.92 & 0.91 & 0.76 & 1.00 \\ 
$ \theta_{freq} \in \widehat{CI}_{freq} $ & 0.95 & 0.00 & 0.95 & 0.93 & 0.90 & 0.83 & 1.00 \\ 
$ \theta_{time} \in \widehat{CI}_{time} $ & 0.94 & 0.00 & 0.95 & 0.94 & 0.92 & 0.86 & 1.00 \\ [0.5em] 
$\widehat{se}_{cost} $ & 0.02 & 0.02 & 0.30 & 0.23 & 0.17 & 0.04 & 0.49 \\ 
$\widehat{se}_{freq} $ & 0.04 & 0.04 & 0.51 & 0.33 & 0.22 & 0.06 & 1.72 \\ 
$\widehat{se}_{time} $ & 0.03 & 0.02 & 0.26 & 0.18 & 0.14 & 0.05 & 1.26 \\ [0.5em] 
Bias$_{cost} $ & -0.00 & 0.60 & -0.04 & -0.03 & -0.06 & -0.04 & -0.03 \\ 
Bias$_{freq} $ & 0.00 & 0.53 & -0.08 & -0.06 & -0.08 & -0.05 & -0.02 \\ 
Bias$_{time} $ & -0.00 & 0.78 & -0.01 & -0.00 & -0.03 & -0.03 & -0.02 \\ [0.5em] 
Rej. $\theta_{cost}=0$ & 1.00 & 1.00 & 0.89 & 0.91 & 0.93 & 0.98 & 1.00 \\ 
Rej. $\theta_{freq}=0$ & 1.00 & 1.00 & 0.62 & 0.75 & 0.84 & 0.96 & 0.00 \\ 
Rej. $\theta_{time}=0$ & 1.00 & 1.00 & 0.95 & 0.96 & 0.97 & 0.99 & 0.94 \\ [0.5em] 
$MSE(\Lambda)^{Train}$ & {.}  & {.}  & 8.80 & 8.83 & 8.89 & 9.22 & {.}  \\ 
$MSE(\Lambda)^{Test}$ & {.}  & {.}  & 8.87 & 8.87 & 8.90 & 9.23 & {.}  \\ 
Share Outlier & 0.00 & 0.00 & 0.00 & 0.00 & 0.00 & 0.00 & 0.00 \\ 
  \bottomrule  
 \end{tabular}

 \begin{minipage}{0.825\textwidth} %
		{   \footnotesize \begin{singlespace}
\textit{Note:}	The table reports the median of the variables $\hat{se}_i$, $BIAS_i$,  $MSE(\Lambda)^{Train}$, and $MSE(\Lambda)^{Test}$  and the average of the variables $ \theta_{i} \in \hat{CI}_{i}$, Rej. $\theta_{i}=0$, and Share Outlier, $i \in \{cost,\, freq,\, time\}$, over all Monte Carlo replicates for the conditional logit using the true specification (Oracle), the conditional logit using the three variables of interest for the estimation (Basic),  the influence function approach, using five different values of $\lambda$ for the estimation of  $\Lambda_s(\bm{w})$, and the neural network (NN), which uses robust standard erros and does not rely on the influence function approach. 
			\end{singlespace}
			\par}
	\end{minipage}
\end{table}

\newpage

\begin{figure}[H]%
	\centering
	\captionsetup{justification=centering}
	\caption{Density of Estimated $t$-Statistic of $\hat{\theta}_{cost}$ for Different Estimators and Large Data}
	\subfigure[No repeated sample splitting ($R=1$)]{\includegraphics[width=1\textwidth]{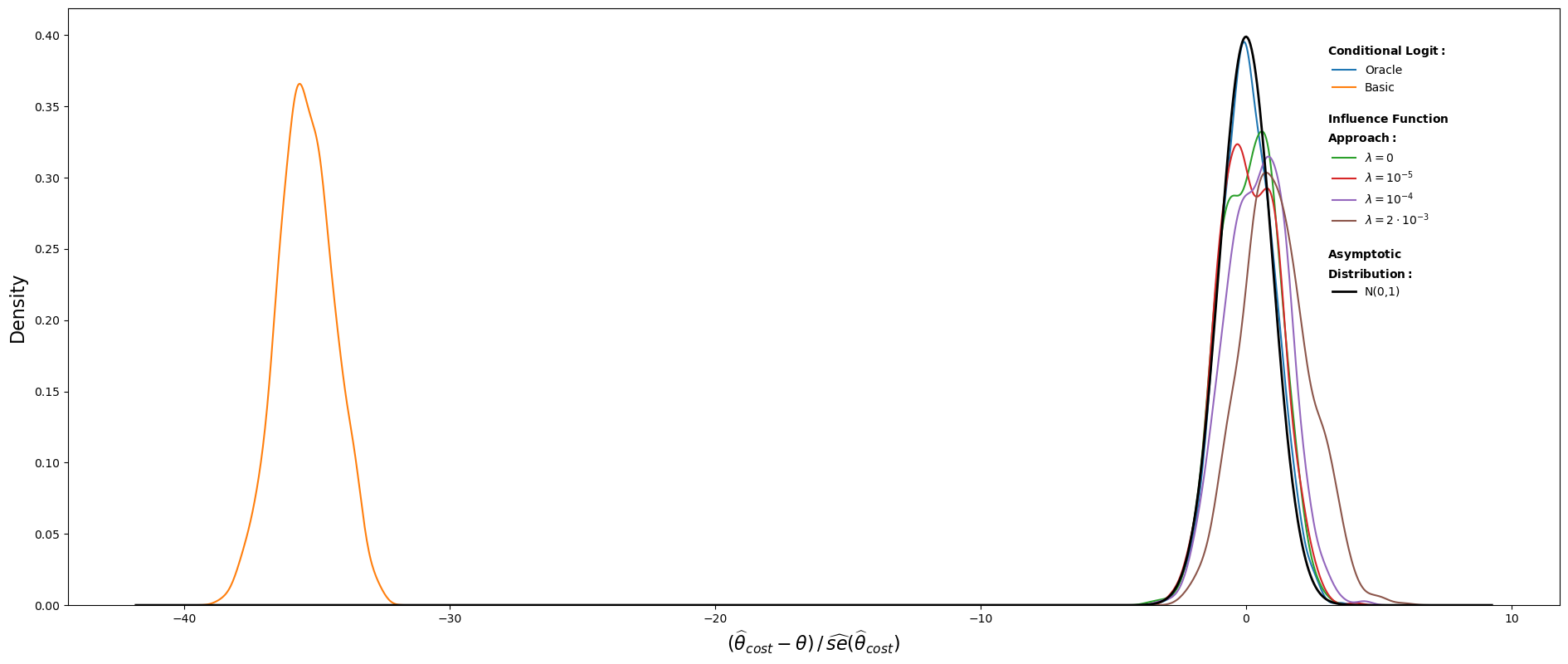}}
	\subfigure[Repeated sample splitting ($R=5$)]{\includegraphics[width=1\textwidth]{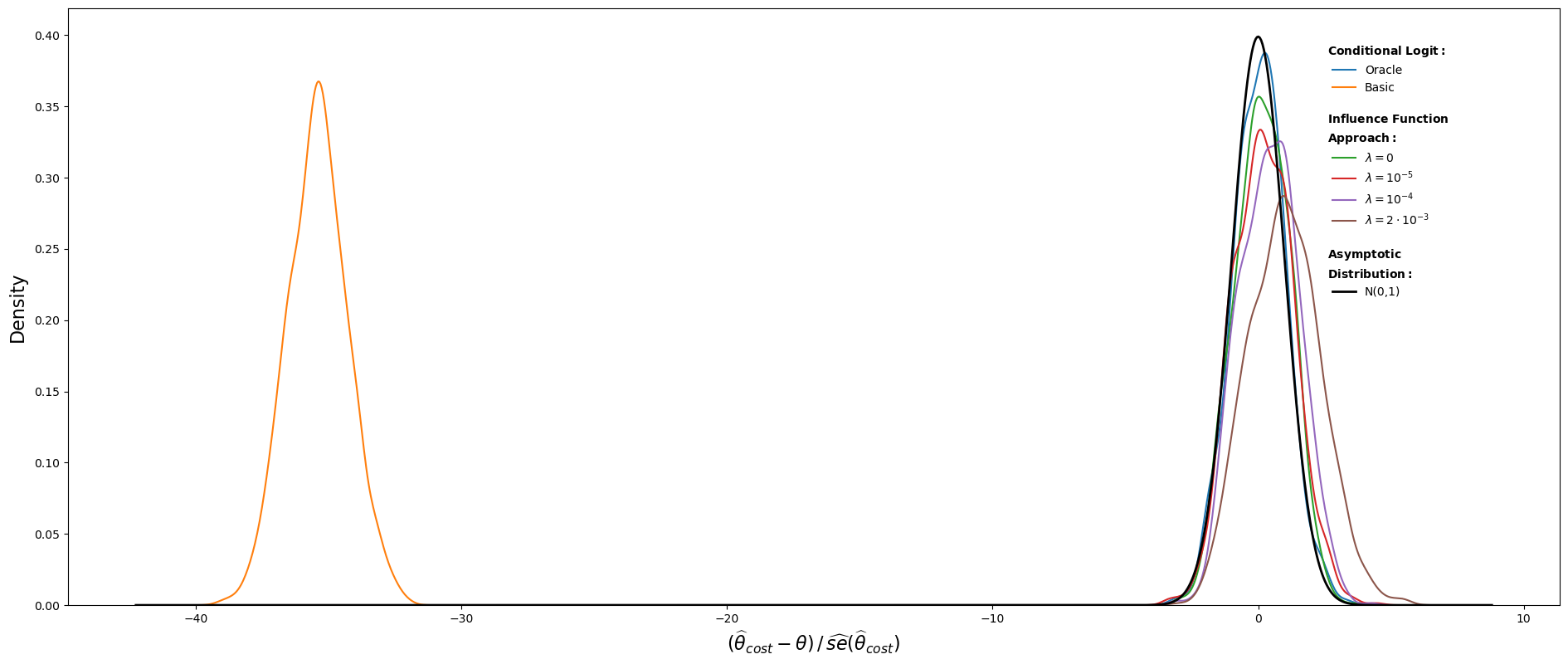}}
	\label{fig:asymptDistr}
	 \begin{minipage}{1\textwidth} %
		{   \footnotesize \begin{singlespace}
\textit{Note: The plot shows kernel density estimates of the estimated $t$-statistic for the conditional logit using the true specification (Oracle), the conditional logit using the three variables of interest for the estimation (Basic),  the influence function approach, using four different values for $\lambda$ for the estimation of  $\Lambda_s(\bm{w})$. Additionally, the standard normal distribution is included.}
			\end{singlespace}
			\par}
				\end{minipage}
\end{figure}


\begin{figure}[H]%
	\centering
	\captionsetup{justification=centering}
	\centering
	\captionsetup{justification=centering}
	\caption{Histograms of  Estimated Coefficient Functions for Influence Function Approach  and Neural Network}
	\subfigure[Cost (IFA)]{\includegraphics[width=0.49\textwidth, height=0.13\textheight]{img/TablesApplication_SwissMetro/Hist_cost_influ_Medium}} 
		\subfigure[Cost (NN)]{\includegraphics[width=0.49\textwidth, height=0.13\textheight]{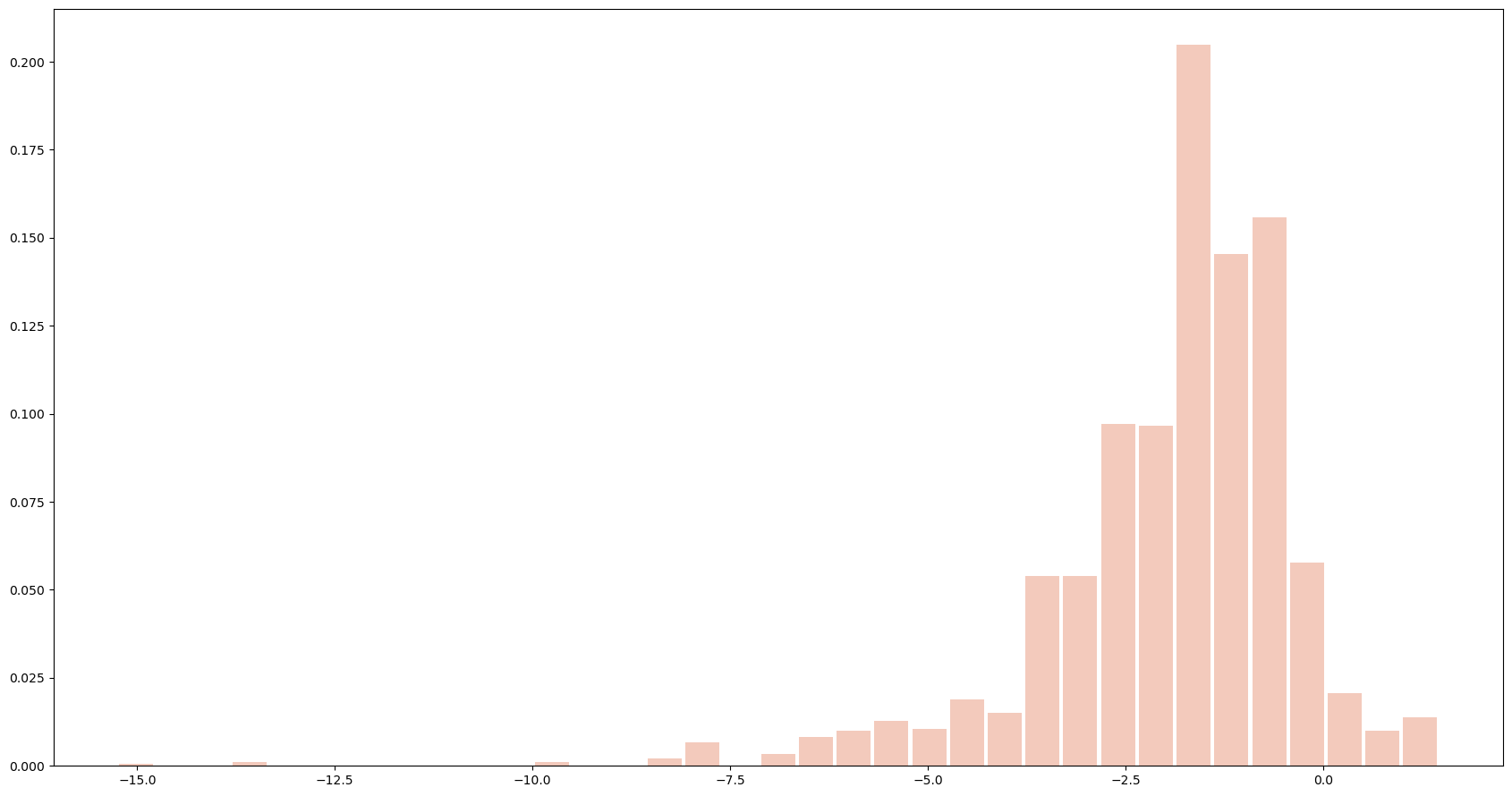}} 
	\subfigure[Frequency (IFA)]{\includegraphics[width=0.49\textwidth, height=0.13\textheight]{img/TablesApplication_SwissMetro/Hist_freq_influ_Medium}} 
		\subfigure[Frequency (NN)]{\includegraphics[width=0.49\textwidth, height=0.13\textheight]{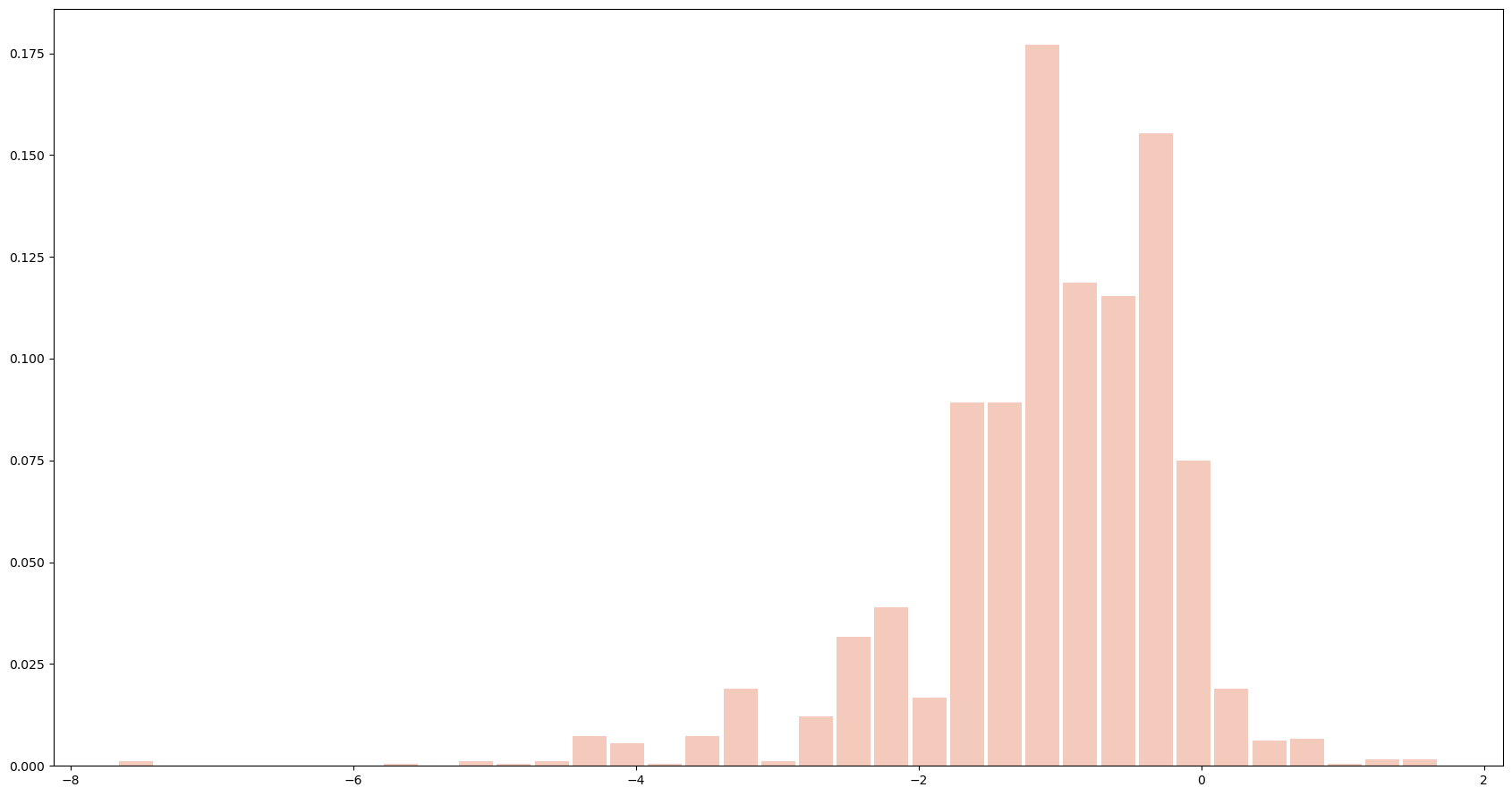}} 
	\subfigure[Travel Time (IFA)]{\includegraphics[width=0.49\textwidth, height=0.13\textheight]{img/TablesApplication_SwissMetro/Hist_traveltime_influ_Medium}} 
		\subfigure[Travel Time (NN)]{\includegraphics[width=0.49\textwidth, height=0.13\textheight]{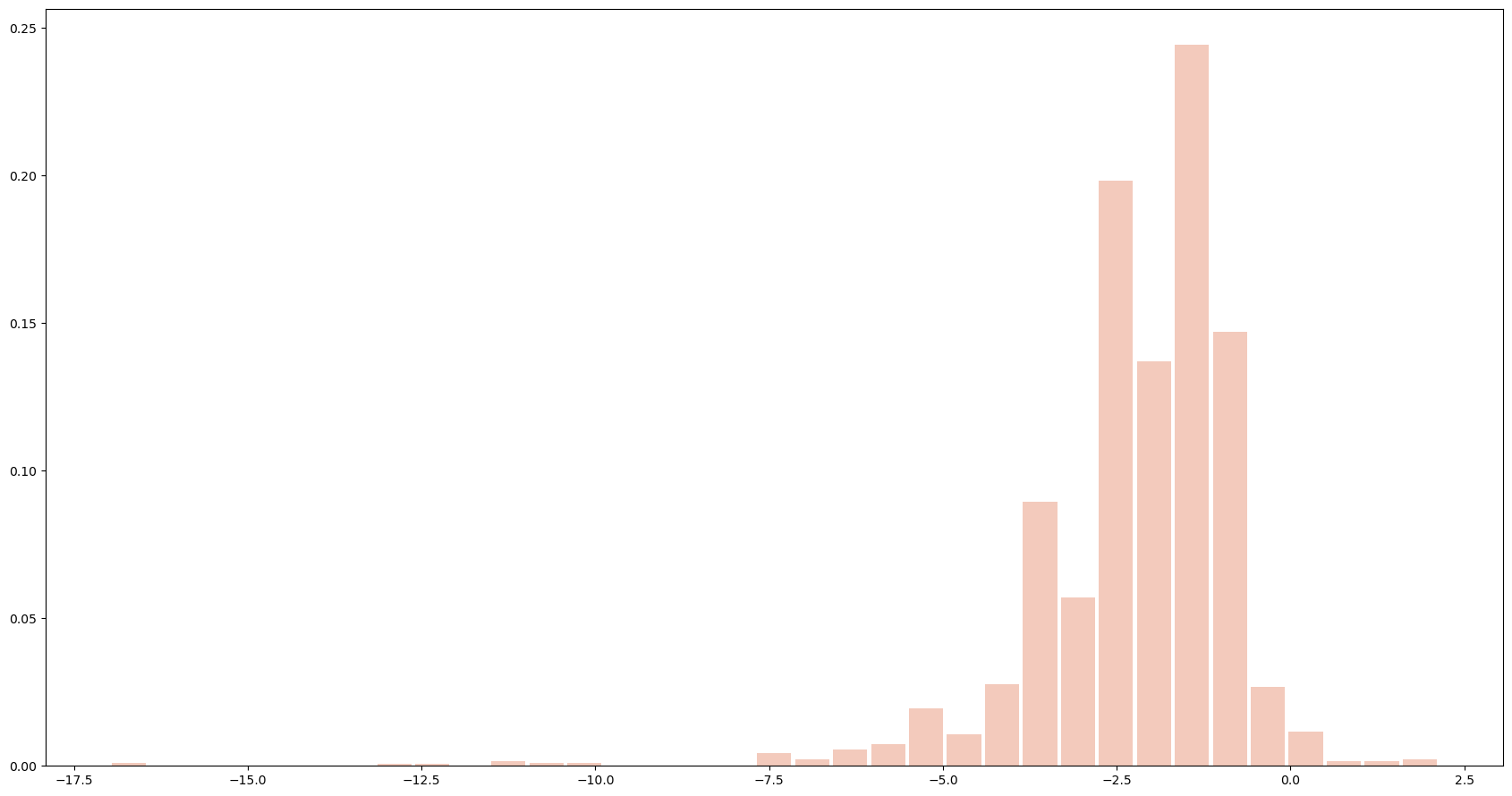}} 
			\subfigure[Intercept Train (IFA)]{\includegraphics[width=0.49\textwidth, height=0.13\textheight]{img/TablesApplication_SwissMetro/Hist_alpha_Train_influ_Medium}} 
		\subfigure[Intercept Train (NN)]{\includegraphics[width=0.49\textwidth, height=0.13\textheight]{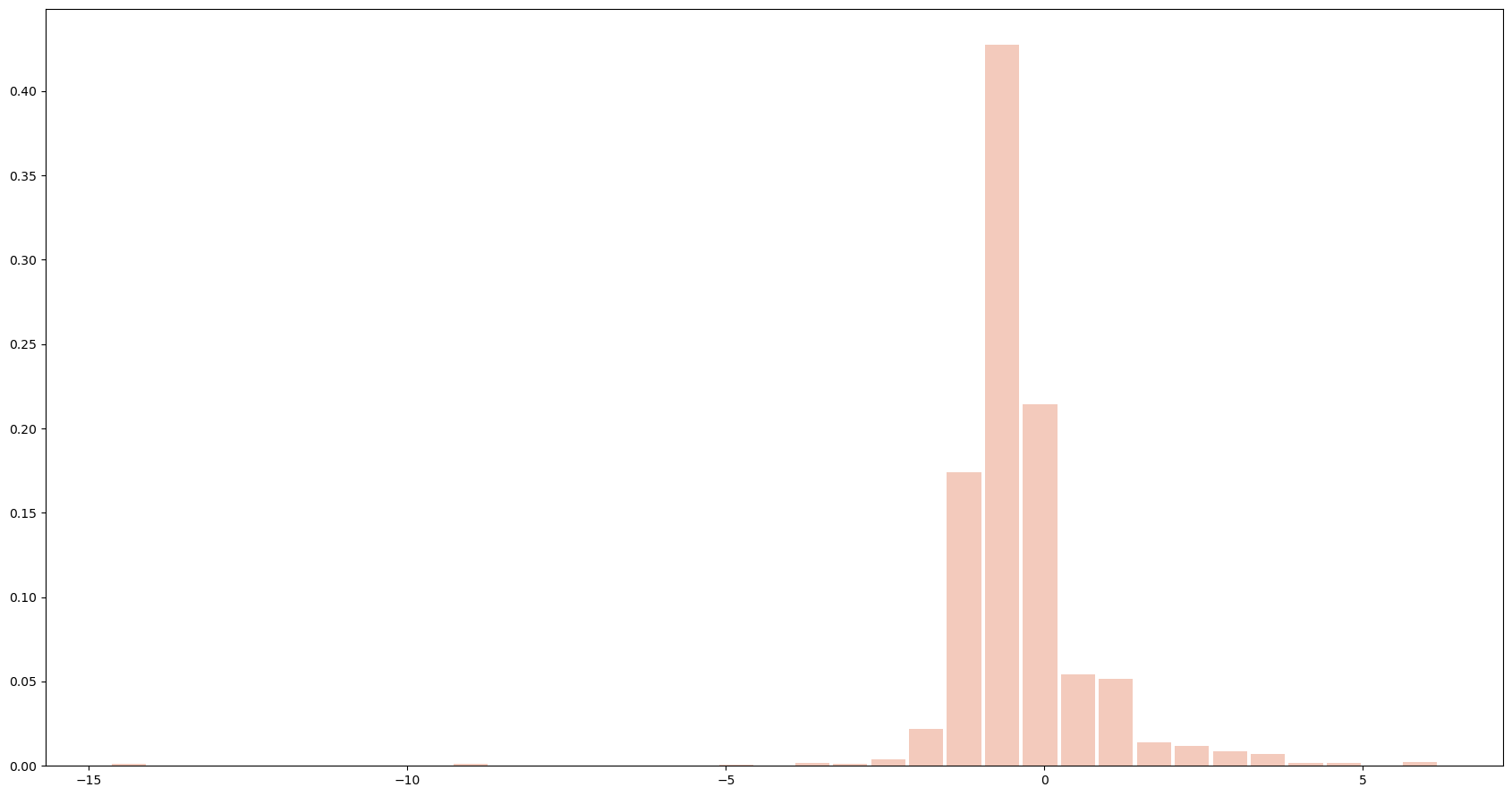}} 
					\subfigure[Intercept SM (IFA)]{\includegraphics[width=0.49\textwidth, height=0.13\textheight]{img/TablesApplication_SwissMetro/Hist_alpha_SM_influ_Medium}} 
		\subfigure[Intercept SM (NN)]{\includegraphics[width=0.49\textwidth, height=0.13\textheight]{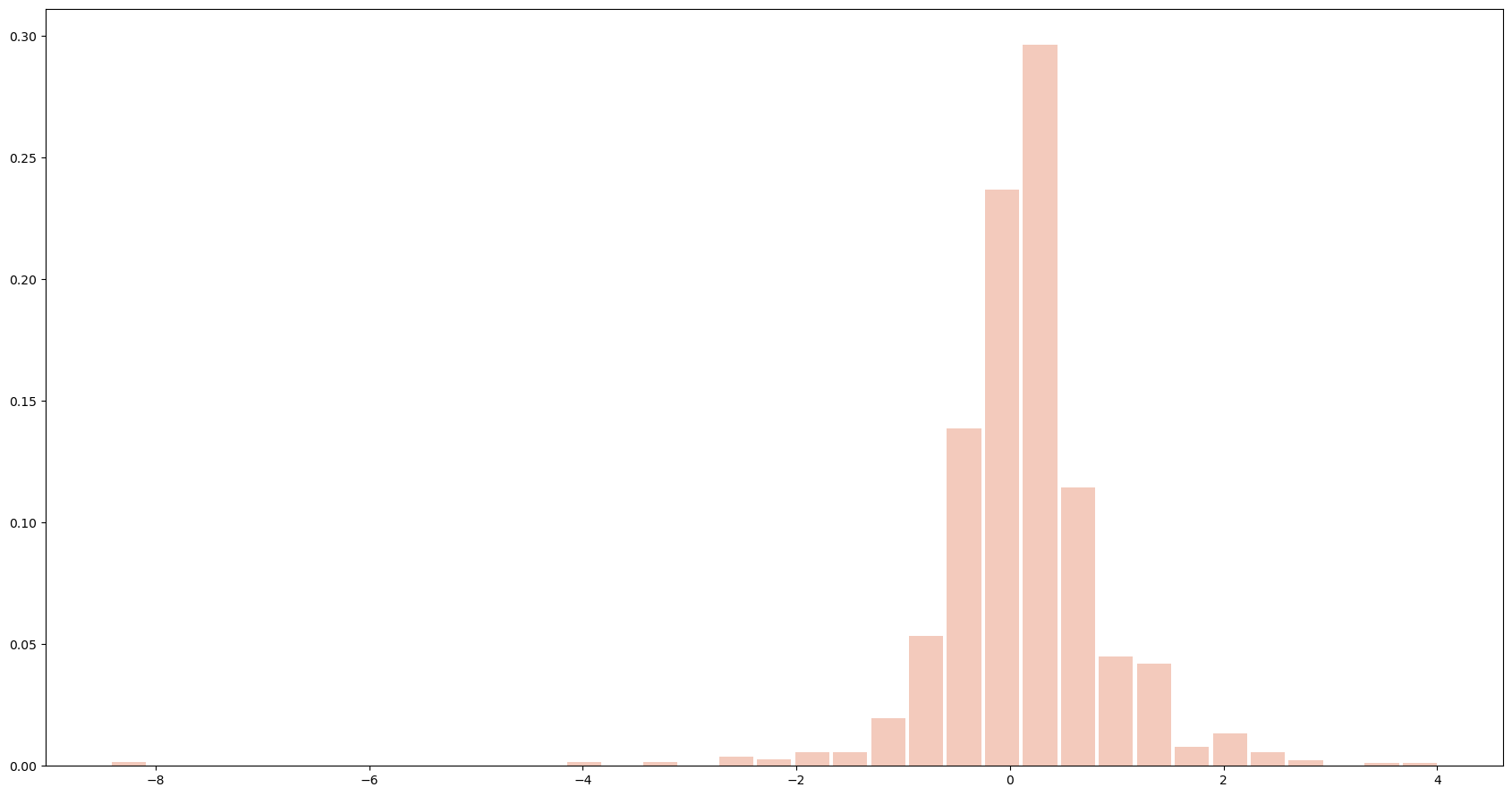}} 
	\label{figDL:HistsLargeSM_Appendix}
	 \begin{minipage}{1\textwidth} %
		{   \footnotesize \begin{singlespace}
\textit{Note:} The orange bars represent the heterogeneous coefficients in the test set predicted with the neural network (NN) model (without the influence function approach), and the blue bars the heterogeneous coefficients in the test set predicted with the influence function approach (IFA) with repeated sample splitting with $R=5$.
			\end{singlespace}
			\par}
				\end{minipage}
\end{figure}

\begin{table}[!htbp] \centering 
  \caption{} 
  \label{} 
    \resizebox{0.95\textwidth}{!}{
\begin{tabular}{@{\extracolsep{5pt}}lcccc} 
\\[-1.8ex]\hline 
\hline \\[-1.8ex] 
\\[-1.8ex] & \multicolumn{2}{c}{Conditional Logit:}& \multicolumn{2}{c}{Nested Logit:} \\ 
\cline{2-3}\cline{4-5} 
\\[-1.8ex] & (1) & (2) & (1) & (2)\\ 
\hline \\[-1.8ex] 
 Const SM & 1.248$^{***}$ (0.183) & 0.721$^{*}$ (0.424) & 1.490$^{***}$ (0.225) & 1.109$^{**}$ (0.548) \\ 
  Const Train & $-$1.110$^{***}$ (0.417) & $-$0.191 (0.748) & $-$1.013$^{**}$ (0.411) & $-$0.007 (0.971) \\ 
  Cost & $-$0.878$^{***}$ (0.042) & $-$0.984$^{**}$ (0.429) & $-$0.976$^{***}$ (0.046) & $-$1.258$^{**}$ (0.500) \\ 
  Freq & $-$0.735$^{***}$ (0.115) & $-$2.307$^{*}$ (1.190) & $-$0.778$^{***}$ (0.122) & $-$2.603 (1.918) \\ 
  Time & $-$1.216$^{***}$ (0.051) & $-$2.710$^{***}$ (0.586) & $-$1.449$^{***}$ (0.048) & $-$3.138$^{***}$ (0.657) \\ 
  Age$_\text{sm}$ & $-$0.234$^{***}$ (0.030) & $-$0.198$^{***}$ (0.045) & $-$0.262$^{***}$ (0.036) & $-$0.262$^{***}$ (0.059) \\ 
  AGE$_\text{train}$ & 0.040 (0.047) & 0.020 (0.087) & 0.035 (0.046) & 0.003 (0.088) \\ 
  Income$_\text{sm}$ & 0.015 (0.030) & $-$0.009 (0.043) & 0.036 (0.034) & 0.006 (0.051) \\ 
  Income$_\text{train}$ & $-$0.279$^{***}$ (0.041) & $-$0.150$^{*}$ (0.079) & $-$0.288$^{***}$ (0.043) & $-$0.164$^{*}$ (0.087) \\ 
  Who1$_\text{sm}$ & $-$0.347$^{**}$ (0.161) & $-$0.012 (0.408) & $-$0.430$^{**}$ (0.197) & $-$0.198 (0.530) \\ 
  Who1$_\text{train}$ & 1.305$^{***}$ (0.390) & 0.029 (0.714) & 1.317$^{***}$ (0.402) & $-$0.030 (0.957) \\ 
  Who2$_\text{sm}$ & 0.047 (0.166) & 0.497 (0.415) & 0.024 (0.200) & 0.448 (0.536) \\ 
  Who2$_\text{train}$ & 1.160$^{***}$ (0.398) & 0.080 (0.730) & 1.175$^{***}$ (0.411) & 0.062 (0.971) \\ 
  Who3$_\text{sm}$ & $-$0.072 (0.181) & 0.904$^{**}$ (0.426) & $-$0.128 (0.214) & 0.875 (0.547) \\ 
  Who3$_\text{train}$ & 1.199$^{***}$ (0.418) & $-$0.437 (0.762) & 1.184$^{***}$ (0.431) & $-$0.644 (0.998) \\ 
  Male$_\text{sm}$ & $-$0.322$^{***}$ (0.077) & $-$0.302$^{***}$ (0.111) & $-$0.327$^{***}$ (0.084) & $-$0.354$^{***}$ (0.137) \\ 
  Male$_\text{train}$ & $-$0.428$^{***}$ (0.115) & $-$0.206 (0.213) & $-$0.423$^{***}$ (0.114) & $-$0.133 (0.219) \\ 
  Luggage$_\text{sm}$ & 0.132$^{**}$ (0.052) & 0.211$^{***}$ (0.076) & 0.129$^{**}$ (0.058) & 0.214$^{**}$ (0.102) \\ 
  Luggage$_\text{train}$ & 0.541$^{***}$ (0.079) & 0.350$^{**}$ (0.144) & 0.562$^{***}$ (0.088) & 0.346$^{**}$ (0.165) \\ 
  Cost*Age &  & $-$0.429$^{***}$ (0.050) &  & $-$0.531$^{***}$ (0.047) \\ 
  Freq*Age &  & 0.088 (0.113) &  & 0.089 (0.115) \\ 
  Time*Age &  & $-$0.065 (0.055) &  & $-$0.127$^{**}$ (0.050) \\ 
  Cost*Income &  & 0.098$^{**}$ (0.042) &  & 0.098$^{*}$ (0.052) \\ 
  Freq*Income &  & $-$0.153 (0.098) &  & $-$0.134 (0.109) \\ 
  Time*Income &  & $-$0.085 (0.054) &  & $-$0.116$^{*}$ (0.070) \\ 
  Cost*Who1 &  & 1.018$^{**}$ (0.419) &  & 1.362$^{***}$ (0.494) \\ 
  Freq*Who1 &  & 1.747 (1.154) &  & 1.961 (1.905) \\ 
  Time*Who1 &  & 1.739$^{***}$ (0.568) &  & 2.156$^{***}$ (0.661) \\ 
  Cost*Who2 &  & 1.028$^{**}$ (0.420) &  & 1.327$^{***}$ (0.495) \\ 
  Freq*Who2 &  & 1.543 (1.171) &  & 1.740 (1.916) \\ 
  Time*Who2 &  & 1.768$^{***}$ (0.574) &  & 2.141$^{***}$ (0.671) \\ 
  Cost*Who3 &  & 1.234$^{***}$ (0.433) &  & 1.582$^{***}$ (0.519) \\ 
  Freq*Who3 &  & 1.779 (1.208) &  & 2.060 (1.939) \\ 
  Time*Who3 &  & 3.099$^{***}$ (0.574) &  & 3.837$^{***}$ (0.673) \\ 
  Cost*MALE &  & $-$0.536$^{***}$ (0.097) &  & $-$0.644$^{***}$ (0.119) \\ 
  Freq*Male &  & $-$0.053 (0.277) &  & $-$0.124 (0.285) \\ 
  Time*Male &  & $-$0.394$^{***}$ (0.139) &  & $-$0.601$^{***}$ (0.156) \\ 
  Cost*Luggage &  & 0.399$^{***}$ (0.075) &  & 0.525$^{***}$ (0.096) \\ 
  Freq*Luggage &  & 0.081 (0.189) &  & 0.095 (0.230) \\ 
  Time*Luggage &  & 0.459$^{***}$ (0.093) &  & 0.611$^{***}$ (0.121) \\ 
  iv:train &  &  & 0.805$^{***}$ (0.039) & 0.738$^{***}$ (0.048) \\ 
  iv:car &  &  & 0.872$^{***}$ (0.039) & 0.761$^{***}$ (0.048) \\ 
 \hline \\[-1.8ex] 
Observations & 7,234 & 7,234 & 7,234 & 7,234 \\ 
R$^{2}$ & 0.123 & 0.148 & 0.124 & 0.150 \\ 
Log Likelihood & $-$5,683.250 & $-$5,520.814 & $-$5,676.610 & $-$5,512.196 \\ 
LR Test & 1,599.627$^{***}$ (df = 19) & 1,924.500$^{***}$ (df = 40) & 1,612.908$^{***}$ (df = 21) & 1,941.736$^{***}$ (df = 42) \\ 
\hline 
\hline \\[-1.8ex] 
\textit{Note:}  & \multicolumn{4}{r}{$^{*}$p$<$0.1; $^{**}$p$<$0.05; $^{***}$p$<$0.01} \\ 
\end{tabular} 
}
\end{table}
\end{appendix}

\end{document}